\documentclass[sigconf]{acmart}

\newcommand{\AUTHORS}{Y, L, H}
\newcommand{\TITLE}{Locality-Sensitive Sketching for Resilient Network Flow Monitoring}
\newcommand{\KEYWORDS}{sketch; network measurement; non-intrusive; pub-sub}
\newcommand{\CONFERENCE}{}
\newcommand{\PAGENUMBERS}{yes}       
\newcommand{\COLOR}{yes}
\newcommand{\showComments}{no}

\long\def\IGNORE#1{}
\newcommand{\beq}{\begin{equation}}
\newcommand{\eeq}{\end{equation}}
\newcommand{\benq}{\begin{eqnarray}}
\newcommand{\eenq}{\end{eqnarray}}
\newcommand{\co}[1]{}
\newcommand{\subsubsubsection}[1]{\textbf{#1}}

\usepackage[T1]{fontenc}
\usepackage[utf8]{inputenc}
\usepackage{bm}                        
\usepackage{courier}

\usepackage{ifthen}

\usepackage{amsmath}
\usepackage{amsthm}
\usepackage{epsfig}
\usepackage{times}
\usepackage{subfigure}
\usepackage{multirow}
\usepackage{url}

\usepackage[ruled,boxed,vlined]{algorithm2e} 

\usepackage{color}

\usepackage{enumitem}
\setlist{itemsep=0pt,parsep=0pt}             

\ifthenelse{\equal{\PAGENUMBERS}{yes}}{%
  \pagestyle{plain}
}{%
  \pagestyle{empty}
}

\newtheorem{theorem}{Theorem}

\usepackage{booktabs}
\usepackage{color}
\usepackage{colortbl}
\usepackage{float}                           

\ifthenelse{\equal{\COLOR}{yes}}{%
  \usepackage{hyperref}
}{%
  \usepackage{hyperref}
}
\usepackage{url}

\hypersetup{%
pdfauthor = {\AUTHORS},
pdftitle = {\TITLE},
pdfsubject = {\CONFERENCE},
pdfkeywords = {\KEYWORDS},
bookmarksopen = {true}
}

\definecolor{placeholderbg}{rgb}{0.85,0.85,0.85}

\usepackage{soul}
\soulregister\cite7
\soulregister\ref7
\soulregister\pageref7

\usepackage[absolute]{textpos}

\newfloat{acmcr}{b}{acmcr}

\newcommand{\note}[2]{
    \ifthenelse{\equal{\showComments}{yes}}{\textcolor{#1}{#2}}{}
}

\date{}
\title{{\TITLE}}
\author{Yongquan Fu, 
              Dongsheng Li,
              Siqi Shen, 
              Yiming Zhang, 
              Kai Chen}

\begin{document}



\ifthenelse{\equal{\PAGENUMBERS}{yes}}{%
  \thispagestyle{fancy}
}{%
  \thispagestyle{empty}
}

\authornote{
              Yongquan Fu (yongquanf@nudt.edu.cn), 
              Dongsheng Li (dsli@nudt.edu.cn),
              Siqi Shen (siqishen@gmail.com), 
              Yiming Zhang (sdiris@gmail.com) are with  Science and Technology Laboratory of Parallel
              and Distributed Processing, College of Computer Science,
              National University of Defense Technology,  
             Kai Chen (kaichen@cse.ust.hk) is with  SING Lab, Hong Kong University of Science and
              Technology
              }

\begin{abstract}
Network monitoring is vital in modern clouds and data center networks for traffic engineering, network diagnosis, network intrusion detection, which need diverse traffic statistics ranging from flow size distributions to heavy hitters. To cope with increasing network rates and massive traffic volumes, sketch based approximate measurement has been extensively studied to trade the accuracy for memory and computation cost, which unfortunately, is sensitive to hash collisions.  In addition, deploying the sketch involves fine-grained performance control and  instrumentation.   

 This paper presents a locality-sensitive sketch (LSS) to be resilient to hash collisions. LSS proactively minimizes the estimation error due to hash collisions with  an autoencoder based optimization model, and reduces the estimation variance by keeping similar network flows to the same bucket array. To illustrate the feasibility of the sketch, we develop a disaggregated monitoring application that supports  non-intrusive sketching deployment and native network-wide analysis.  Testbed shows that the framework adapts to line rates and provides  accurate query results. Real-world trace-driven simulations show that LSS remains stable performance under wide ranges of parameters and dramatically outperforms state-of-the-art sketching structures, with over $10^3$ to  $10^5$ times reduction in relative errors for per-flow queries as the ratio of the  number of buckets to the number of network flows reduces from 10\% to 0.1\%.  
 \end{abstract}
 

\maketitle

\section{Introduction}

Network measurement is  of paramount importance for traffic engineering, network diagnosis, network forensics, intrusion detection and prevention in clouds and data centers, which need a variety of traffic measurement, such as flow size estimation,  flow distribution, heavy hitters. Recently, the self-running network proposal \cite{DBLP:journals/corr/abs-1710-11583} highlights an automatic management loop for  large-scale networks with  timely and accurate data-driven network statistics as the driving force for machine learning techniques. 

Network-flow monitoring is challenging due to ever increasing  line rates, massive traffic volumes, and large numbers of active flows.  Traffic statistics tasks require advanced data structures and traffic statistical algorithms \cite{DBLP:conf/globecom/ChoiPZ04,DBLP:conf/sigcomm/MoshrefYGV14,DBLP:conf/sigcomm/GuoYXDHMLWPCLK15}. Many space- and time-efficient approaches \cite{DBLP:journals/tocs/EstanV03,DBLP:conf/imc/KrishnamurthySZC03,DBLP:conf/vldb/MankuM02,DBLP:conf/imc/LiBCDGIL06,DBLP:conf/sigmetrics/LuMPDK08,DBLP:conf/sigmetrics/KumarSXW04,DBLP:conf/icdt/MetwallyAA05,DBLP:conf/imc/SchwellerGPC04,DBLP:conf/sigmetrics/LallSOXZ06,DBLP:journals/ton/YoonLCP11,DBLP:conf/sigcomm/LiuMVSB16,DBLP:journals/cacm/Cormode17,DBLP:conf/sigcomm/Ben-BasatEFLW17,DBLP:conf/sigcomm/HuangJLLTCZ17,DBLP:conf/sosr/SivaramanNRMR17} have been studied, e.g., traffic sampling, traffic counting, traffic sketching.   Compared to other approaches, the sketch  has received extensive attentions due to their competitive trade off between space resource consumption and query efficiency. Further, multiple sketch structures can be composed for joint traffic analytics.
 
Generally, a sketch builds a dimensional-reduction representation  to approximately capture traffic counters. Its physical structure  is a memory-efficient and constant-speed bucket array to accumulate incoming flow counters.  Existing sketch structures  \cite{DBLP:conf/icalp/CharikarCF02,DBLP:conf/latin/CormodeM04,DBLP:journals/cacm/CormodeH09} hash incoming packets to randomly chosen buckets and take the accumulated counter in these buckets as the estimator. Recently, OpenSketch \cite{DBLP:conf/nsdi/YuJM13}, UnivMon \cite{DBLP:conf/sigcomm/LiuMVSB16}, SketchVisor \cite{DBLP:conf/sigcomm/HuangJLLTCZ17}, ElasticSketch \cite{DBLP:conf/sigcomm/0003JLHGZMLU18}, and  SketchLearn \cite{DBLP:conf/sigcomm/HuangLB18} further extend the generality of the sketch structure to support diverse monitoring tasks. 

 The sketch based monitoring approach faces two weaknesses. First, the estimation is sensitive to hash collisions, i.e., multiple keys are mapped to the same bucket, as this noisy bucket  no longer returns exact results for any  of inserted keys. Existing approaches typically aggregate multiple independent bucket arrays in order to relieve the degree of hash collisions. However, as we show in Figure \ref{MotiEq}, existing sketch structures such as count-sketch CS \cite{DBLP:conf/icalp/CharikarCF02} and count-min sketch CM \cite{DBLP:conf/latin/CormodeM04} are sensitive to hash collisions, where noisy estimators become the majority as the sketch becomes more compressive with respect to the number of inserted keys. Recently, ElasticSketch \cite{DBLP:conf/sigcomm/0003JLHGZMLU18} and SketchLearn \cite{DBLP:conf/sigcomm/HuangLB18} track large flows with a  hash table and  separate large flows from the sketch structure. Unfortunately, the hash table needs to  allocate dedicated space for new items, thus it is less efficient than the sketch with a constant-size bucket structure. 

Second, the sketch based monitoring system needs fine-grained performance control and  instrumentation. The need of coping with line-rate packets increases the resource contentions of the sketch structure with colocated deployed systems \cite{DBLP:conf/sigcomm/HuangJLLTCZ17}. Further, modifying the sketching based monitoring applications introduces complicated debugging and performance issues.

To address the first weakness of existing sketches,  we present a new class of sketch called locality-sensitive sketch (LSS for short) that is resilient to hash collisions.  Our key observation is that: \textit{ if a noisy bucket contains similar values, then the average should approximate the original value well}. To that end, LSS approximately minimizes the estimation error based on the equivalence relationship between a sketch and a linear autoencoder model; furthermore, LSS reduces the variance of the estimation error  by clustering similar key-value pairs based on a lightweight K-means clustering process \cite{Jain:1999:DCR:331499.331504}.

We present two  optimization techniques to make LSS practical for streamed monitoring requirements. First, the clustering process should be  online to adapt the streamed flows. We exploit the temporally self-similar nature of the network traffic \cite{DBLP:journals/ton/LelandTWW94}, by  training an offline cluster model with  traffic traces  and  mapping online flow records with trained cluster centers.  Second,  the insertion process should deal with incremental flow counters, since the flow size grows as packets are delivered. We cache the flow size in a Cuckoo filter \cite{DBLP:conf/conext/FanAKM14}, and remap the flow to the nearest cluster center. 

We address the second weakness by presenting a disaggregated monitoring application in Section \ref{DSSec} that implements the LSS sketch in a non-intrusive approach and allows for native network-wide analytics. The framework decouples the line-rate packet streams from the sketching process for scalability purpose. An ingestion component  at server or middlebox splits line-rate  packet streams to flowlets \cite{DBLP:journals/ccr/KandulaKSB07,DBLP:conf/sigcomm/AlizadehEDVCFLMPYV14,DBLP:conf/sigcomm/ZhuKCGLMMYZZZ15} and aggregates real-time flowlet counters to reduce the monitoring traffic, and publishes them to a publish/subscribe framework. The flowlet-counter stream are subscribed by the sketch maintenance component that dynamically keeps the LSS sketch  in a sliding window model. Streamed LSS sketches are subscribed by the query component to perform the network-wide analysis. 



In Section \ref{EvaSec}, testbed shows that the framework adapts to line rates and provides accurate query results.  Real-world  trace-driven simulation confirms that  LSS  dramatically reduces the estimation error under the same memory footprint. LSS remains stable performance under wide ranges of parameters and dramatically outperforms state-of-the-art sketching structures, with over $10^3$ to  $10^5$ times reduction in relative errors for per-flow queries as the ratio of  the  number of buckets to the number of network flows reduces from 10\% to 0.1\%.



 \section{Motivations}
\label{BackSec}

Each flow is typically represented as a key-value pair, where where the key is defined by a combination of packet fields, e..g, the 5-tuple representation, and the value summarizes the flow's statistics, e.g., packet numbers or byte counts. Existing sketch based monitoring applications work at the packet streams.  For each incoming packet,  a sketch based monitor inspects the packet header to extract the key and calculate the packet's value,   then insert this record to the sketch data structure, which incrementally accumulates the value of the given key with one or multiple hash based bucket arrays.  To estimate the accumulated value of a key,  the monitor queries the sketch with the input key, which returns an approximate value over the shared bucket arrays for all inserted keys. We illustrate the insertion and query processes of existing sketches in details in Appendix \ref{BoundAppendix}.

A sketch based monitoring application typically comprises an \textit{ingestion}  component that intercepts incoming  packets from the physical network interface and generates  key-value input for the sketch,  a \textit{sketching} component that feeds the key-value input to a stream of  sketch structures, where each sketch keeps a fixed number of key-value pairs, and  a \textit{query}  component that transforms  monitoring tasks to query primitives on the sketch.

\subsection{Resilience}

The sketch structure should remain fairly accurate under a wide range of parameter configurations. Unfortunately,  a sketch is sensitive to hash collisions where multiple keys are mapped to the same bucket. 

We  quantify the expected number of noisy buckets that have hash collisions. We take {Count-sketch} (CS for short) \cite{DBLP:conf/icalp/CharikarCF02} and count-min sketch (CM) \cite{DBLP:conf/latin/CormodeM04} as examples, as both are probably two of the most popular sketch structures. We bound the expected number of buckets that suffer from hash collisions in Lemma \ref{noisyVal}.

 \begin{lemma}\label{noisyVal}
Assume that each key is mapped to a bucket in each bank uniformly at random.   Let $m$ denote the number of buckets, $N$ the number of unique keys.  For a sketch with $c$ banks of bucket arrays, where each bucket array is of size $\frac{m}{c}$, the expected percent of noisy buckets is $1-{ e }^{ -cN/m }-\frac { cN }{ m } \cdot { e }^{ -c\left( N-1 \right) /m }$.
 \end{lemma}
 
\begin{figure}[!t]
  \centering
        \subfigure[Noisy buckets vs. $\frac{m}{N}$.]{ \label{fig:noisy}
          \setlength{\epsfxsize}{.2\textwidth}
          \epsffile{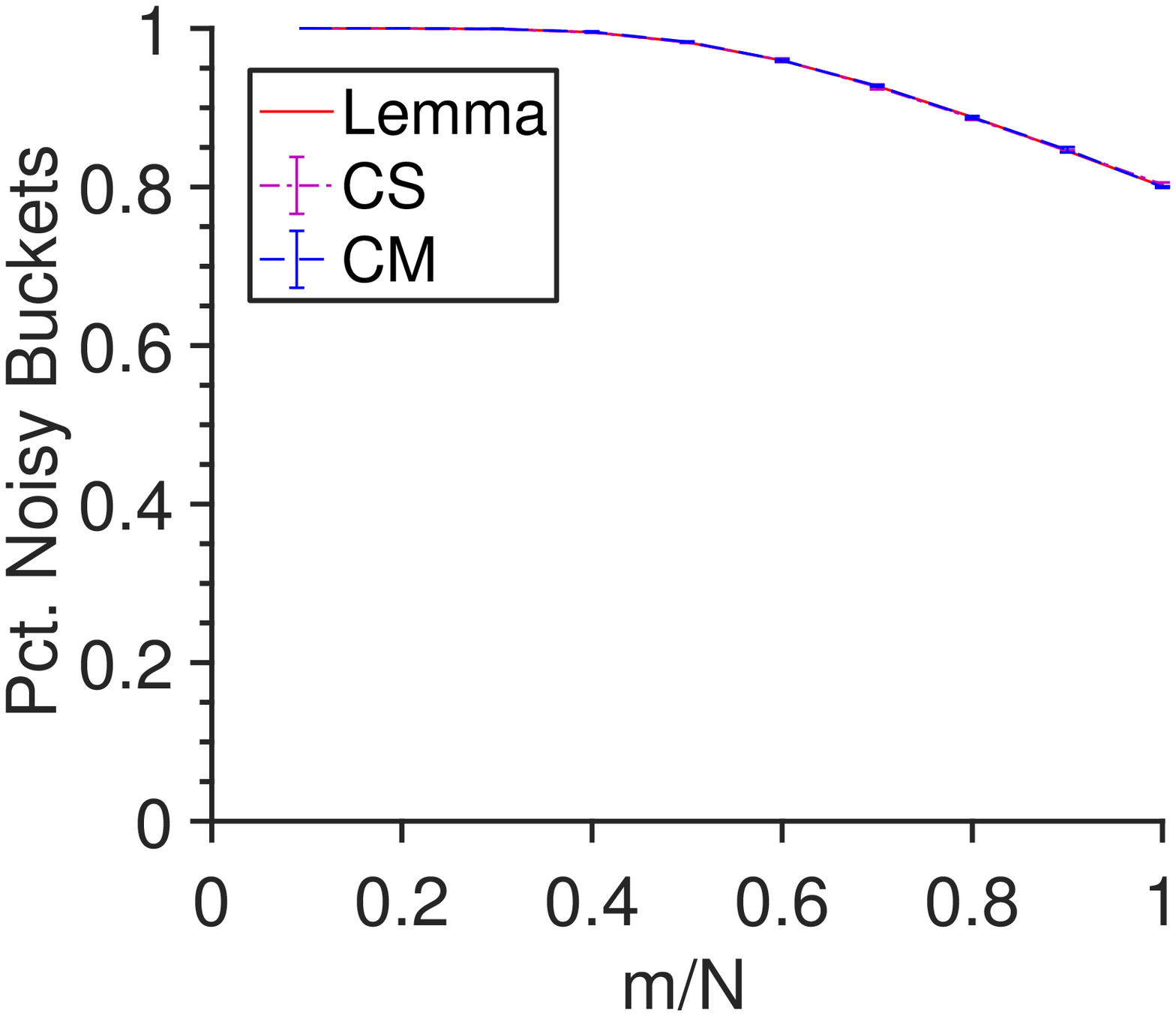}
         }
         \subfigure[Noisy estimator  vs. $\frac{m}{N}$.]{ \label{fig:noisyBuckets}
          \setlength{\epsfxsize}{.2\textwidth}
          \epsffile{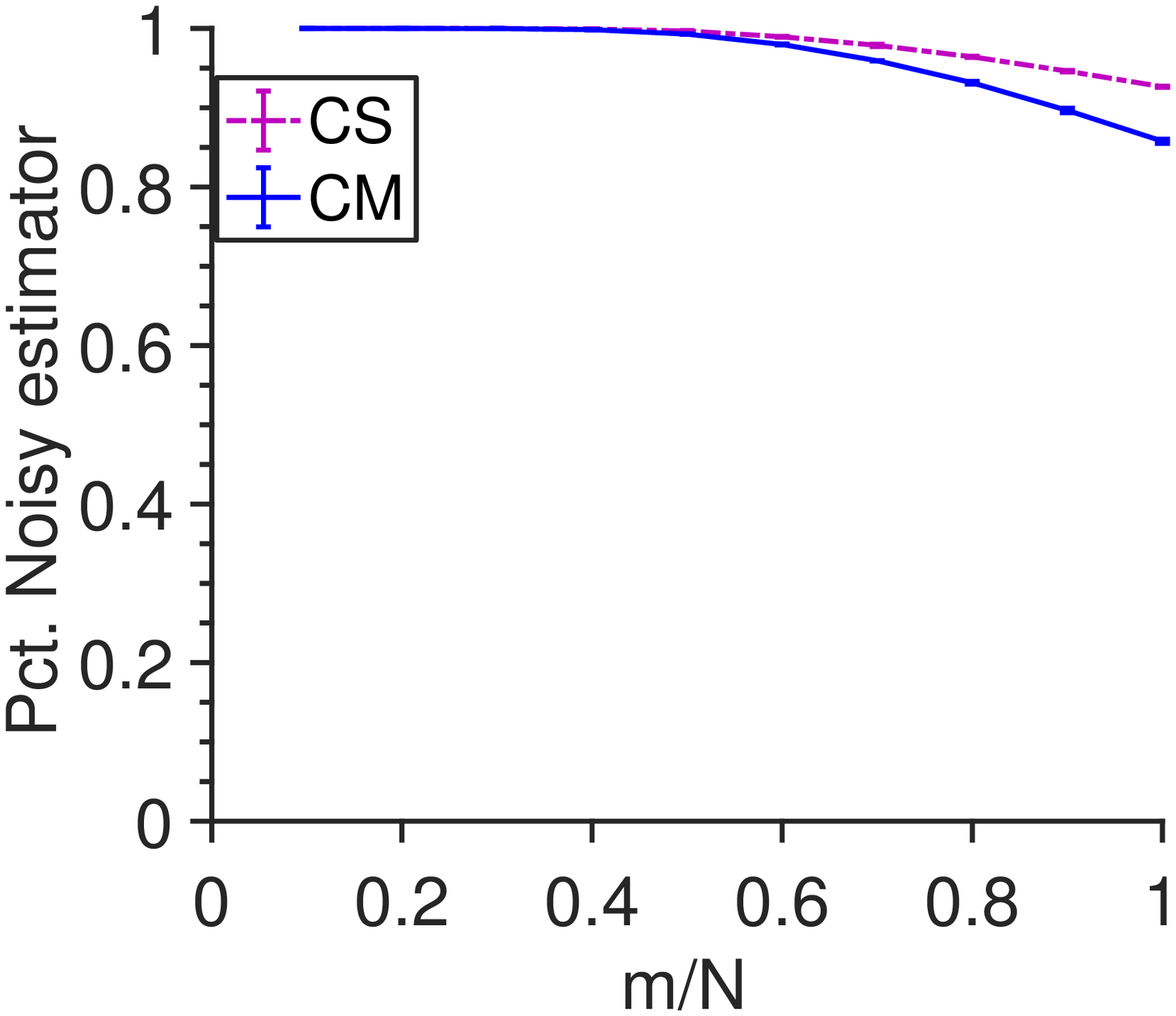}
         }        
     \caption{The  percents of noisy buckets as well as noisy estimators  for count-sketch CS  \cite{DBLP:conf/icalp/CharikarCF02} and count-min sketch CM \cite{DBLP:conf/latin/CormodeM04} as a function of the ratio of the number $m$ of buckets to the number $N$ of flows. The number $c$ of banks is set to three based on recommended parameters. We plot the theoretical expected values and the empirical values based on a CAIDA trace (the statistics is introduced in  section \ref{simSubsec}). }
     \label{MotiEq}
\end{figure}

The proof is due to the ball-bin model \cite{DBLP:journals/siamcomp/AzarBKU99} that characterizes the expectation of the number of keys per bucket. The details are put in the Appendix \ref{LemmaAppendix}.  We illustrate the hash collisions in Figure \ref{fig:noisy}, the theoretical results match with empirical hash collisions. We can see that, the probability of hash collisions increases fast with decreasing ratios between  the number of buckets and the number of unique keys. Figure \ref{fig:noisyBuckets} also confirms that hash collisions significantly degrade the effectiveness of CM and CS. 

In addition, researchers have bounded the prediction accuracy based on the accumulation sum of the inserted items (See the Appendix \ref{BoundAppendix} for a detailed introduction). Unfortunately, the performance bounds are  proportional to the accumulated sum of all items. Since network flows are typically highly skewed (refer to Figures \ref{fig:testbedTraffic} and \ref{fig:Dist.}),  the sum of all items is still orders of magnitude larger than a single item. 

\subsection{Application}

Having presented the hash-collision problem, we next  discuss the implementional challenges faced by the sketch based monitoring applications for modern network management tasks. 


\noindent (i) \textbf{Scalability challenge}. Although a sketch only produces flow-level estimation results,  existing monitoring applications feed packet-granularity streams  to the sketch. As network is getting faster from 10 Gbps to 40 Gbps and beyond, more packets must be inspected for the same amount of time, which implies that the sketch's  space and time  complexity must be tightly controlled. Given $n$ packets being in the same flow, the sketch still needs $O(n \times k)$ hash-function evaluations, where $k$ denotes the number of bucket arrays in the sketch.  

\noindent  (ii) \textbf{Fine-instrument challenge}.  For heterogeneous and multi-tenant cloud data center networks, network monitoring desires to be  non-intrusive and modular. It would maximize  the generality  under diverse environments.  Existing sketching based solutions integrates these components into the deployed platform. Developers  to perform fine-grained instrumentation to the operating system and programs, introducing complicated debugging and performance issues.  Moreover, it is difficult to modify the sketching algorithm after deployment, due to the tight coupling with the program.  However, the sketch component and the query component may undergo frequent updates, as the monitoring application has to meet diverse network management needs. Also, the network ingestion component requires developers  to perform fine-grained instrumentation. 

\subsection{High-level Overview of Our Work}

To minimize the effects of hash collisions while simultaneously keeping the simplicity of the hashing based data structures, this paper turns from passively tolerating noisy buckets to proactively recovering the noisy buckets. Suppose that all items are of the same value, we can see that the average of a bucket's accumulation returns the correct result for each of inserted items. Thus, in order to recover the noisy bucket, we need to map similar key-value pairs to the same bucket array, and recover the noisy bucket by averaging its accumulation counter. Combining these insights, we present a new class of sketch called locality-sensitive sketching or LSS for short.   LSS  has two distinct merits: (i) \textbf{Resilient to hash collisions}. LSS  averages the bucket's counter to produce the estimator that is equivalent to optimize an autoencoder framework. (ii) \textbf{Locality-sensitive to reduce the variance of the estimator}. LSS learns the cluster structure of network flows based on transferred learning from offline traces, and clusters similar network flows to the same bucket array in order to minimize the estimation variance.

\begin{figure}[!t]
  \centering
{
  {\includegraphics[width=.43\textwidth]{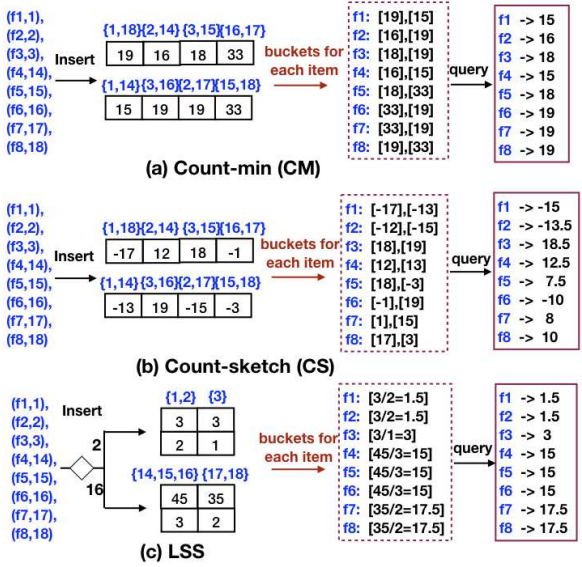}}
  }
     \caption{Sketch estimation with count-min sketch CM, count-sketch CS and our work LSS. Each sketch consists of two bucket arrays.  The leftmost column  represents the sketch after inserting a sequence of  items (items that are mapped to each bucket are listed on the top of the bucket.); the middle column  represents the value of each mapped bucket for each item (For LSS, we list the average of  the mapped bucket for each item.); the rightmost column  contrasts the original value with the estimated value for each kind of sketch.}
     \label{fig:example}
\end{figure}

\textbf{Example}:  Figure \ref{fig:example} illustrates the difference between the count-min sketch CM \cite{DBLP:conf/latin/CormodeM04}, count-sketch CS \cite{DBLP:conf/icalp/CharikarCF02} and LSS. CM and CS insert each item to a random bucket in each bucket array. While LSS clusters similar items to the same bucket array and maps each item to only one bucket. From the query result in the rightmost column, we can see that LSS significantly reduces the estimation error compared to CM and CS. This is because LSS groups similar items together to reduce the estimation variance, and averages the bucket's counter to repair the  prediction error. While CM and CS passively  tolerate hash collisions.

\textbf{Monitoring Application}: To illustrate the feasibility of LSS and overcome the deployment hurdles, we present  a disaggregated  monitoring framework in Section \ref{DSSec} that implements LSS in a  non-intrusive and modular monitoring application. The framework disaggregates the sketch components  from the ingestion components, so as to allow for smooth modifications of sketch structures and coping with the underlying physical environments.  

We reduce the monitoring traffic with a network ingestion component that splits real-time packet streams to flowlets \cite{DBLP:journals/ccr/KandulaKSB07,DBLP:conf/sigcomm/AlizadehEDVCFLMPYV14}.  We  temporally accumulate flowlet counters with a  high-performance hash table based on Trumpet \cite{DBLP:conf/sigcomm/MoshrefYGV16}. The hash table  aggregates online packets by flow identifiers and accumulates flow counters,  and reduces packet-processing delay by cache prefetching and batch processing. 
Suppose that the average packet size is 1,000 bytes,  a flowlet has 100 flows and each flow has 100 packets on average, and a  key-value pair size is 8 bytes. The traffic of a flowlet will be 100 $\times $ 100 $\times $ 1,000 = $10^7$ bytes. While the size of  100 flow counters will be 100 $\times$ 8 = 800 bytes. The monitoring traffic results in over $10^4$ times reduction in volume.

\section{Locality-sensitive Sketch}
\label{LSSThySec}

We present a new class of sketch called LSS that provides accurate approximations  in a compact space.  Table \ref{tab:notations} lists key notations.

\begin{table}
 \centering
\begin{footnotesize}
\caption{Key notations. \label{tab:notations}} {
\begin{tabular}{| l | l |} \hline
Notation & Meaning \\ \hline

$N$  & Number of unique keys \\\hline

$X$ & Key-value streams \\\hline

$\hat X$ & Estimated key-value streams  \\\hline

$A$ & Indicator matrix \\\hline

$\left\{ { C }_{ i } \right\} $ & Cluster centers  \\\hline

$I$ & Bucket array \\\hline

$k$ & Number of cluster centers  \\\hline

$m$ & Number of buckets   \\\hline

\end{tabular}}
\end{footnotesize}
\end{table}%

\subsection{Framework}
\label{AutoencoderSec}

 \subsubsection{Autoencoder based Recovery} 
 \label{AutoSubsec}
 
Having shown that hash collisions are inherent in any hashing based sketches, we next present an autoencoder guided  approach to  proactively minimizes the estimation error of noisy buckets.

We model the hash process of a sketch structure that randomly maps incoming items to a bucket array uniformly at random. Assume that a sketch consists of one bucket array for ease of analysis. Suppose that a sketch structure  randomly maps incoming items to a bucket array uniformly at random. Let $X: N\times 1$ denote the vector of the streaming key-value sequence from the network ingestion component. Let $A: N\times m$ denote the indicator matrix of mapping the vector $X$ to a bucket array $I$ of size $m  \times  1$. Let $A(i,j) = 1$ iff the $i$-th item $X_i$ is mapped to the $j$-th bucket $I_j$, and $A(i,l) = 0$ for $l \neq j, l \in \left[ 1,m \right] $.

 \begin{figure}[!t]
  \centering
{  \setlength{\epsfxsize}{.2\textwidth}
          \epsffile{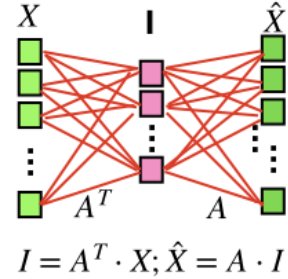}
         }
     \caption{Illustration of the autoencoder and the sketching process for Theorem \ref{acLemma}.}
     \label{autoencoderExample}
\end{figure} 

 We next show in Theorem \ref{acLemma}  that,  the sketch is mathematically equivalent to a linear autoencoder\footnote{An autoencoder \cite{DBLP:journals/pami/BengioCV13} is a neural network that takes a vector $x$ at the input, and reconstructs the input vector at the output layer. An autoencoder consists of an encoder that maps the input to a hidden layer $f = \sigma _e (W_e x + b_e)$, and a decoder that reconstructs the input as $\hat x = \sigma _d (W_d f + b_d)$, where $b_e$, $b_d$ serve as bias variables, $W_e$ and $W_d$ are weight matrices that map the input to the hidden layer, and the hidden layer to the output, respectively. Generally, an autoencoder enforces parameter sharing  $W_e = W_d^T$ (called weight tying) to avoid overfitting.} :  \textit{The insertion process corresponds to the encoding phase of the autoencoder; the query process corresponds to  the decoding phase of the autoencoder}.    

\begin{theorem}\label{acLemma}
A sketch with one bucket array is equivalent to a linear autoencoder: the insertion process corresponds to an encoding phase $I = \left( { A }^{ T }X \right)$, while the query phase corresponds to  a decoding phase $\hat { X } =A\cdot I$.
\end{theorem}

The proof is due to the algebraic transformation of the insertion and the query process of the sketch. The details are put in the Appendix \ref{ACAppendix}. Figure \ref{autoencoderExample} illustrates the mathematical equivalence between the sketching process and the autoencoder. The immediate result is that we can formulate an optimization framework for the sketch. Assume that the mapping matrix is a variable, we can formulate an optimization problem:
\begin{equation}\label{AAT}
\min _{A}  \left\| \hat { X } -X \right\| =\left\| \left( A{ A }^{ T } \right) X-X \right\| 
\end{equation} 
To derive optimized solution for Eq \eqref{AAT}, assuming that the mapping matrix is a random variable,  the  Principle Component Analysis (PCA) \cite{DBLP:journals/pami/BengioCV13}  finds a dimensional-reduction hyperplane with the smallest reconstruction error for a one hidden-layer autoencoder. Unfortunately, it will require to keep the whole stream $X$, which is infeasible for network monitoring context. Moreover, PCA calculates a dense matrix $A$, while the sketch enforces the matrix $A$ to be ultra-sparse. 
 
 Although we cannot derive the optimized mapping matrix $A$ based on the autoencoder,  \textit{we can asymptotically minimize the reconstruction error by relaxing the mapping matrices for the insertion and query processes}. Specifically, let $A$ denote the mapping matrix for the insertion matrix, let $B$ denote the mapping matrix for the query phase, we find an optimized matrix $B$ with respect to $A$ in Eq \eqref{AAT}:

\begin{equation}\label{opt1}
\min _{ B }{ \left\| X-B{ A }^{ T }X \right\|  } =\min _{ B }{ \left\| \left( I-B{ A }^{ T } \right) X \right\|  } 
\end{equation}

\begin{figure}[!t]
  \centering
{
  {\includegraphics[width=.3\textwidth]{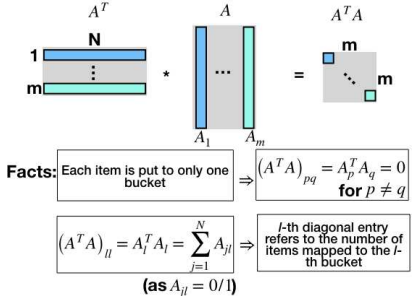}}}
     \caption{ The product ${ A }^{ T }A$ yields a diagonal matrix where non-diagonal entries are all zeros, and diagonal entries refer to the numbers of key-value pairs mapped to the corresponding buckets. }
     \label{ATA}
\end{figure}

We can derive a closed-form solution of $B$ for Eq \eqref{opt1}   as: 
\begin{equation}\label{optSolution}
B=A{ \left( { A }^{ T }A \right)  }^{ -1 } 
\end{equation}
Although the mapping matrix $A$ is still random, the product $C = { A }^{ T }A $  is a diagonal matrix where  the $i$-th diagonal entry counts the number of items mapped to the $i$-th bucket, as illustrated in Figure \ref{ATA}.

Suppose that we keep the diagonal matrix $C$  with $O(m)$ space, we obtain an approximately optimization representation: 
\begin{equation}\label{optApprox}
\hat X = A{ \left( C \right)  }^{ -1 } { A }^{ T }X 
\end{equation}
Eq \eqref{optApprox} can be formulated as an encoder-decoder phase: 
\textit{(a) encoder: $I = { A }^{ T }X$; (b) decoder: $\hat X  = A{ \left( C \right)  }^{ -1 } I$}. This formulation inspires a new class of sketch structure that proactively minimizes the estimation error, by averaging each bucket in the bucket array $I$ with the numbers of keys mapped to this bucket. 

\subsubsection{Variance Minimization}
 \label{AnalysisSec}
 
Having presented the autoencoder inspired sketch-recovery process, we next present a lightweight approach to minimize the variance of the prediction error because of  the skewed distributions of items.   Our key insight is that,  \textit{to minimize the variance, we have to group keys with similar values to the same bucket array}.  The grouping requirement belongs to an unsupervised learning problem that clusters items to minimizes the intra-cluster variance, and maximizes the inter-cluster dissimilarity.  We choose the well-studied K-means clustering method \cite{Jain:1999:DCR:331499.331504} to automatically find  cluster centers, which represents clusters with a list of cluster centers. Then, we map flow records to the nearest cluster centers. More details of the K-means model can be found in the Appendix \ref{KMAppendix}. 

Specifically, we can bound the estimation variance of the autoencoder based estimator with clustered inputs. For a bucket $j$, let the items mapped to this bucket be represented as  a set of independent and identically distributed variables: $\left\{ { X }_{ i }^{ j } \right\} $. Let $\mu$ denote the expectation of the variable $\mu =  E\left[   { X }_{ i }^{ j }  \right]  $.  Let $n_j$ denote the number of items inserted to the $j$-th bucket, let $Y_j=\frac {\sum _{ i }{ { X }_{ i }^{ j } }  }{ n_j } $ denote the average based estimator. Suppose that we have grouped items by the similarity of values,   we  assume that the difference $ \left| { X }_{ i }^{ j }-\mu  \right|$ is bounded by a positive constant $M$ for any variable ${ X }_{ i }^{ j }$.  We next bound the range between the average estimator $Y_j$ and the ground-truth value $X_i^j$ for the $j$-th bucket as follows:
  
\begin{theorem}
$Y_j$ is an unbiased estimator for any variable $ { X }_{ i }^{ j }$.    $Pr\left( \left| Y_j -\mu  \right| \ge a \right) \le \frac { { M }^{ 2 } }{ { a }^{ 2 }{ n_j }^{ 2 } } $  for a positive constant $a$. Moreover, $Pr\left( \left| Y_j-{ X }_{ i }^{ j } \right| \ge a \right)  \le \frac { { M }^{ 2 } }{ { \left( a-M \right)  }^{ 2 }{ n_j }^{ 2 } } $. 
\end{theorem}

The proof is due to the concentration bounds of the Chebyshev's inequality, whereas the details are put in the Appendix \ref{Thm3Appendix}. We can see that after clustering the input to similar groups, the average estimator  not only produces an unbiased result, but also keeps the estimator bounded with probability proportional to the squared cluster's interval $M$. 


\subsubsection{Sketch Membership}
\label{cuckooIntro}

A sketch structure does not keep the key-value membership itself. However, querying a non-existing key is meaningless, thus in practice,  a sketch is usually combined with a  membership-representation data structure, e.g., Bloom filter \cite{DBLP:conf/sigcomm/SongDTL05}, d-left hash table \cite{DBLP:conf/sigcomm/BonomiMPSV06} or a Cuckoo Filter \cite{DBLP:conf/conext/FanAKM14}. The cuckoo hash table that  inserts items based on cuckoo hashing,  is shown to be more efficient than the Bloom filter at low false positives \cite{DBLP:conf/conext/FanAKM14,DBLP:conf/conext/ZhouFLKA13,DBLP:conf/sigmetrics/DaiZLWL16}.  Thus we keep the membership with a  cuckoo hash table\footnote{Briefly, the Cuckoo Filter inserts items based on cuckoo hashing, which uses multiple hash functions to map each item to candidate buckets. For an incoming item, if one of candidate buckets has empty slots, then we calculate the digest of this item and put the digest to one empty slot; otherwise, we pick one nonempty slot and displace its digest to its alternative candidate bucket,  then we put the new item to this slot. The displaced digest may further ``kick out'' other digests until no displacements of existing digests, or reaching a maximum number of  displacements.} that  supports efficient insertion and deletion of items. We set the number of hash functions to two and the number of slots per bucket to four in order to fit each bucket to a cache line (denoted as a (2,4) filter) \cite{DBLP:conf/conext/FanAKM14,DBLP:conf/conext/ZhouFLKA13,DBLP:conf/sigmetrics/DaiZLWL16}. For a f-bit digest, the upper bound of the false positive rate of an item is approximately $\frac{4*2}{2^{f}}$. We choose  a 16-bit fingerprint with a  false positive rate  at $0.012\%$, which practically provides nearly-exact query. 

\subsection{LSS Structure and Operations}

For ease of presentation, we present the basic idea of the LSS  with simplifications.  We  assume that the input is represented as a list of unique key-value pairs, thus no duplication exists for any pair of keys. In the next subsection, we propose practice approaches to  deal with duplicated items.

An  LSS is organized as a number $k$ of bucket arrays. A bucket array consists of a number of buckets, where each bucket has two fields: (i) A ValSum field that records the sum of values; (ii) A KeyCount filed that records the number of unique keys inserted to this bucket.  Each bucket array corresponds to a cluster of similar items. We represent the clusters with $k$ cluster centers. LSS maps each item to only one bucket array that corresponds to the nearest cluster center for this item. 

As shown in Figure \ref{fig:example}~(c), for each incoming key-value item, we  select the nearest cluster center with respect to the value,  choose the corresponding bucket array, and insert the key-value item to a bucket indexed by the hash of the key. The bucket's ValSum counter is incremented by the incoming value, and the KeyCount increments by one iff the key is a new one. 

\co{Figure \ref{fig:ExampleLSS} plots an illustration of  LSS data structure. Given an incoming key-value pair $(f5,60)$, we compare  $f5$'s value  with two  cluster centers and select the second bucket array denoted as array-2, since  $f5$'s value  is closer to 80. Then, we map $f5$ to the second  bucket, and increment the bucket from $(83,1)$ to $(143,2)$. The average of this bucket is $\frac{143}{2} = 71.5$, which approximates the value of $f5$ with a relative error 0.19.  


 \begin{figure}[!t]
  \centering
{
  {\includegraphics[width=.4\textwidth]{e1.eps}}} 
     \caption{Insert a key-value pair $(f5,60)$ to an LSS instance.  }
     \label{fig:ExampleLSS}
\end{figure}
}

\co{Let the cluster centers be 15 and 80, respectively. Assume that we have inserted five key-value pairs $(f0,83)$, $(f1,78)$, $(f2, 10)$, $(f3,18)$, $(f4,17)$ to the instance: $f0$ and $f1$ are closer to the second cluster center 80, so both keys were mapped to array-2; while other keys are closer to the first cluster center 15, thus they were all mapped to array-1. The first bucket in array-1 contains two keys $f3$ and $f4$, thus it incurs a hash collision. However, the average of this bucket is $\frac{35}{2} =17.5$, which approximates the values of $f3$ (18) and $f4$ (17) quite well, since each bucket array contains similar key-value pairs.}

An LSS provides  \textit{group}, \textit{insert}, and  \textit{query} operators to support the  dimension-reduction representation of key-value streams. 

\subsubsection{Group}

LSS groups similar items together, by calculating a list of cluster centers as clustering reference points for  items. As discussed in the above subsection, we choose the well-known K-means clustering method to find cluster centers due to its simplicity and competitive performance, although more complex clustering methods may  achieve slightly better performance.   

 Grouping flows to clusters should cope with online streams.  Training the cluster model for packet streams is infeasible, in contrast, we need to perform one-pass processing for online network flows: we initialize  cluster centers a priori, and map streamed network flows with initialized cluster centers.  Fortunately,   it is well known that the  flow-size distributions are self-similar \cite{DBLP:journals/ton/LelandTWW94,DBLP:conf/imc/BensonAM10}, thus the cluster structure is transferrable. Our experiments in section \ref{EvaSec} also confirm this observation. Therefore,  we find clustering patterns on packet traces in an offline manner, then group online flows with obtained clustering patterns in the offline phase.

\subsubsubsection{Offline Training}: We obtain flow traces and train the K-means clustering mode in an offline manner. We tune the number of clusters in order to obtain a fine-grained grouping model for the flow size distribution, which bounds the variance within each group in order to control the error variance of the average estimator. 


\subsubsubsection{Online Mapping}: To speed up the mapping process, we sort the cluster centers a priori, which takes $O(k \log k)$ time, where $k$ denotes the number of cluster centers. Then,  for each online key-value pair, we directly map this pair to the nearest cluster center with a binary-search process on sorted cluster centers in time $O( \log k)$.  Finally, based on the index of the cluster center, we map this key-value  pair to the corresponding bucket array in the LSS sketch.  

\subsubsection{Insert}

LSS  maps a key-value pair to the nearest cluster center, and accumulates the value to the corresponding bucket array.  Algorithm \ref{alg:Insert} shows the steps of inserting a new key-value pair into LSS.  First, we locate the bucket array corresponding to the nearest cluster center to the incoming key-value pair. Second, we choose a random bucket by hashing the key with one hash function. Third, we accumulate the key-value pair to  the  bucket: (i) ValSum = ValSum  + value; (ii) KeyCount  = KeyCount + 1(if and only if key has not been hashed into this bucket array). Finally, we store the cluster index of the incoming key for network-wide key-value queries. 

\textbf{Complexity}: We represent the cluster-index field with eight bits that indexes $2^8=256$ clusters in total.  Each key-value pair is  mapped to only one bucket in a LSS sketch, which involves only one hash-function evaluation. Accessing  a (2,4)-filter involves two hash-function evaluation.   

\begin{algorithm}[!t]
{
\begin{small}
\SetKwFunction{Insert}{Insert}
\SetKwInOut{Input}{input}
\SetKwInOut{Output}{output}
\Insert{$\kappa$, $v$}\\
{


${ i }_{\kappa}= argmin _{ i }{ \left\| v - C_i \right\|  } $\;

bucket = $I_{{ i }_{\kappa}} \left[ h\left( \kappa \right)  \right] $\;

bucket.ValSum + =  $v$, bucket.KeyCount +=1\;

Store ($h_{finger} (\kappa) , { i }_{\kappa}$) to the Cuckoo hash table\;
}
\caption{Insert a non-duplicated Key-value pair to LSS. $I$ denotes the bucket array. $h(\cdot)$ denotes the hash function. $C_i$ denotes the $i$-th cluster center. $h_{finger}$ denotes the hash function for the fingerprint calculation}
\label{alg:Insert}
\end{small}
}
\end{algorithm}

\subsubsection{Query}

To query the value of a key on the LSS, we need to locate the bucket array. To that end, we query the Cuckoo hash table with the input key to get the cluster index of this key.    Finally, we return the weighted value $\frac{ValSum}{KeyCount}$ as the approximated result.  Algorithm \ref{alg:query} summarizes the steps for the query process.

\textbf{Time Complexity}: Querying a (2,4)-filter needs two hash-function calculations. Obtaining the LSS bucket's counter needs one hash-function calculation. Thus the time complexity of the query process is the same as that of the insertion process.

\begin{algorithm}[!t]
{
\begin{small}
\SetKwFunction{Query}{Query}
\SetKwInOut{Input}{input}
\SetKwInOut{Output}{output}
\Query{$\kappa$}\\
{



Query the Cuckoo hash table to get the cluster index ${ i }_{\kappa}$ for $\kappa$\;

bucket = $I_{{ i }_{\kappa}} \left[ h\left( \kappa \right)  \right] $\;

\Return $\frac{bucket.ValSum}{bucket.KeyCount}$ \;

}

\caption{Key-value query for a key $\kappa$.}
\label{alg:query}
\end{small}
}
\end{algorithm}

\subsection{Handling Duplicated Online Streams}
\label{DuplicateSubsec}

LSS requires to always keep a flow within the nearest cluster. As the flow size is unknown before it completes,  the ingestion process may publish multiple records for the same flow. Thus we have to efficiently identify the nearest cluster center for a dynamic flow and adjust the cluster mapping for changing flows.  

We propose a duplication adaptive maintenance method to dynamically maintain flow records  towards the nearest cluster. Algorithm \ref{alg:InsertD} summarizes the duplication-adaptive maintenance process.  If the flow has not been inserted to LSS, then we put it into the bucket array corresponding to the closest cluster center;  otherwise, the flow has been mapped to LSS, we locate the mapped cluster of this flow,  select the corresponding bucket array,  and then increment the flow record at the mapped bucket. Finally, we  check whether or not to move the flow to a new cluster: if the flow record is still nearest to the current cluster center, no movement should be made; otherwise, we need to move the flow record to the bucket array corresponding to the nearest cluster center: we delete the flow record from the current bucket array, and insert it to the bucket array corresponding to the nearest cluster center. 

\textbf{Complexity}: For a new  key-value pair, we need two hash-function evaluations to visit the (2,4)-filter, and one hash-function evaluation to access the LSS sketch. To save the hashing complexity, we reuse the hash function across LSS bucket arrays, thus we only need three hash-function evaluations to insert an existing key-value pair. During the insertion process, we  temporally keep a Cuckoo hash table $CH$; while after the sketch terminates the insertion process, we squeeze the Cuckoo hash table to keep only the fingerprint and the cluster index.


\begin{algorithm}[!t]
{
\begin{small}
\SetKwFunction{InsertDuplicate}{InsertDuplicate}
\SetKwInOut{Input}{input}
\SetKwInOut{Output}{output}
\InsertDuplicate{$\kappa $, $v$, $CH$}\\
{

Query  $CH$ with $\kappa $ to get its cluster index ${ i }_{\kappa }$\;

\If{${ i }_{\kappa }$ is NULL}{

${ i }_{\kappa }= argmin _{ i }{ \left\| v- C_i \right\|  } $\;

$I_{{ i }_{\kappa }} \left[ h\left( \kappa  \right)  \right] $.ValSum + =  v, $I_{{ i }_{\kappa }} \left[ h\left( \kappa  \right)  \right] $.KeyCount +=1\;

Store ($h_{finger}(\kappa)$, (${ i }_{\kappa }$, $v$)) to $CH$\;

}\Else{

Accumulate $v$ to the current value $v_{\kappa }$ of $\kappa$ in $CH$\;

$I_{{ i }_{\kappa }} \left[ h\left( \kappa  \right)  \right] $.ValSum + =  $v$\;

Retrieve the value $v_{\kappa }^*$ for the key $\kappa$  in $CH$ \;

Find the nearest cluster center $i^*$ for $v_{\kappa }^*$\;

\If{$i^* \neq { i }_{\kappa }$}{



$I_{{ i }_{\kappa }} \left[ h\left( \kappa  \right)  \right] $.ValSum  - =  $v_{\kappa }^*$, $I_{{ i }_{\kappa }} \left[ h\left( \kappa  \right)  \right] $.KeyCount -=1\;

$I_{i^*} \left[ h\left( \kappa  \right)  \right] $.ValSum  + =  $v_{\kappa }^*$, $I_{i^*} \left[ h\left( \kappa  \right)  \right] $.KeyCount +=1\;


Store ($h_{finger}(\kappa)$, ($i^*$, $v_{\kappa }^*$)) to  $CH$\;

}
}
}
\caption{Duplication-adaptive LSS maintenance.  }
\label{alg:InsertD}
\end{small}
}
\end{algorithm}

 \subsection{LSS Parameters}
\label{baSize}

We next present parameter guidelines in order to trade off the estimation accuracy and the memory footprint. 

\textbf{Bucket-Array Size}: We configure  the size of a bucket array $i$ based on the combination of three factors: (i) \textit{Cluster entropy} $H$: For a cluster covering a short interval,  a small bucket array is enough to achieve a low estimation error. This short cluster contains a low degree of uncertainty. The uncertainty of the cluster entries can be quantified with the \textbf{entropy}, ${ H }_{ i }=-\sum _{ j\in { S }_{ i } }^{  }{ { f }_{ j } } log{ f }_{ j }  \in \left[ 0,1 \right] $, where $S_i$ denotes the set of unique items for the $i$-th cluster, $f_j$ denotes the frequency of item $j$ in this cluster.   (ii) \textit{Cluster center} $\mu$: For a cluster with a large cluster center, it is likely to be the heavy tails of the flow's distribution, which needs more buckets to control the hash collisions. We quantify the cluster center with the ratio of each cluster center against the sum of all cluster centers, i.e., ${ \mu  }_{ i }=\frac { { \mu  }_{ i } }{ \sum _{ j }^{  }{ { \mu  }_{ j } }  }  \in \left[ 0,1 \right] $.  (iii) \textit{Cluster density} $d$: For two clusters with approximately the same cluster uncertainty, a larger cluster need more buckets to reduce the estimation error. We quantify the cluster density with the ratio of the cluster entries to the total number of items, i.e., ${ d }_{ i }=\frac { { d }_{ i } }{ \sum _{ j }^{  }{ { d }_{ j } }  }  \in \left[ 0,1 \right] $.

Let $m$ denote the total number of buckets for LSS, $H_i$ the entropy of the $i$-th cluster, $g _i$ the $i$-th cluster center, and $d_i$ the percent of items for the $i$-th cluster, we allocate  $\frac { { H }_{ i }{ d }_{ i }{\mu  }_{ i } }{ \sum _{ j }{ { H }_{ j }{ d }_{ j }{ \mu  }_{ j } }  } \cdot m$ buckets for the $i$-th bucket array. We derive these parameters through the offline K-means training process. 

\textbf{Number of Clusters}: Finding the optimal number of K-means clusters is known to be NP-hard \cite{Jain:1999:DCR:331499.331504}. Thus we empirically determine the number of clusters based on sensitivity analysis that locates diminishing returns of the prediction accuracy.

 \section{Disaggregated Monitoring Application}
\label{DSSec}

Having presented the LSS sketch, to illustrate the feasibility of the locality-sensitive sketching, we next propose a monitoring application that implements LSS in a modular and non-intrusive framework.   

As the network flows are infinite in essence, recent network flows are usually more important.   In a sliding window model, each packet is sequentially and independently processed in a one-pass manner. For a sequence based window, it processes past $N$ items; while for a time based window, it processes items in a past time period. The framework supports both sliding window models, although   the time based window may ingest too many flows during bursty periods, while the sequence based window is more robust in this case.

The proposed monitoring architecture splits the monitoring application  into non-coherent  ingestion, sketching, query runtime functions that can be horizontally scaled in the data center. A monitoring function atomically defines an intermediate stage in the monitoring process. Each ingestion function colocates with the server or middlebox to aggregate packet streams to flowlet streams. Second, each sketching function maintains the LSS sketch based on ordered  flowlet streams. Third, each query function performs monitoring queries on LSS sketches. Finally, we keep system configuration up to date via a global coordinator. A publish/subscribe (Pub/Sub for short) framework\footnote{The Pub/Sub topic framework provides seamless messaging supports for monitoring functions, which   represents message flows among disaggregated components.  One or multiple producer entities publish messages towards the same topic,  then the Pub/Sub messaging framework  delivers ordered messages to  consumers subscribed to the same topic. } delivers ordered streaming messages across monitoring functions. We choose the Pulsar messaging system originally created at Yahoo \cite{pulsar} as the  Pub/Sub underlay.


\noindent (i) \textbf{Ingestion Stage}: The ingestion stage provides a device-independent key-value intermediate presentation model for network monitoring. It  splits packets at servers or middleboxes at line rates to flowlets, and publishes key-value formatted flowlet-record messages in a batch mode to the Pub/Sub framework.  When a packet arrives, we look up the hash table with a key calculated based on the hash of its 5-tuple information: If the key is in the hash table, then we accumulate the per-flow counter with this packet's information; If the key is not in the hash table and there exists empty entries in the table, then we put the key and the corresponding per-flow counter to the hash table; Otherwise, we  publish all accumulated flow counters in batch, and reset the hash table to accommodate for new entries.  


 \noindent (ii) \textbf{Sketching  Stage}: The sketching component subscribes to one or multiple topics published by the ingestion components, then dynamically keeps an independent  LSS sketch  for each sliding window. For the  sequence based sliding window, each LSS sketch keeps at most $N$ flow records and is emitted to the sketch topic afterwards; while for the time based window, each LSS sketch is emitted after the interval ends.  Upon receiving a flow record from a subscribed topic, the component selects the corresponding LSS sketch,  groups  this record towards the nearest cluster center, and inserts this record to the corresponding bucket array in the LSS sketch. We handle duplicated flow records based on subsection \ref{DuplicateSubsec}.  

\noindent (iii)  \textbf{Query Stage}: LSS supports diverse query tasks similar to existing sketch structures. We list the most representative ones: 

(a) \textbf{Per-flow frequency and entropy query}.  They track the traffic volume of each distinct flow, or count the flow bytes.  LSS directly returns the size of a given flow. To query the size distribution  of each inserted flow, we  iteratively obtain  approximation results with identifiers of inserted flows, then we build a list of approximated flow sizes as the flow size distribution. Similarly, we derive the entropy metric  as  the frequency distribution of approximated flow sizes. 

(b) \textbf{Heavy hitters}.  It finds top-K flows ingesting the most traffic volumes.   For a given heavy-hitter detection threshold,  we obtain  approximated values of inserted flows from the LSS sketch, and select those exceeding the threshold as heavy hitters. Based on heavy hitters, we can also find flows spanning multiple windows that fluctuate beyond a predefined threshold, i.e., the heavy changes.

(c) \textbf{Flow cardinality}. LSS counts the exact  number of distinct  flows, since  LSS maps  each flow to a unique bucket. Therefore, we directly calculate the sum of KeyCount fields for each non-empty buckets, and return the accumulation result as the number of distinct flows. 

Moreover, the framework supports a  network-native query interface.  Each sketching component publishes to the same sketching topic, then a centralized query component subscribes to this sketching topic and performs queries on received sketches. Moreover, some network management tasks may need to  query historical sketches during a time interval. To that end, the query component  stores the received sketch and its arrival timestamp in a persistent storage, and then lists sketches within a given time interval.

 \section{Evaluation}
\label{EvaSec}

\subsection{Experimental Setup}

We ran experiments on a  multi-tenant private cluster to evaluate the locality-sensitive sketching and disaggregated monitoring. The cluster is shared by tens of different clients. We set up the experiments on ten servers in two racks connected by a 10 Gbps switch, each server is configured as 8-core Intel(R) Xeon(R) CPU E5-1620, 47 GB memory, and Intel  10-Gigabit X540-AT2 network card.  We set up the Apache Pulsar 2.2.0 Pub/Sub as a standalone service on a dedicated server. We configure Apache Pulsar with the default setup.  We split nine servers to two groups: (i) Six servers run the network ingestion component to produce flowlet records for port-mirrored traffic from the top-of-the-rack switch based on the Intel DPDK 16.04 interface, and publishes to the Pub/Sub framework; (ii) Three servers run the sketching component to maintain the LSS sketch for each of six ingestion servers. Each LSS sketch is published to the Pub/Sub framework after it accumulates 10,000 flows. 


\textbf{Default LSS Parameters}:  We set the sliding window to consist of 10,000 flows by default. We dimension the total number of buckets with respect to the number of flows in a sliding window.  For a sliding window that consists of $N$ flows, we set the default number $m$ of LSS buckets to $0.1 \times N = 1,000$. For each LSS bucket, we set the storage size to four bytes (two bytes for each field).  We set the default number of clusters to 30. Each cluster center is represented with four bytes. Thus an LSS with 1,000 buckets and 30 cluster centers takes 4.12KB.  The offline traces take 10,000 flow samples, each sample is represented as four bytes, which take 40KB in total. We set the default heavy-hitter threshold to the 90-th percentile of the offline traces. We choose LSS' default parameters based on the diminishing returns via extensive evaluation in Subsection \ref{SensitivitySec}.    

\co{ consists of 1,000  buckets that; We select 10,000 samples to train the K-means cluster centers. }

 
\textbf{Metrics}: We choose three representative monitoring tasks to evaluate the sketch's performance, namely the flow-size query, the flow-entropy query, and the heavy-hitter query. We quantify the performance of the first two tasks with the relative error metric: defined as $\left| x_r - x_e \right| /(x_r)$, where $x_r$ and $x_e$ denoted the ground-truth metric and the estimated metric, respectively, and the last task based on the F1 score  defined as the harmonic mean of the precision and the recall values, where the closer the F1 score towards one, the better the heavy-hitter estimator.

\subsection{Testbed Results}

\begin{figure}[!t]
  \centering
{
  {\includegraphics[width=.2\textwidth]{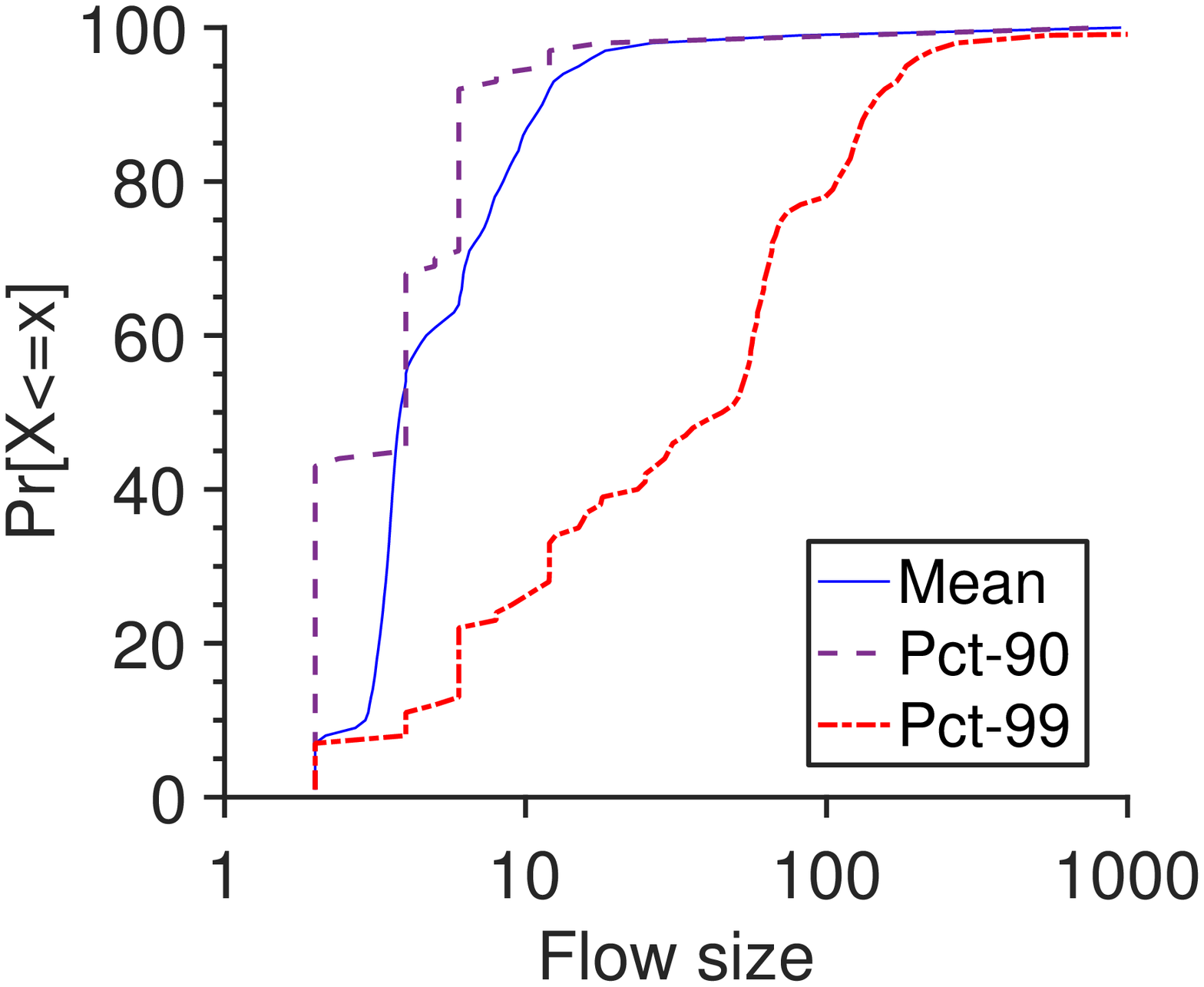}}}
     \caption{Traffic distribution of the testbed.}
     \label{fig:testbedTraffic}
\end{figure}

 
 We summarize the ground-truth distribution of flows captured in the private cluster. Figure \ref{fig:testbedTraffic} plots the cumulative distribution functions (CDF) of network flows in each interval. We see that the mean and the 90-th percentile of the network flows are less than 10 for over 90\% of all traces. However, the 99-th percentiles of traces span over three orders of magnitudes, thus the network flow distribution is highly skewed, which is similar to the CAIDA traffic trace in the next subsection.

 \subsubsection{Disaggregated Performance}
 
\begin{figure}[!t]
  \centering
{
  {\includegraphics[width=.2\textwidth]{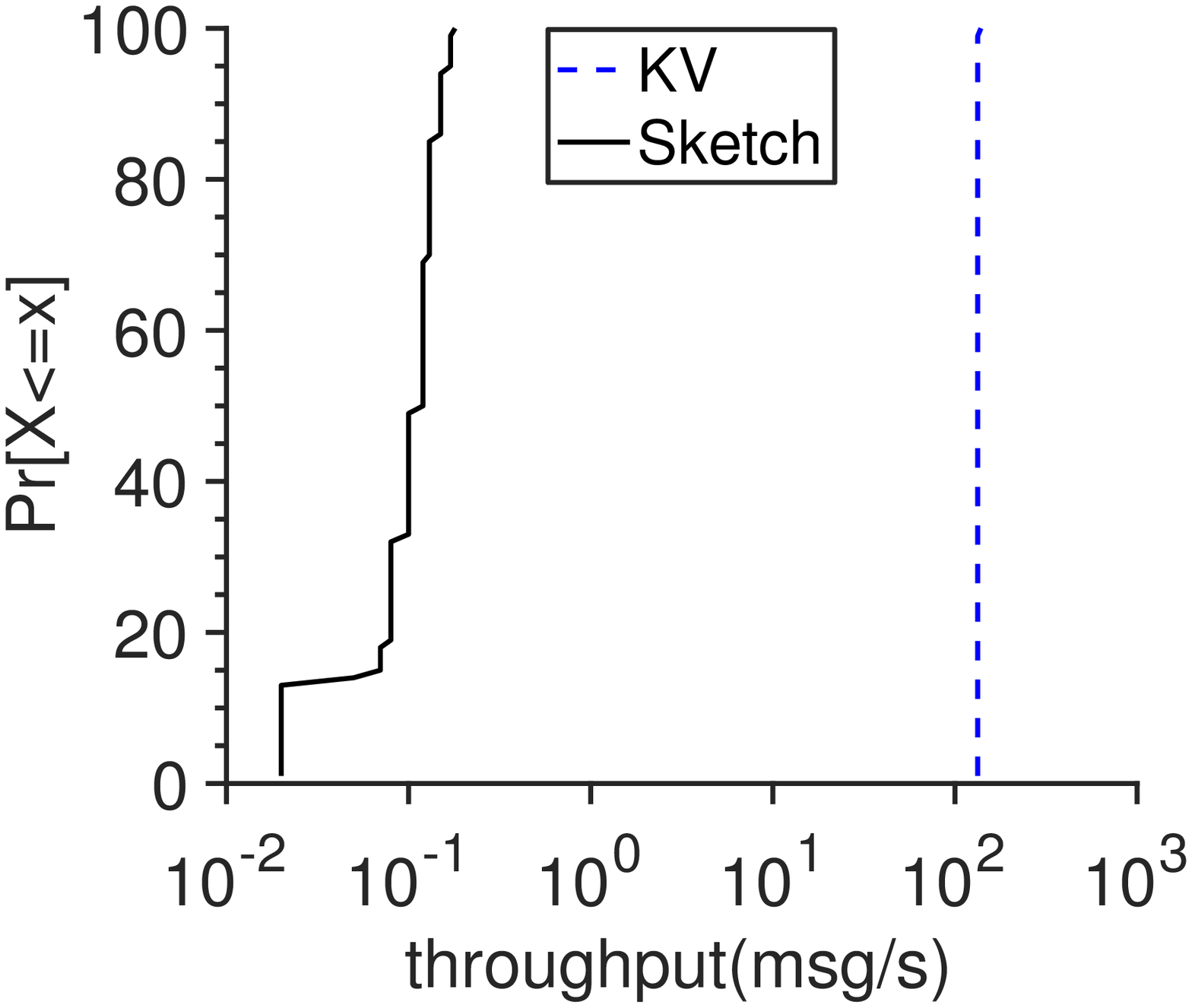}}}
     \caption{Publishing throughput for the ingestion  and  sketching components.}
     \label{fig:testbedpublish}
\end{figure}

We test the publishing throughput for the ingestion and the sketching components. Each flowlet message consists of 1,000 temporary flow counters in the hash table, while each sketching message consists of one LSS sketch. Figure \ref{fig:testbedpublish} plots the CDFs of the message throughputs of the ingestion components and those of the sketching components. We see that  the throughput of the ingestion component is nearly three orders of magnitudes larger than that of the sketching component, since the ingestion component depends on the line rates, while the sketching component depends on the readiness of the sliding window. 
 
 \co{The sketching component emits an LSS sketch to a message whenever it keeps more than 10,000 flows. }

\begin{figure}[!t]
  \centering
{
  {\includegraphics[width=.2\textwidth]{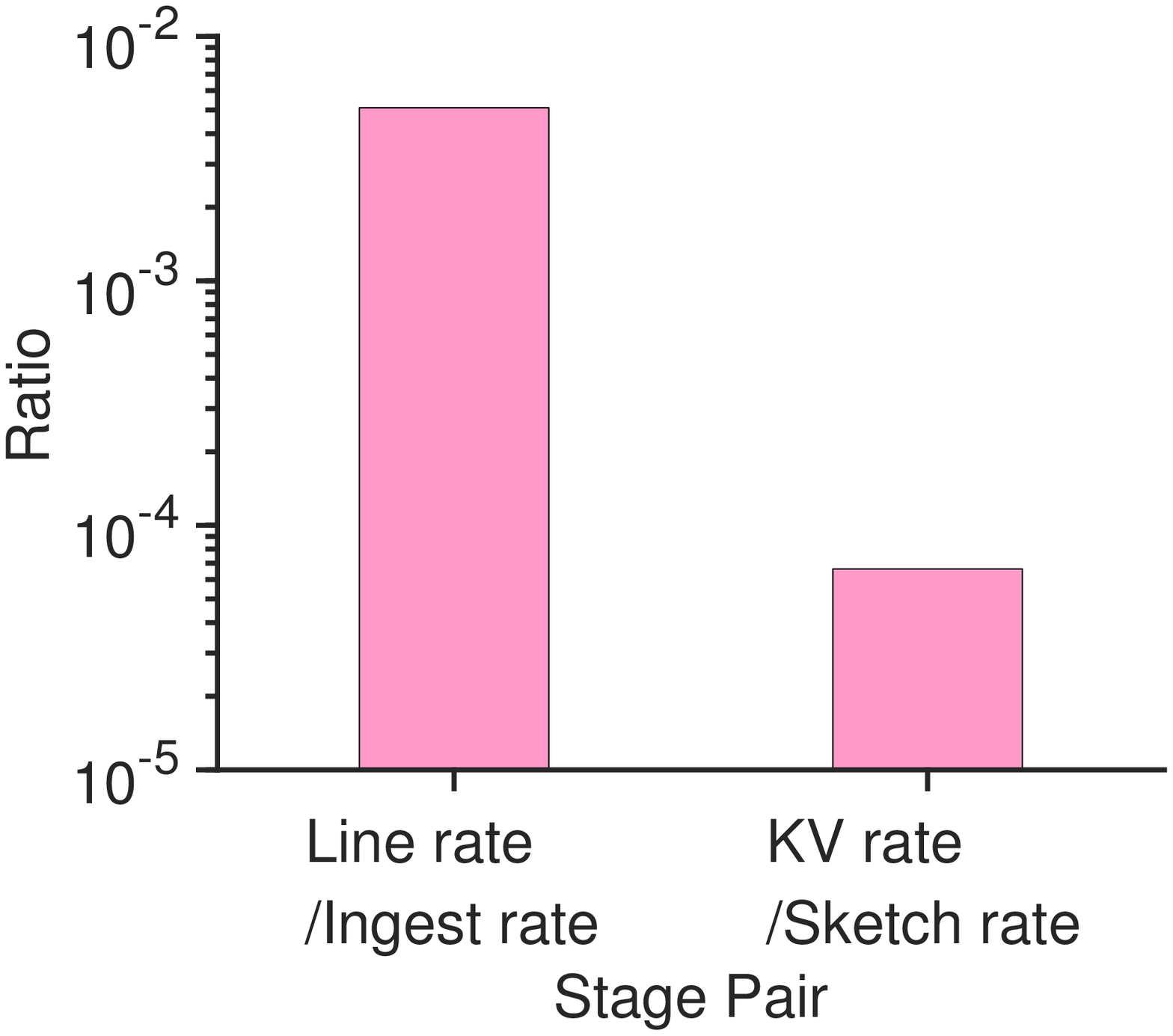}}}
     \caption{Packet rate vs. the ingestion rate, and key-value arrival rate vs. sketch insertion rate.}
     \label{fig:testbedThroughput}
\end{figure}

We next compare the relative performance of the ingestion component and the sketching component.  Figure \ref{fig:testbedThroughput} shows the relative rate between the packet's arrival rate and the ingestion rate, as well as that between the flow-record arrival rate and LSS' insertion rate.  We see that  the arrival rate is orders of magnitude smaller than the corresponding consumption rate for both ingestion component and the sketching component.  Since each component is tuned with respect to the input's arrival rate. We also constrain the size of the ingestion hash table and the LSS sketch in order to avoid CPU's L3-cache misses.


\co{
We next count  the million requests per second (MRPS) of the LSS sketch for all queries.  Figure \ref{fig:testbedQuery} plots the CDFs of the query rates. We see that 70\% of the query rates exceed ten million requests per second, since LSS accesses only one bucket array in constant time for each query. 


\begin{figure}[!t]
  \centering
{
  {\includegraphics[width=.2\textwidth]{testbed/QueryRateTestbed0108output.eps}}}
     \caption{CDF of the query rate for the LSS sketch instance.}
     \label{fig:testbedQuery} 
\end{figure}
}

 \subsubsection{Sketching Performance}

 \begin{figure}[!t]
  \centering
\subfigure[Flow-size query]{ \label{fig:testbedFlowErr}
  {\includegraphics[width=.14\textwidth]{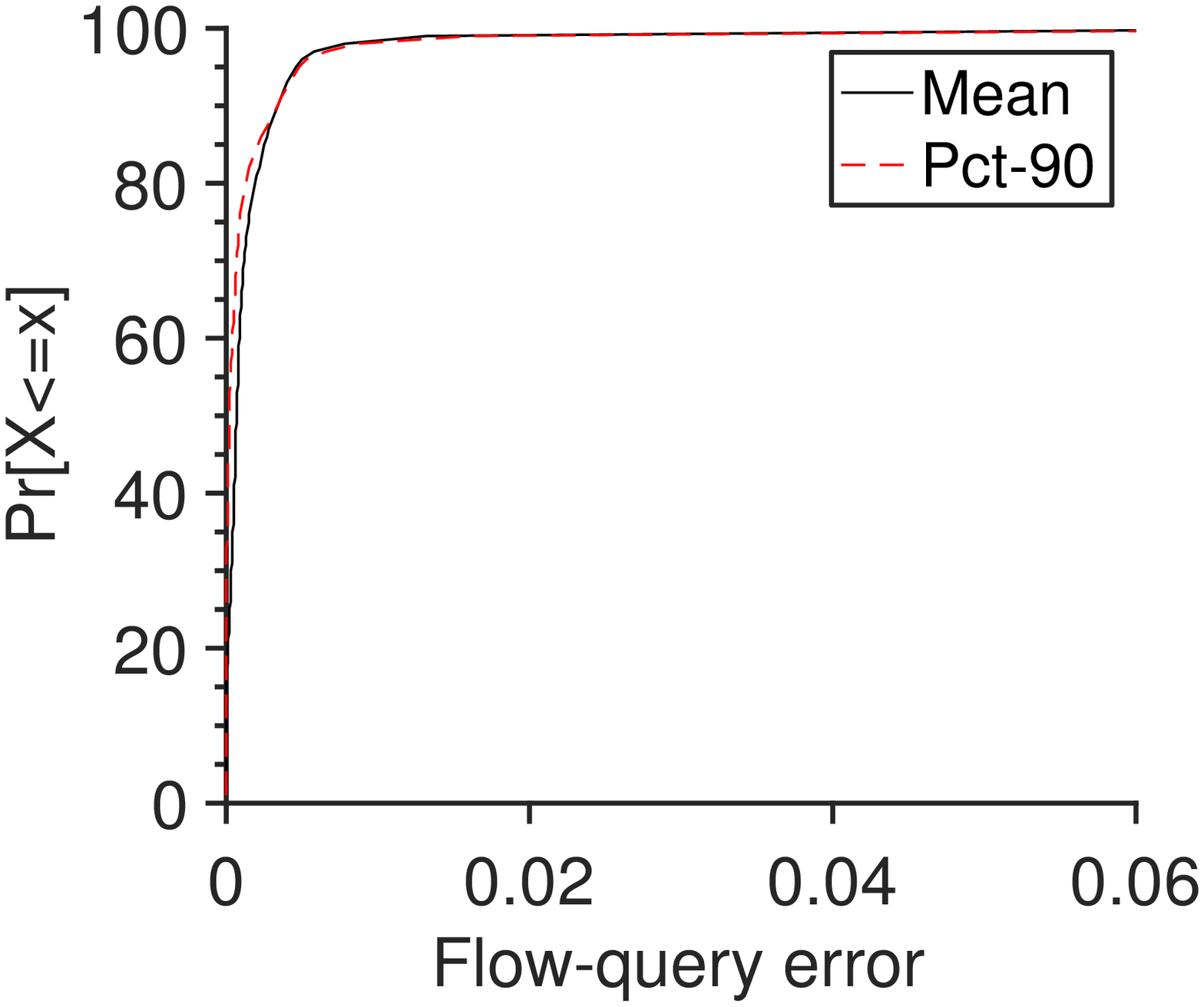}}} 
  \subfigure[Flow entropy]{ \label{fig:testbedEntropy}
  {\includegraphics[width=.14\textwidth]{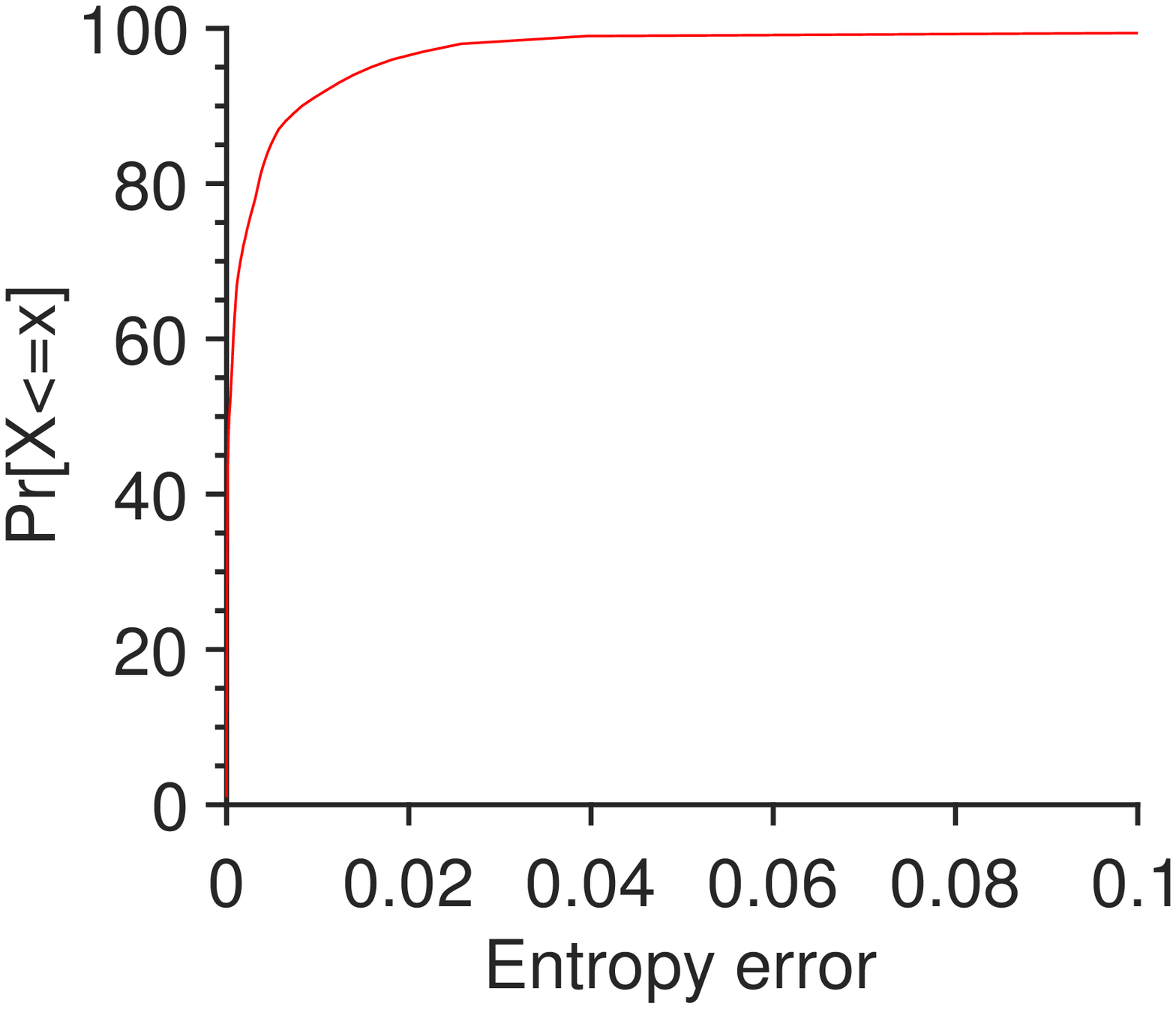}}}
  \subfigure[Heavy hitters]{ \label{fig:testbedHHDetect}
  {\includegraphics[width=.14\textwidth]{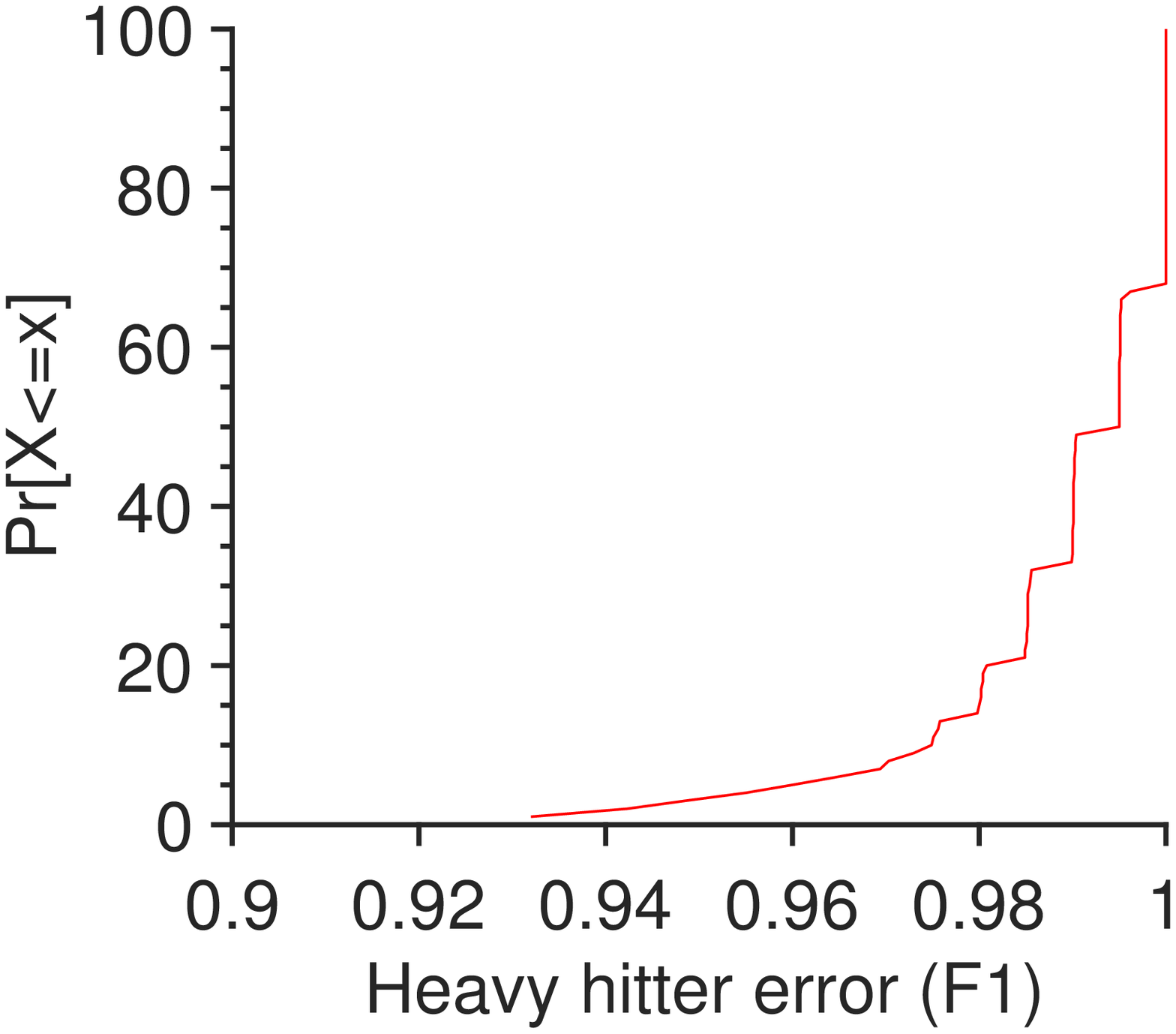}}}
     \caption{Performance of representative monitoring tasks on the testbed.}
     \label{fig:testbedPerf}
\end{figure}

(i) \textbf{Flow-size query}: Next, we evaluate the relative error of estimated flow sizes. For each flow in each interval, we compare the estimated flow size against the ground-truth flow size. Figure \ref{fig:testbedFlowErr} plots the CDFs of the mean relative errors. We see that the relative errors of over 90\% of all estimations are smaller than 0.01. Since LSS accurately captures skewed flows  with clustered bucket  arrays. 


(ii)  \textbf{Flow-entropy query}: We next evaluate the accuracy of the entropy of the flow distribution for each interval. Figure \ref{fig:testbedEntropy}  plots the CDFs of the relative errors of estimated flow entropies. We see that over 90\% of  estimations are smaller than 0.06, because of accurate estimations of  flow sizes.

(iii)  \textbf{Heavy hitter query}: Having shown that the flow entropy is accurately estimated, we next test the accuracy of estimated heavy hitters by calculating the F1 scores. Figure \ref{fig:testbedHHDetect} plots the CDFs of F1 scores. We see that over 90\% of tests are greater than 0.95. As  LSS captures fine-grained flow distributions with clustered bucket arrays.

\begin{figure}[!t]
  \centering
\subfigure[Server 1]{ 
  {\includegraphics[width=.14\textwidth]{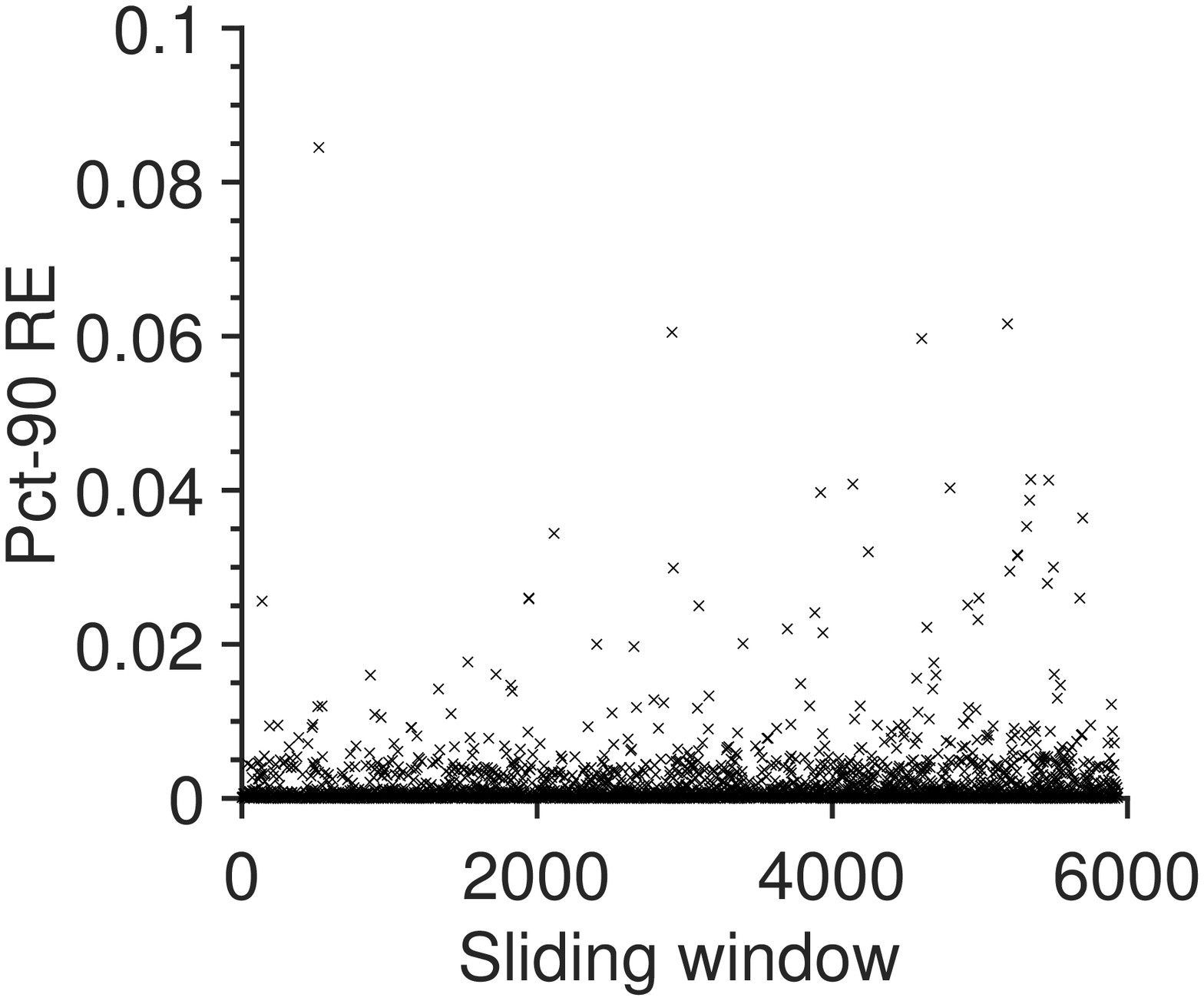}}} 
  \subfigure[Server 2]{ 
  {\includegraphics[width=.14\textwidth]{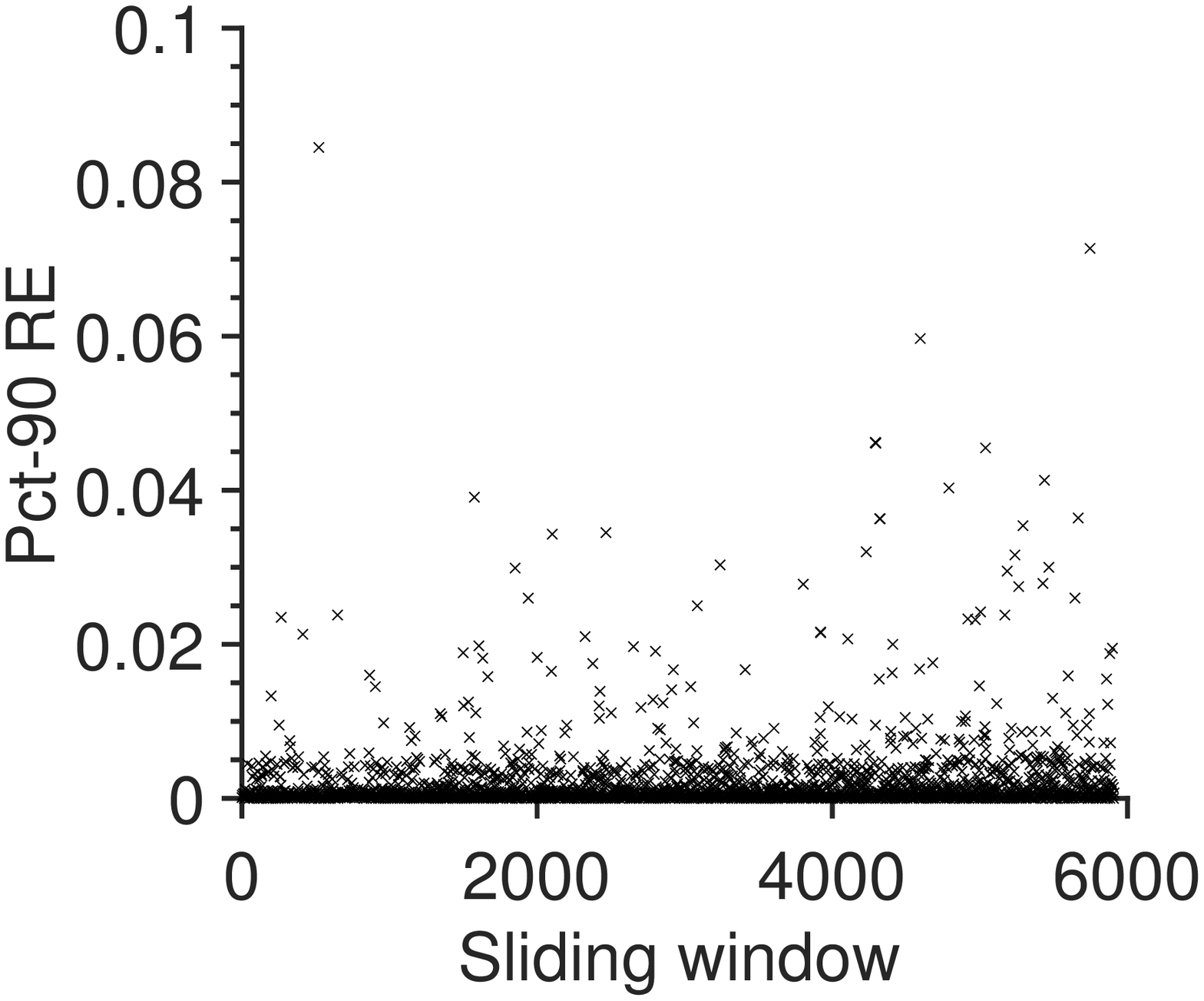}}}%
  \subfigure[Server 3]{ 
  {\includegraphics[width=.14\textwidth]{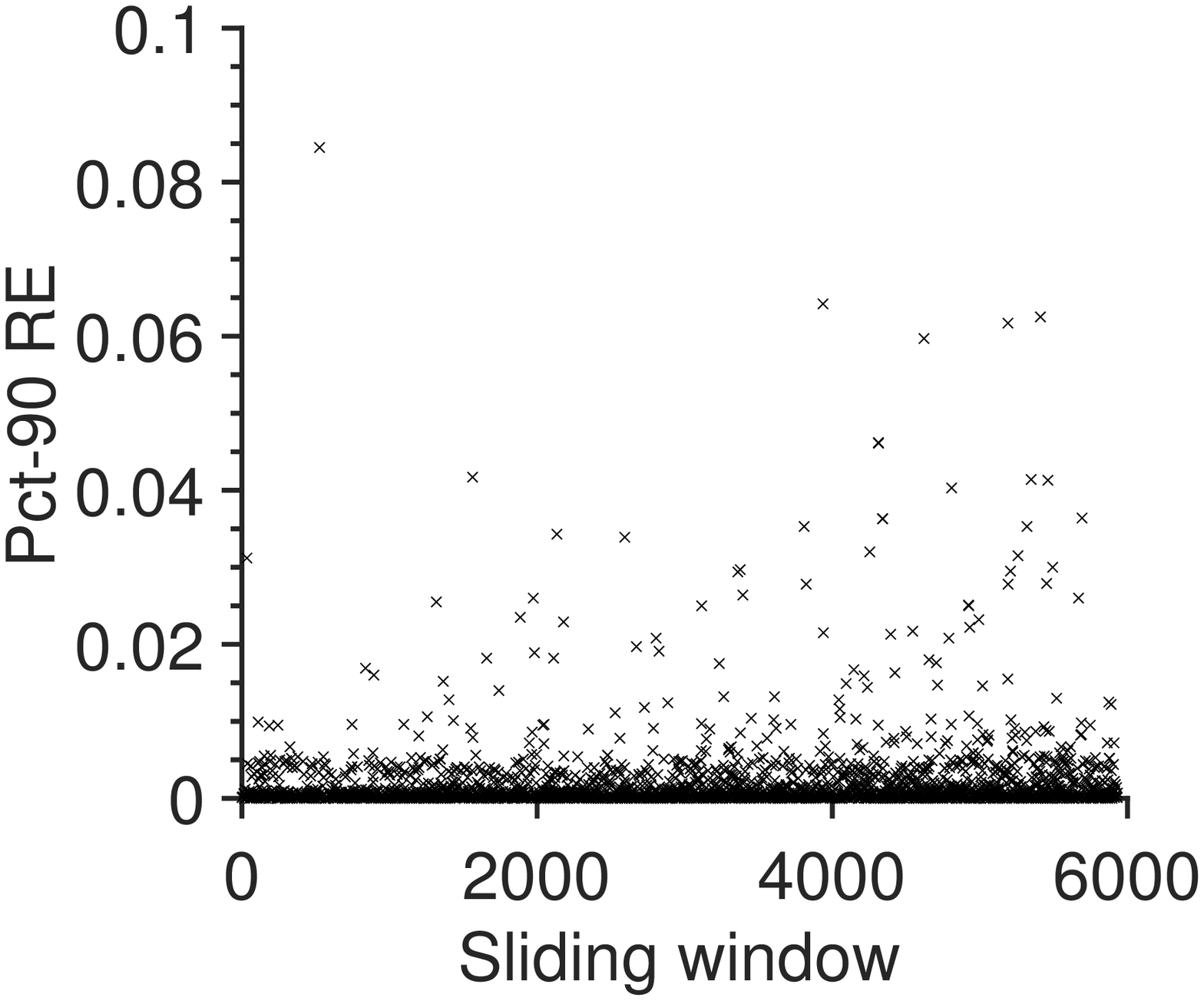}}}
     \caption{90-th percentile of flow-query relative errors on three servers running the query component.}
     \label{fig:testbedPerfDyna}
\end{figure}

 (iv)  \textbf{Estimation stability}: We next test the estimation stability on the testbed.   Figure \ref{fig:testbedPerfDyna} shows the 90-th percentiles of the flow-query relative errors of three query components. We see that most of the 90-th percentiles are zeros, while non-zero entries are smaller than 0.01 in most cases. Thus the estimation remains stably accurate across sliding windows.

\co{(iv) \textbf{Network-wide query}: We next evaluate the network-wide performance. A query entity subscribes to all sketching topics, and query the flow subset that spans multiple Test servers.   
}

\subsection{Trace-driven Simulation}
\label{simSubsec}

\begin{figure}[!t]
  \centering
{
  {\includegraphics[width=.2\textwidth]{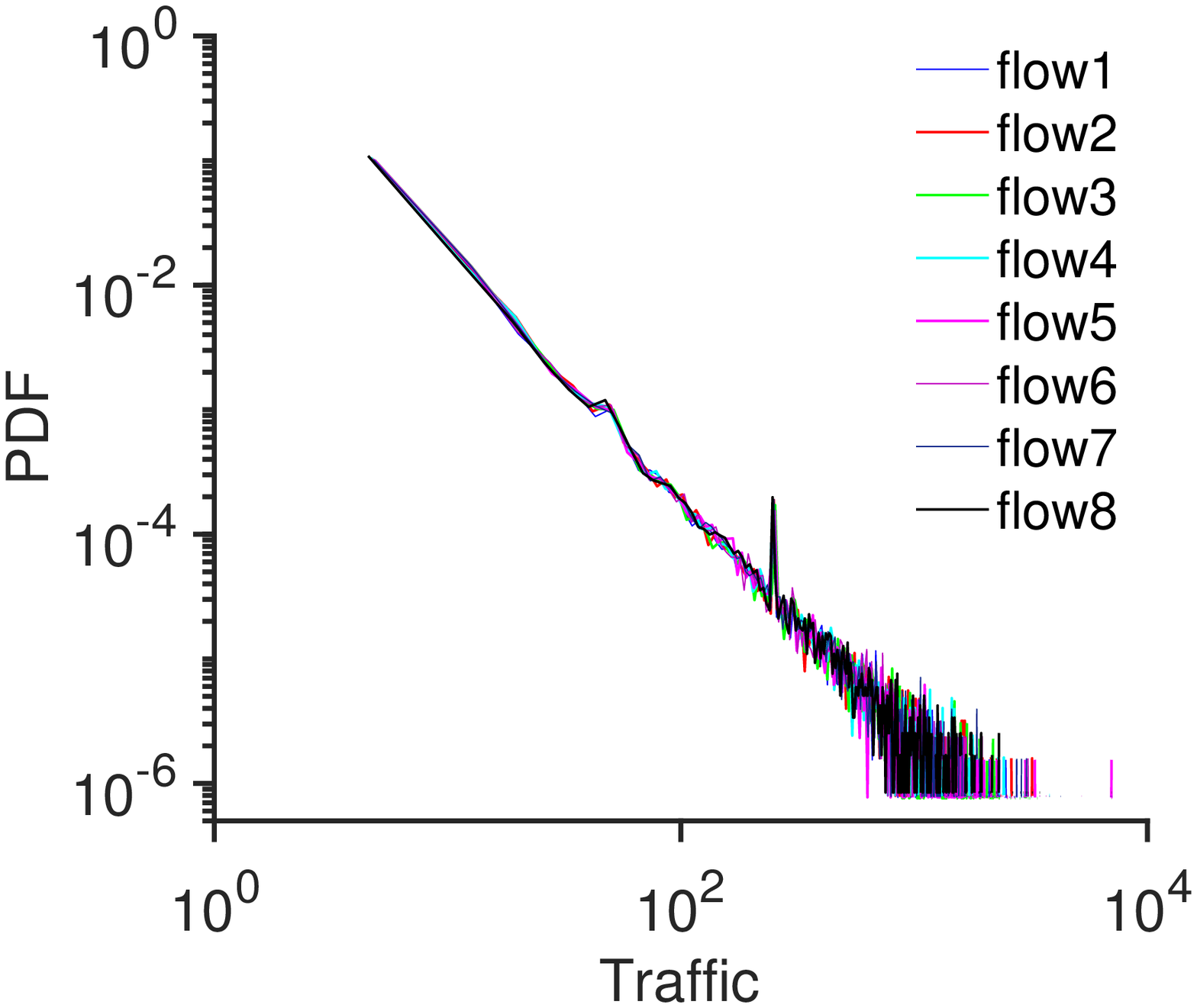}}}
     \caption{The probability density distributions (PDFs) of flow sizes in the CAIDA data set. We separate records to eight contiguous epochs denoted from flow1 to flow8, and plot the PDFs for each epoch. We see that the PDFs of different parts match well with each other.}
     \label{fig:Dist.}
\end{figure}

Our testbed is limited by the server scale. Therefore, we perform a real-world trace-driven experiment study. We replay network traces collected on February 18, 2016 at the Equinix-Chicago monitor by CAIDA \cite{DBLP:conf/sigcomm/0003JLHGZMLU18},  and feed to the Apache Pulsar Pub/Sub software framework.  We follow the default parameters of the testbed study. We calculate K-means cluster centers by randomly sampling 10,000 flows from the trace. Figure \ref{fig:Dist.} shows that different traces share nearly identical power-law flow-size distributions, thus the flow size distribution is not only skewed, but also temporally self-similar across epochs. This is because the self-similarity is a latent property in the network traffic \cite{DBLP:journals/ton/LelandTWW94,DBLP:conf/imc/BensonAM10}.

\subsubsection{Comparison}

(i) \subsubsubsection{Vary Memory}: We compare LSS with  count-min (CM) \cite{DBLP:conf/latin/CormodeM04}, count-sketch (CS) \cite{DBLP:conf/icalp/CharikarCF02}, and Elastic Sketch (ES) \cite{DBLP:conf/sigcomm/0003JLHGZMLU18} that are most related with our work.  CM and CS are commonly used to find heavy hitters and perform flow queries \cite{DBLP:conf/conext/MoshrefYGV15,DBLP:conf/sigcomm/LiuMVSB16,DBLP:conf/sigcomm/HuangJLLTCZ17}.   We set the same memory footprint for all compared sketch structures. We follow the recommended parameter configuration for CM \cite{DBLP:conf/latin/CormodeM04}, CS  \cite{DBLP:conf/icalp/CharikarCF02} and ES  \cite{DBLP:conf/sigcomm/0003JLHGZMLU18}. 

\begin{figure}[!t]
  \centering
\subfigure[Flow-size query]{
  {\includegraphics[width=.14\textwidth]{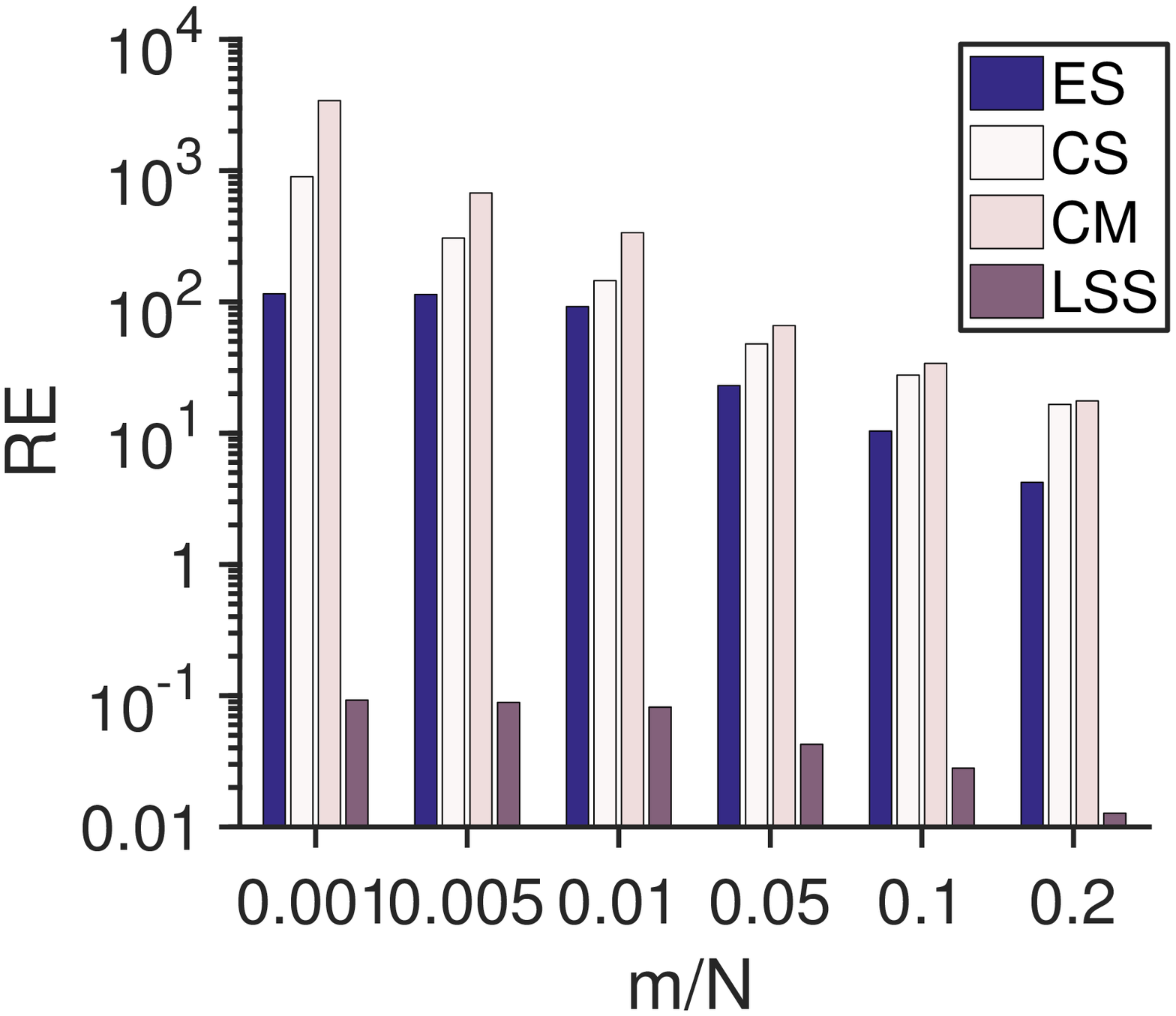}}}
  \subfigure[Flow entropy]{
  {\includegraphics[width=.14\textwidth]{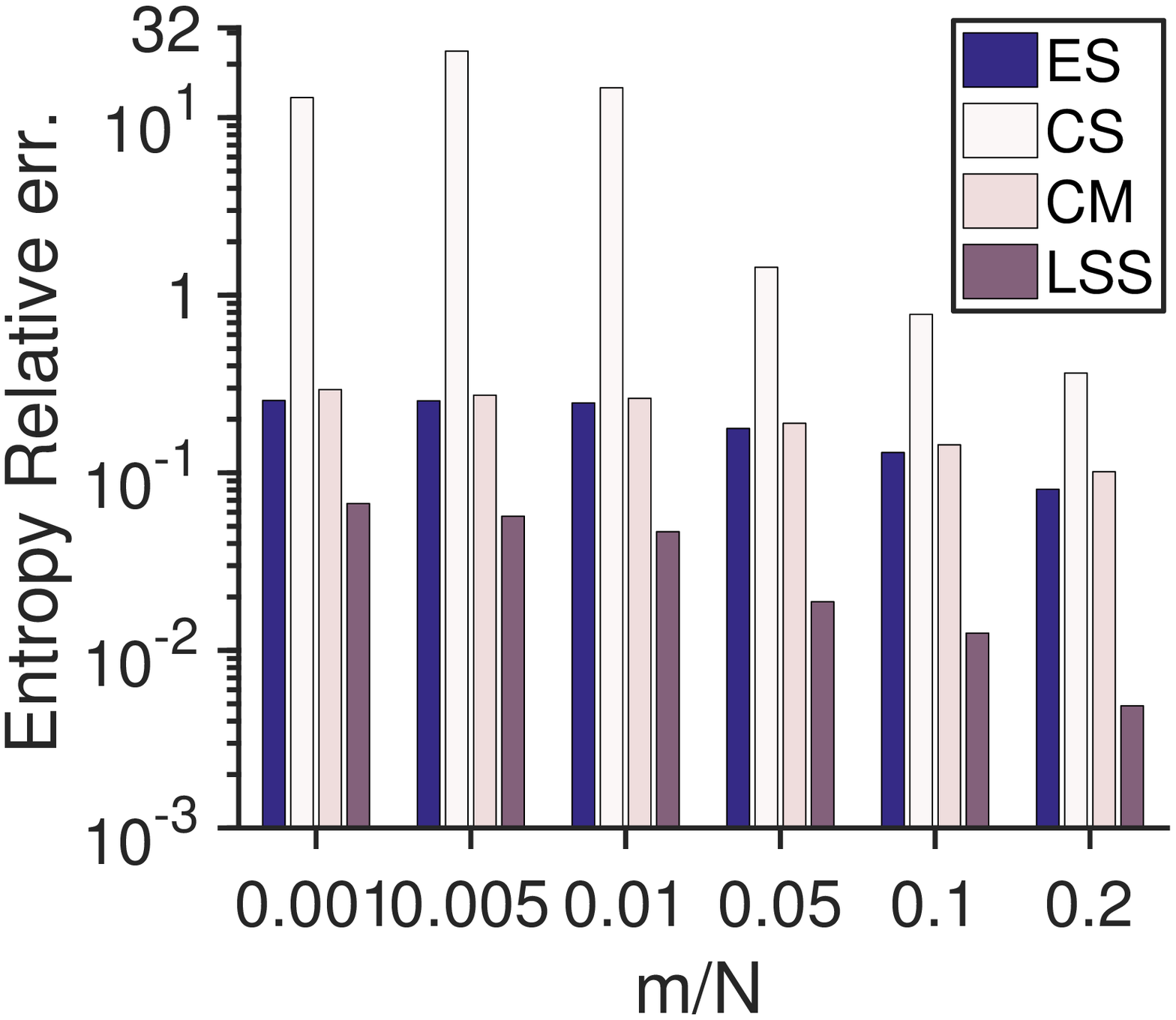}}}
  \subfigure[Heavy hitters]{
  {\includegraphics[width=.14\textwidth]{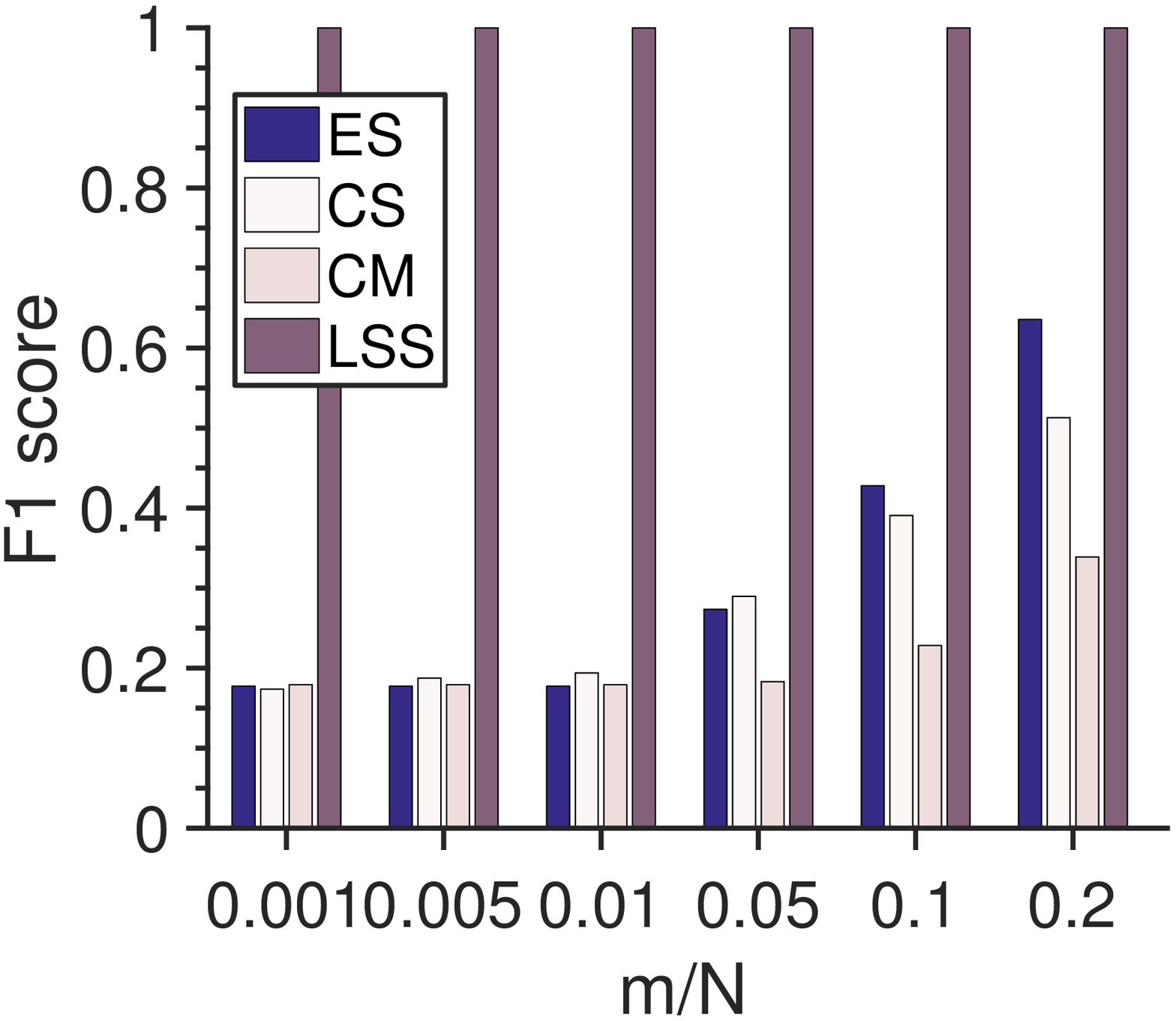}}}
     \caption{Accuracy of  LSS and CM, CS, ES  in terms of the ratios of  the number of LSS buckets to the number of flows.}
     \label{fig:comparePerf}
\end{figure}

Figure \ref{fig:comparePerf} plots the performance of the flow-size, flow-entropy, and heavy-hitter query tasks,  as we vary the ratio between the number of buckets in LSS and the number of unique flows. We see that LSS significantly outperforms other sketch structures in all cases. 

For the flow-size query tasks, LSS'  relative error  is  over $10^3$ to  $10^5$ times less than those of CS, CM and ElasticSketch, as the ratio between the number of buckets and the number of key-value pairs decrements from 10\% to 0.1\%. This is because LSS adapts to skewed flows with locality sensitivity and autoencoder based error minimization.

For the flow-entropy task, LSS' relative error is 4.3 to 13 times smaller than that of ElasticSketch,  4.8 to 14 times smaller than that of CM, and 70 to 200 times smaller than that of CS. ElasticSketch's accuracy is similar to that of CM in most cases, while CS has a much larger relative error than other methods. We can see that the flow-entropy task is less sensitive to flow-size errors, since the entropy depends on the frequency of each estimated value.  

For the heavy-hitter task, LSS is close to optimal compared to the other methods, since LSS accurately estimates the size of each flow with an autoencoder based recovery mechanism.  ElasticSketch's accuracy is similar to CM and CS when the ratio $\frac{m}{N}$ is not greater than 0.1, and has a better F1 score than CS and CM afterwards, since  ElasticSketch needs to keep large flows with the hash table and stores other flows to the count-min sketch.  

 \co{
(ii) \subsubsubsection{Varying Flow Fields}: We compare the stability of  LSS with CS and CM, all of which  do not filter heavy hitters like ElasticSketch.   As shown in Figure \ref{fig:flowFields}, all sketch structures marginally improve the accuracy as we  increment the key's input from one field to all five fields.  LSS remains to be  the most accurate sketch, since LSS captures fine-grained flow distributions with data-driven clustered buckets.

\co{Further, we can see that the performance on the source-only key is relatively worse than those on other kinds of keys, since the source-only key aggregates
they aggregate flows sent from the same source address, thus they introduce more significant  heavy tails. }
 
 \begin{figure}[!t]
  \centering
\subfigure[Flow query]{
  {\includegraphics[width=.14\textwidth]{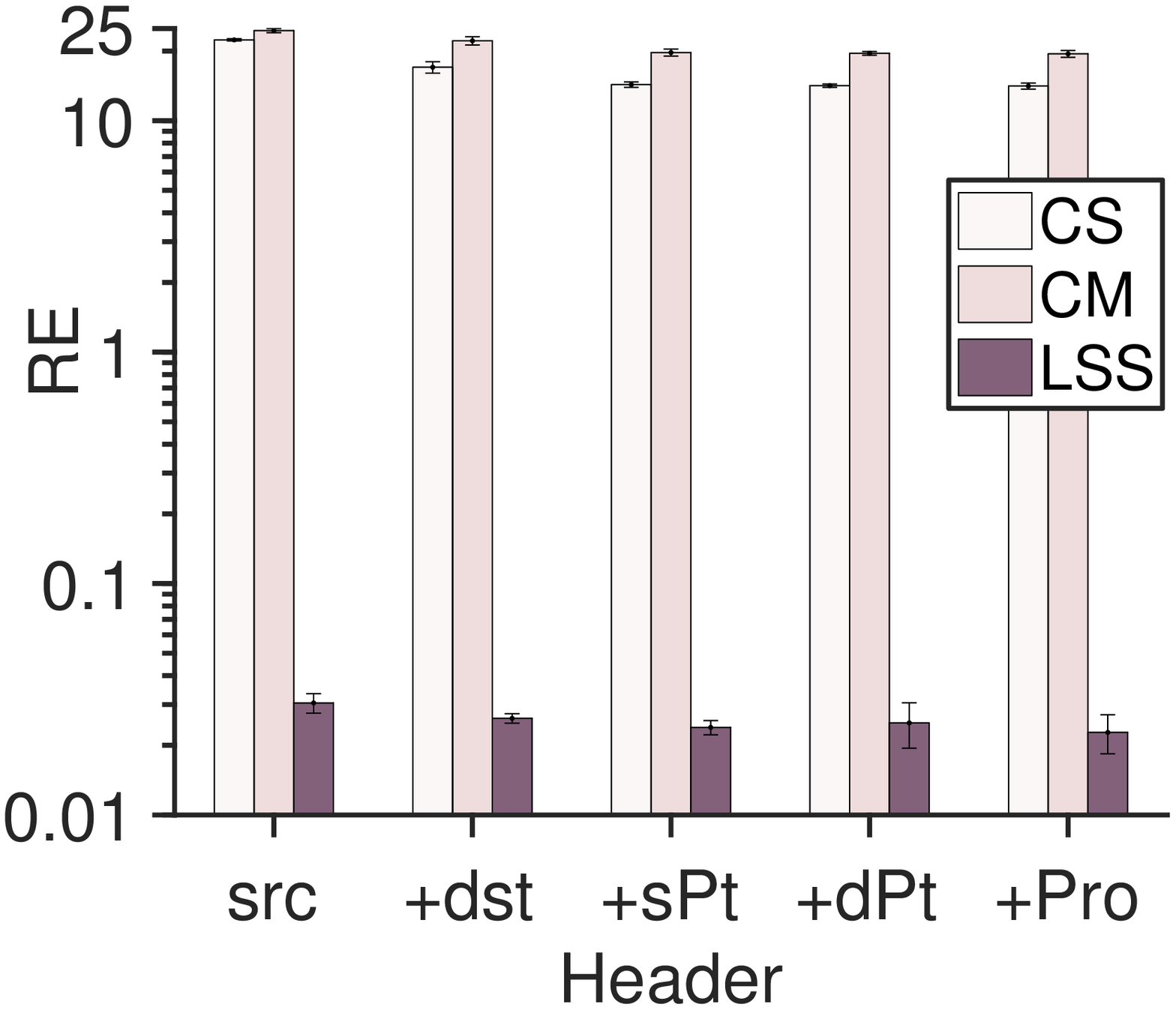}}}
  \subfigure[Entropy]{
  {\includegraphics[width=.14\textwidth]{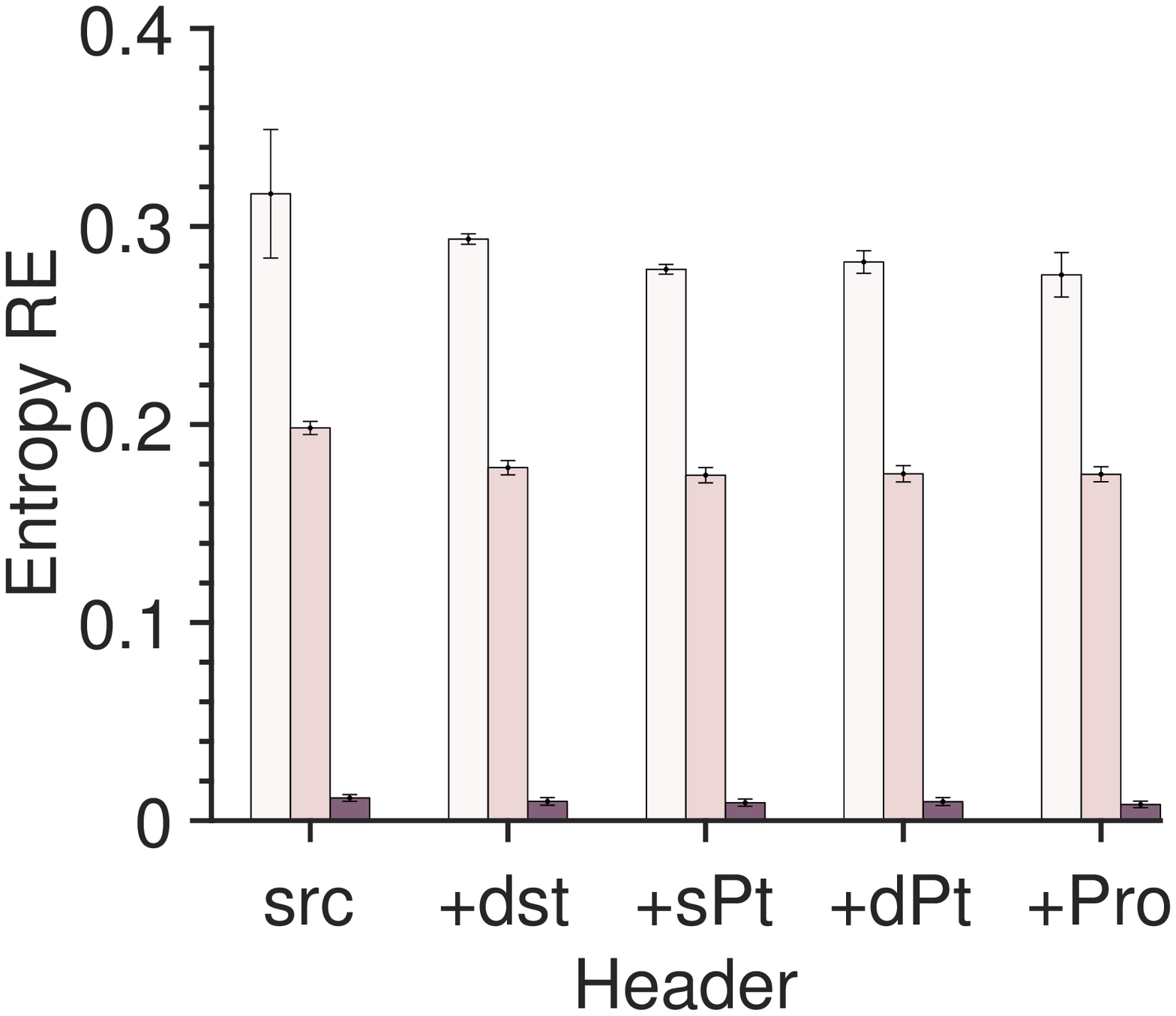}}}
  \subfigure[Heavy hitters]{
  {\includegraphics[width=.14\textwidth]{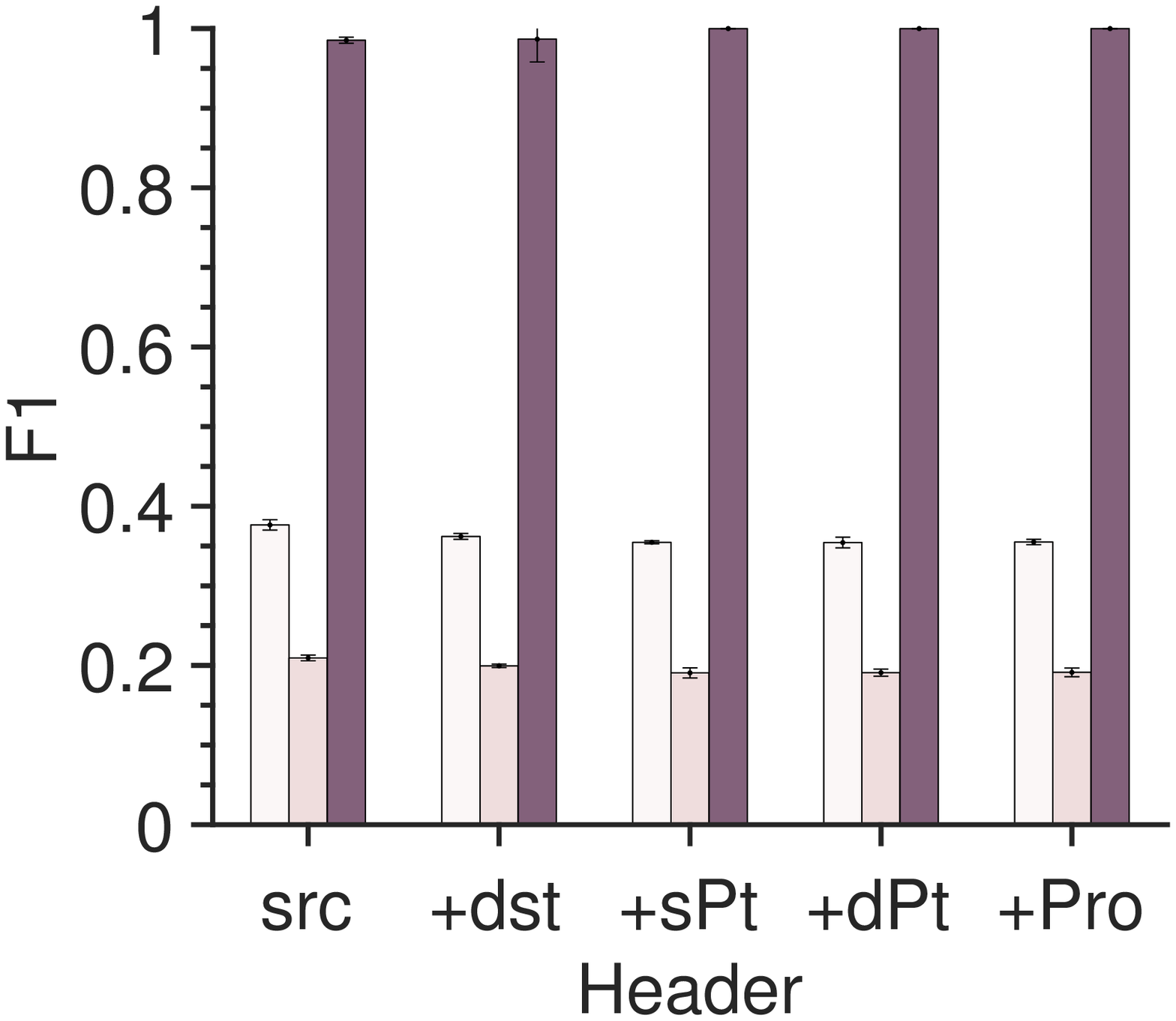}}}
     \caption{Accuracy of CS, CM and LSS as we expand the fields based on the 5-tuple model.}
     \label{fig:flowFields}
\end{figure} 
 }

\begin{figure}[!t]
  \centering
\subfigure[Flow query]{
  {\includegraphics[width=.14\textwidth]{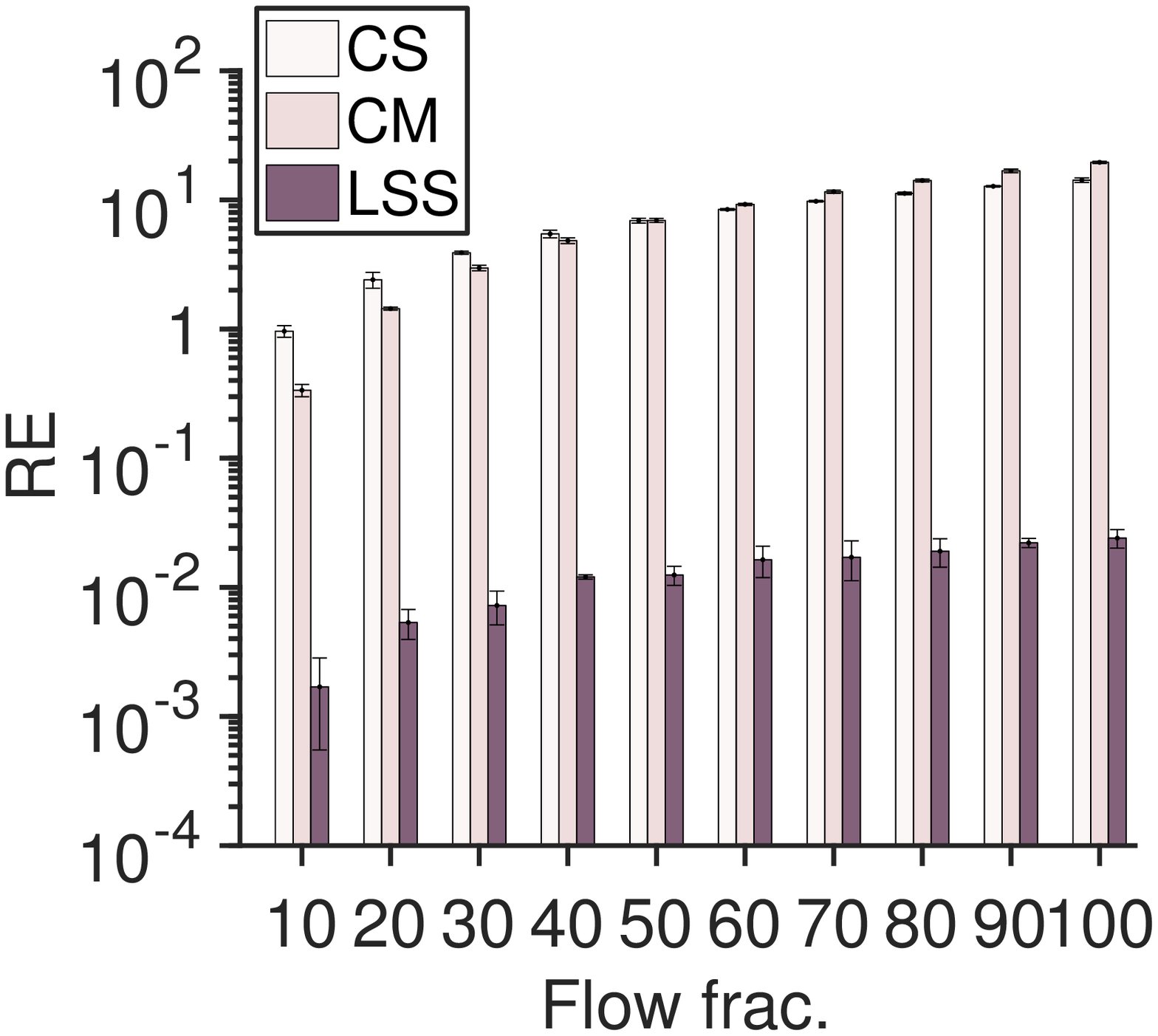}}}
  \subfigure[Entropy]{
  {\includegraphics[width=.14\textwidth]{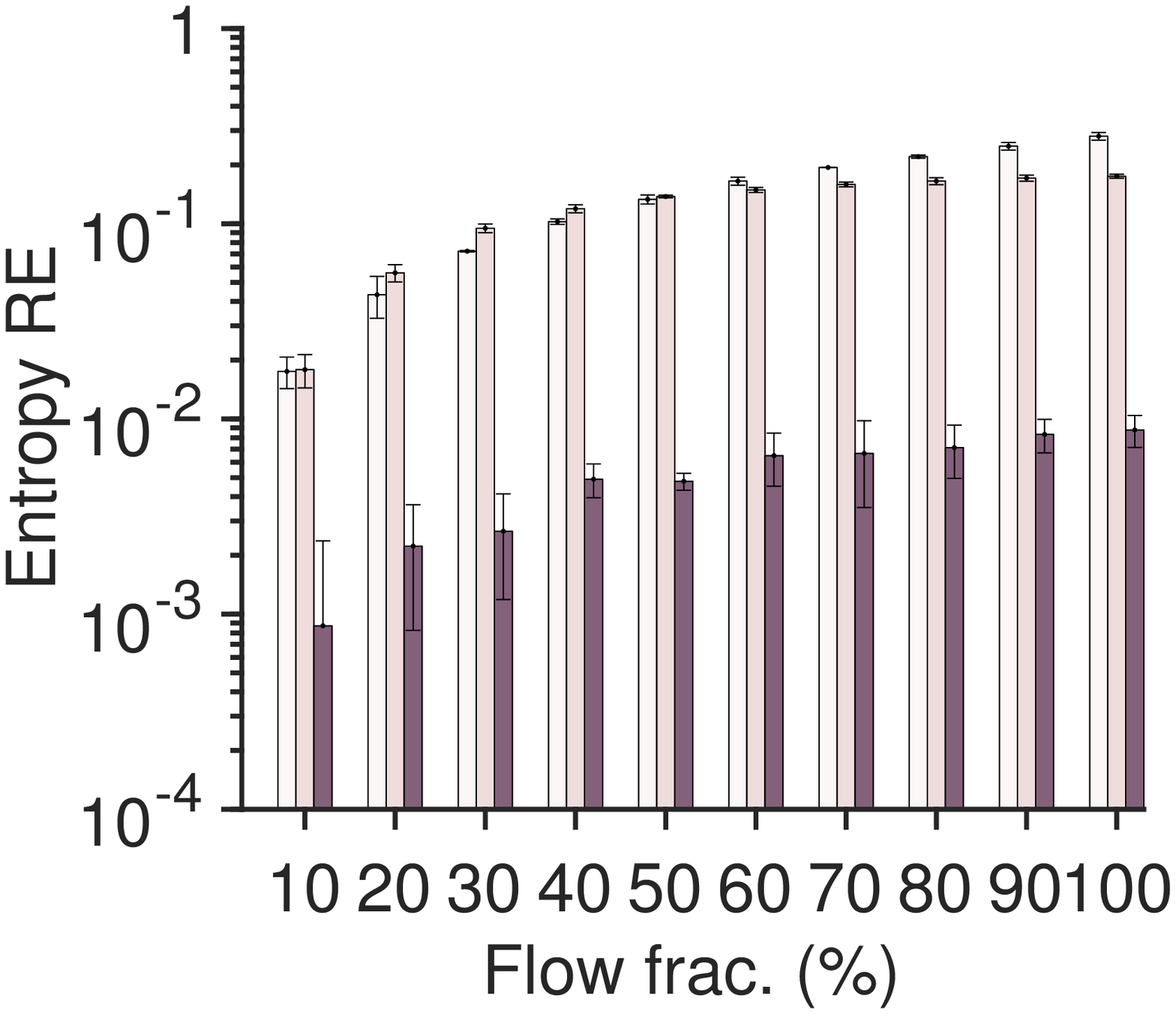}}}
  \subfigure[Heavy hitters]{
  {\includegraphics[width=.14\textwidth]{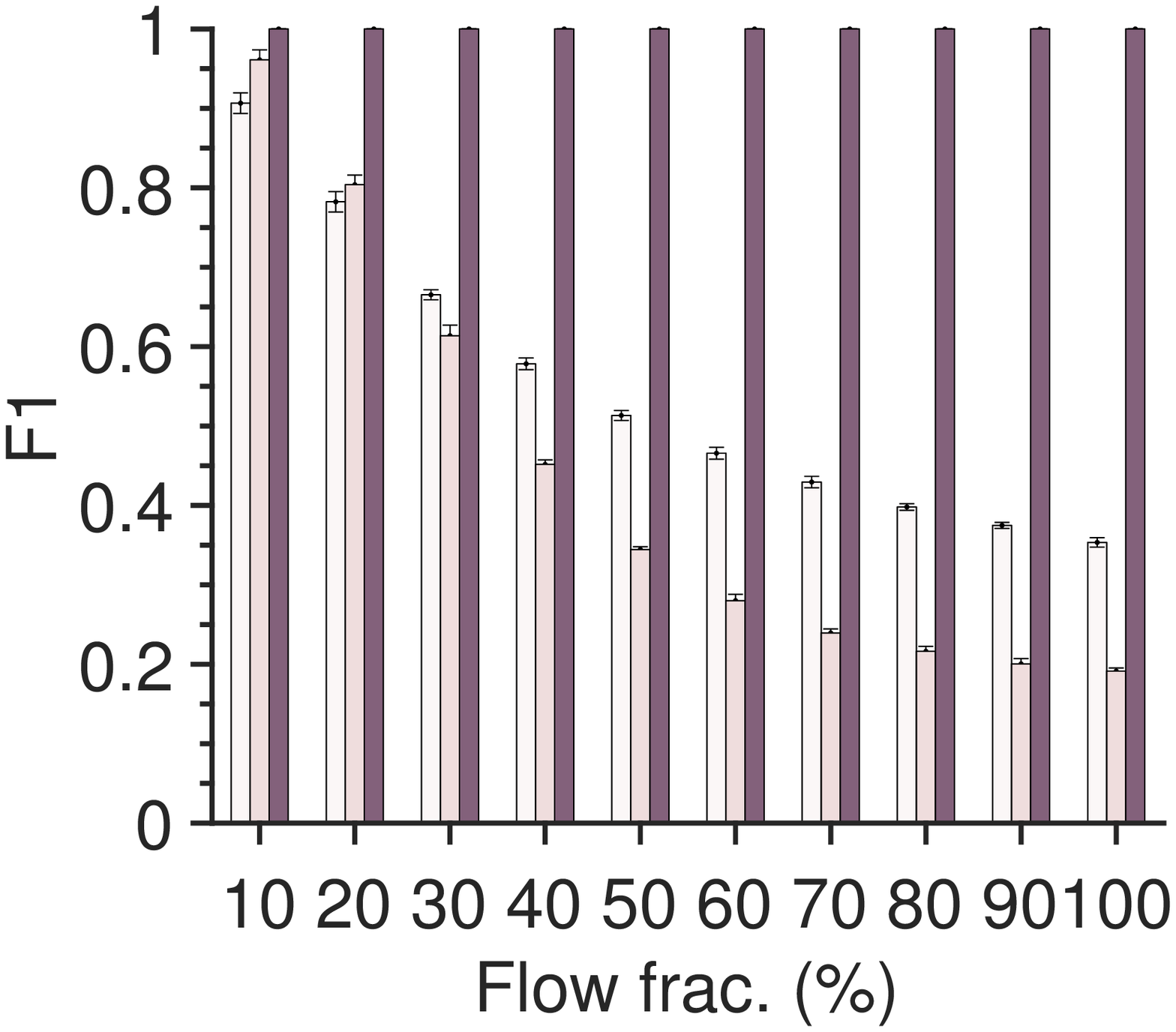}}}
     \caption{Performance  of CS, CM and LSS  as we vary the fractions of inserted flows.}
     \label{fig:FlowDynamic}
\end{figure}

(ii)  \subsubsubsection{Varying Flows}:  Having shown that filtering large flows from the sketch is  less effective than  an autoencoder based recovery of the locality-sensitive bucket arrays,  we next compare  CS, CM and LSS that do not filter flows with hash tables. Figure \ref{fig:FlowDynamic} shows that LSS remains fairly accurate across configurations, as we progressively add more flows in an epoch to the sketch.  While CS and CM are severely affected due to hash collisions. Since LSS clusters similar flows to the same bucket array, and performs the error minimization for each bucket array.

\subsubsection{Sensitivity}
\label{SensitivitySec}

Having shown that LSS remains fairly accurate across different memory footprints, we next evaluate the sensitivity of LSS.  We fix all but one parameters to the default configuration for the Testbed evaluation,  and study the performance variation as we change a specific parameter. 

\begin{figure}[!t]
  \centering
\subfigure[Flow query]{
  {\includegraphics[width=.14\textwidth]{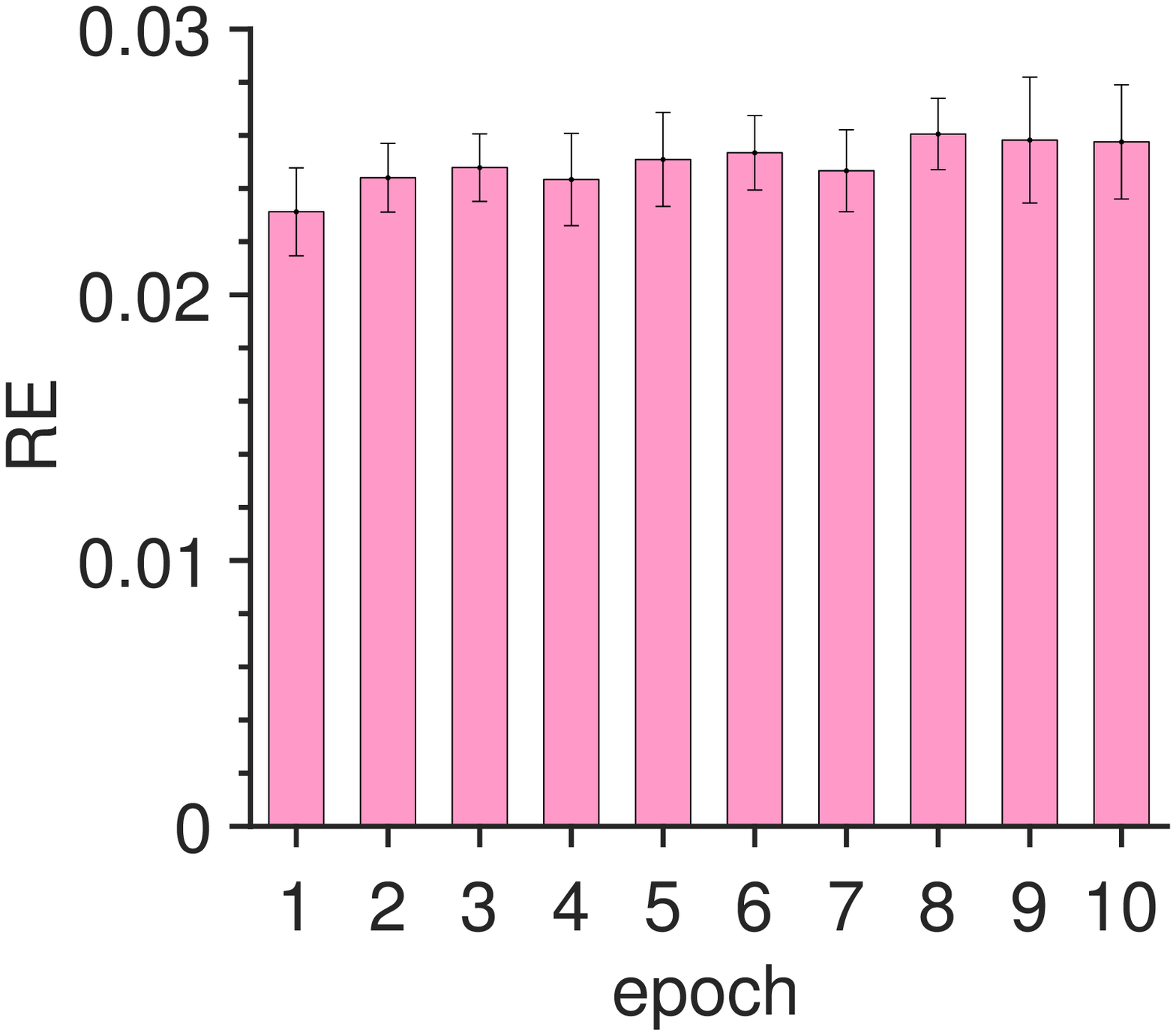}}}
  \subfigure[Entropy]{
  {\includegraphics[width=.14\textwidth]{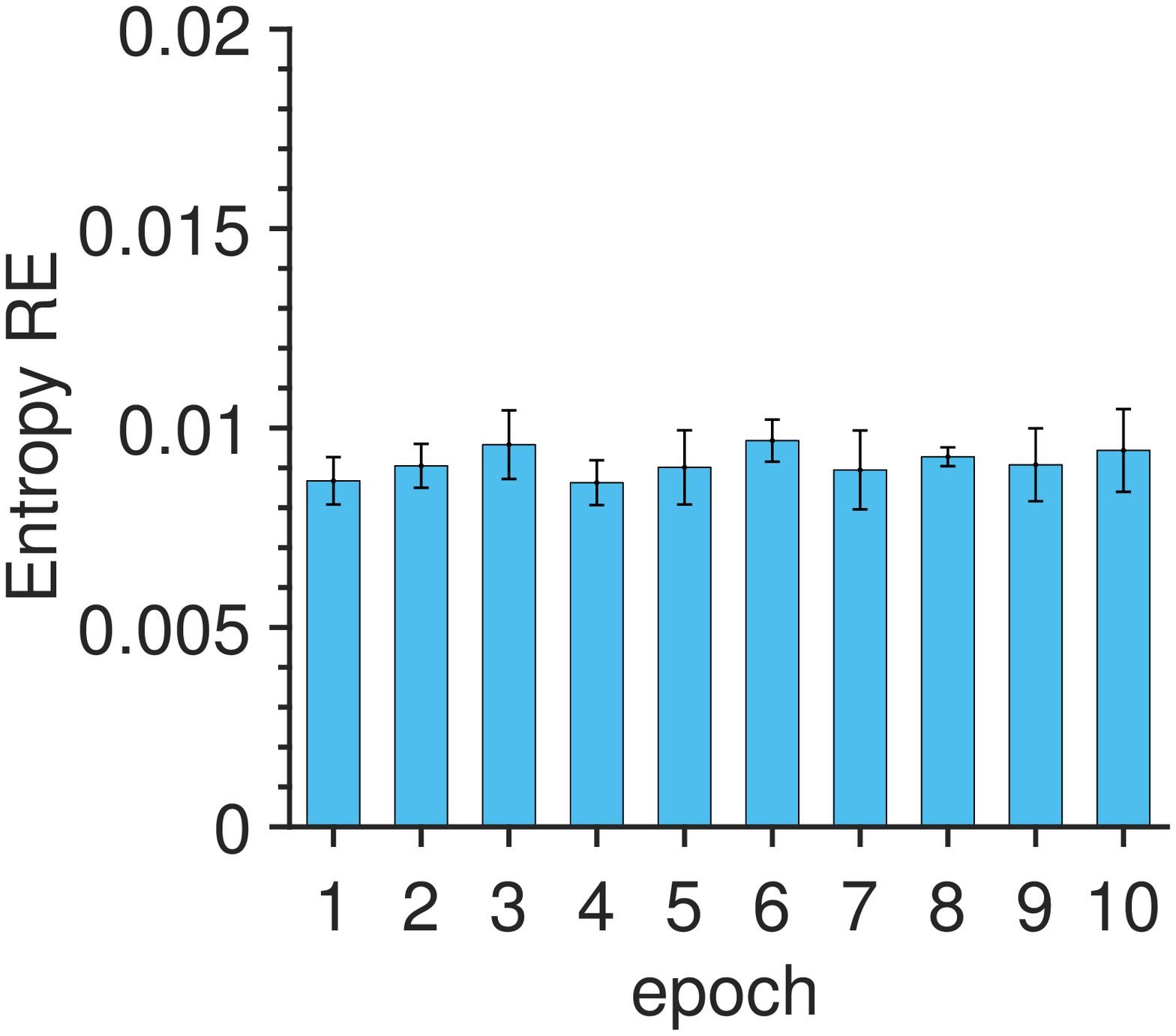}}}
  \subfigure[Heavy hitters]{
  {\includegraphics[width=.14\textwidth]{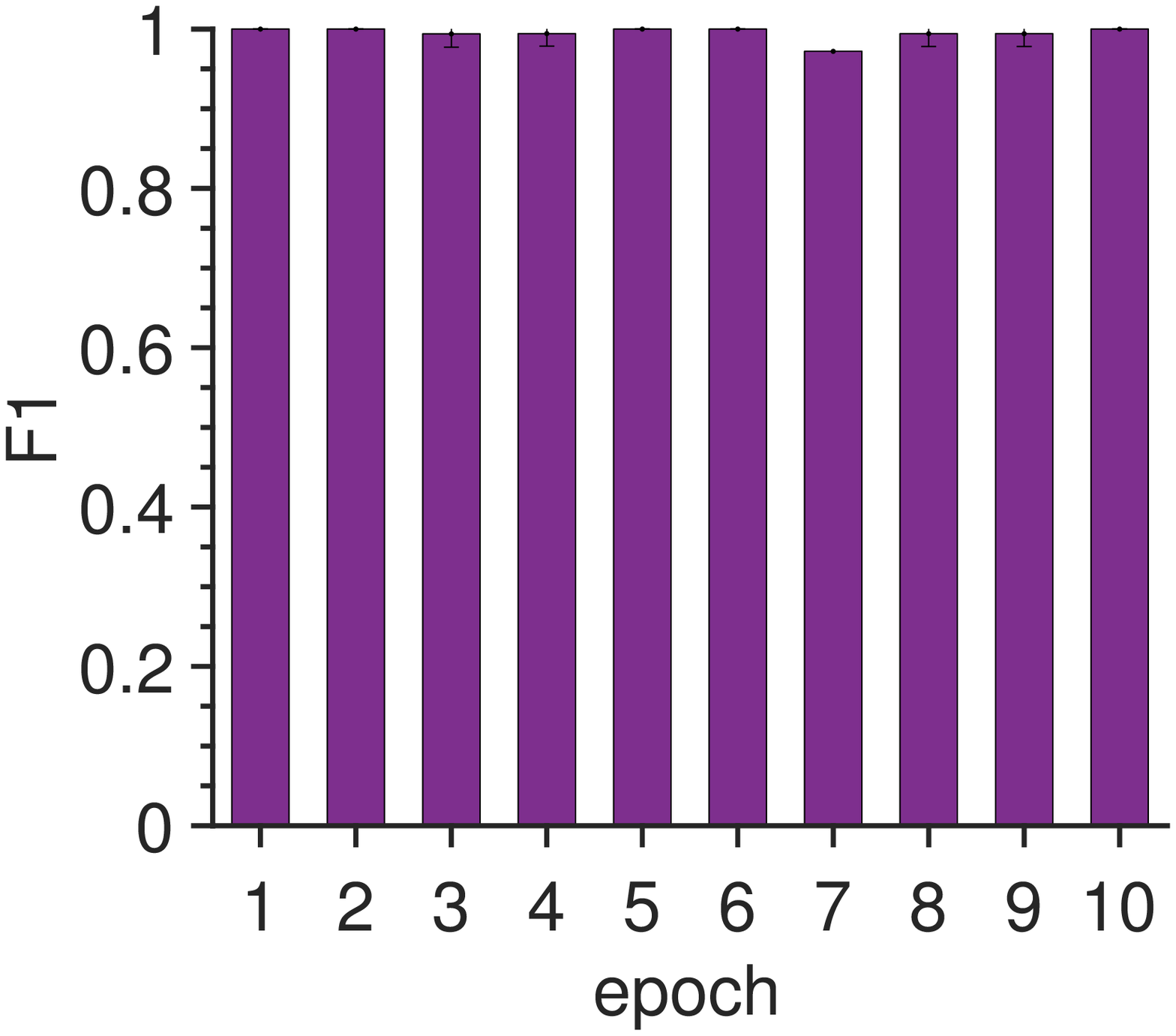}}}
     \caption{LSS performance of different epochs by reusing the offline cluster model of the first epoch.}
     \label{fig:ClusterStable}
\end{figure}

(i)  \subsubsubsection{Offline Cluster-model Stability}: We  tested  LSS'  sensitivity to offline clustering models by reusing the cluster centers that are trained with respect to the first epoch.  Figure \ref{fig:ClusterStable} shows that  three monitoring tasks remain fairly accurate across epochs. Since the cluster model captures the global structure of the flow distribution.

\co{Further, increasing the number of buckets significantly alleviates tails' sensitivity, since the estimation is close to exact for all flows. }

\begin{figure}[!t]
  \centering
\subfigure[Flow query]{
  {\includegraphics[width=.14\textwidth]{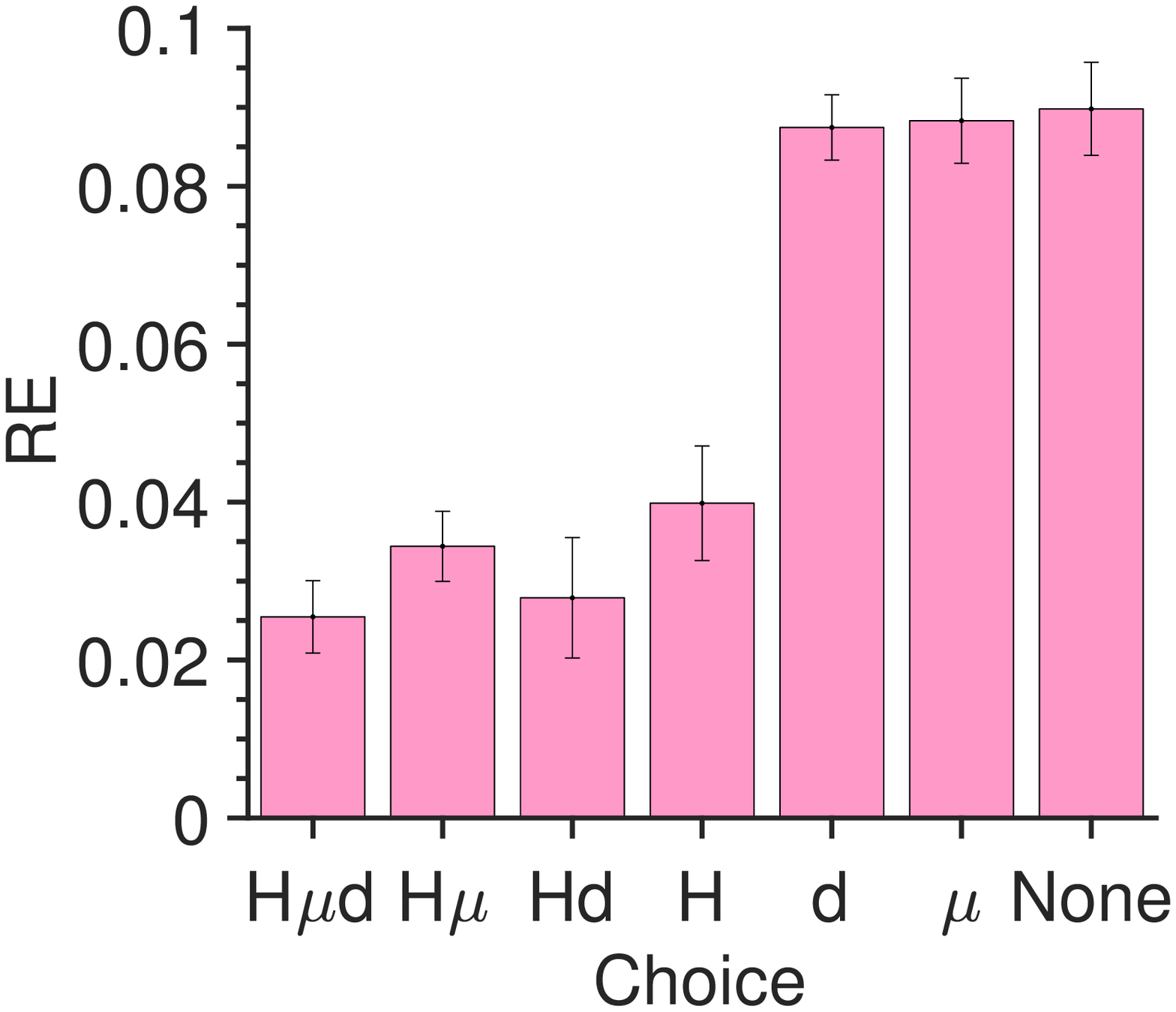}}}
  \subfigure[Entropy]{
  {\includegraphics[width=.14\textwidth]{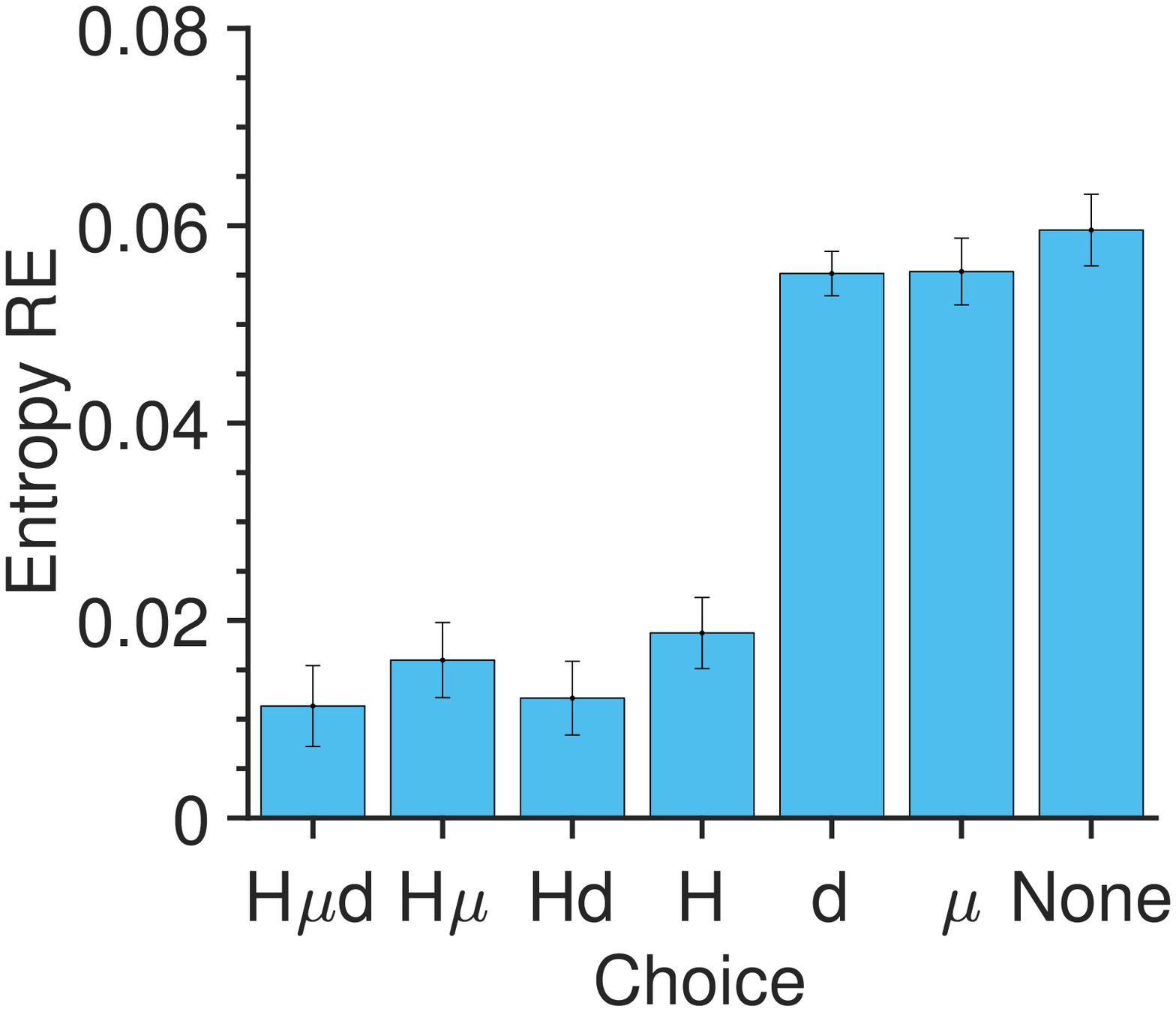}}}
  \subfigure[Heavy hitters]{
  {\includegraphics[width=.14\textwidth]{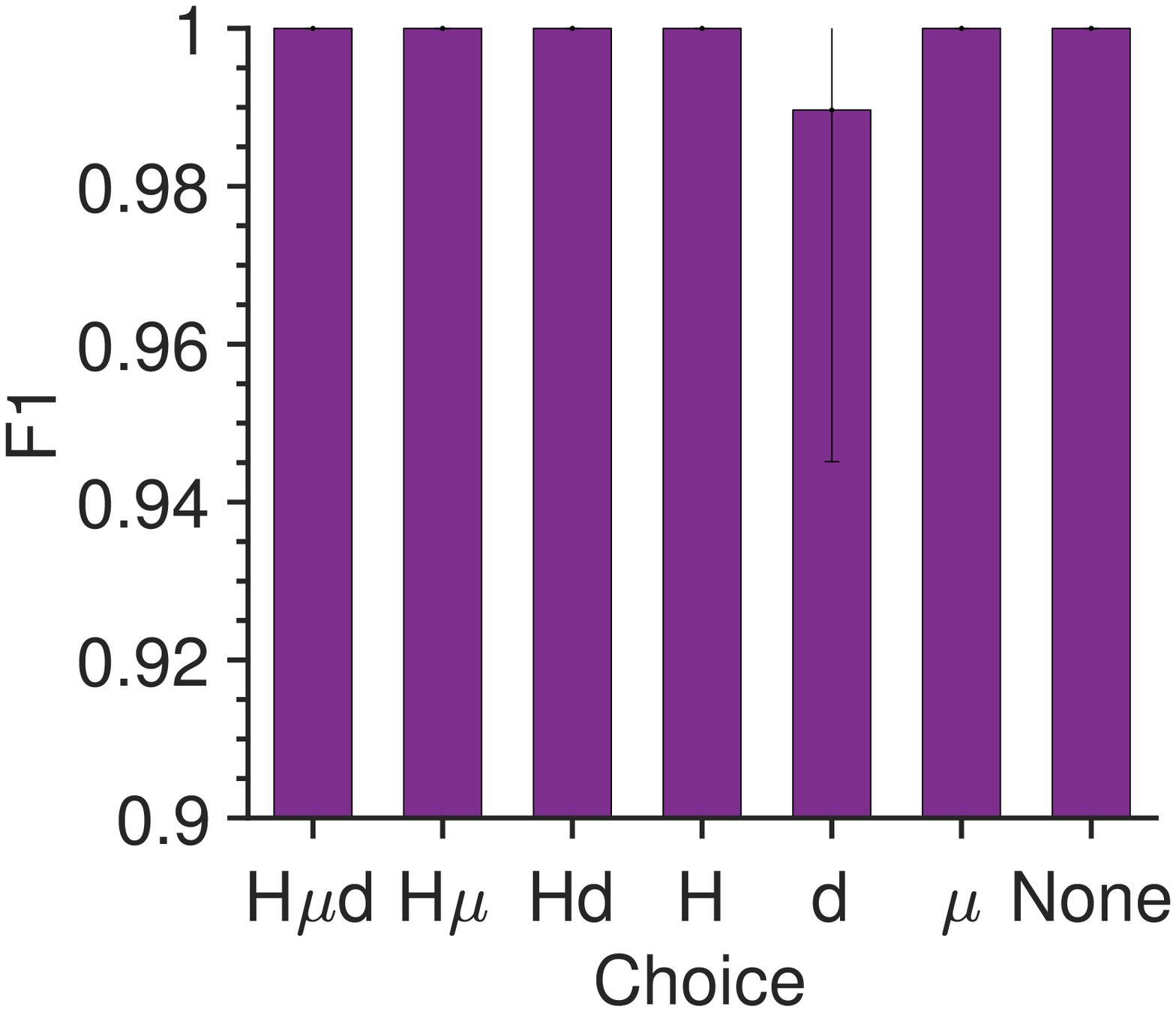}}}
     \caption{LSS performance vs. bucket-array policies.}
     \label{fig:Policy}
\end{figure}

(ii)  \subsubsubsection{Varying Bucket-array Policy}: We next test the effectiveness of the heuristics to configure the size of  bucket arrays. Figure \ref{fig:Policy} shows that the combination of the cluster uncertainty ($H$), the cluster center ($\mu$) and the cluster density ($d$) achieves high accuracy for three query tasks. We see that the cluster uncertainty is the most important metric, as removing the cluster uncertainty significantly degrades prediction accuracy.

\co{
(ii)  \subsubsubsection{With Or Without Clustering}: We next test the effectiveness of the clustering process for LSS. We skip the clustering process and map each flow record to all bucket arrays (denoted as LSS-c). We apply the clustering process to CM and CS (denoted as CM+c, CS+c, respectively).  Figure \ref{fig:DesignChoice} shows that LSS outperforms LSS-c by several times, thus the clustering is vital for LSS' performance. The clustering is  useless for CM and CS, as both CM+c and CS+c degrades the prediction accuracy.

 \begin{figure}[!t]
  \centering
\subfigure[Flow query]{
  {\includegraphics[width=.14\textwidth]{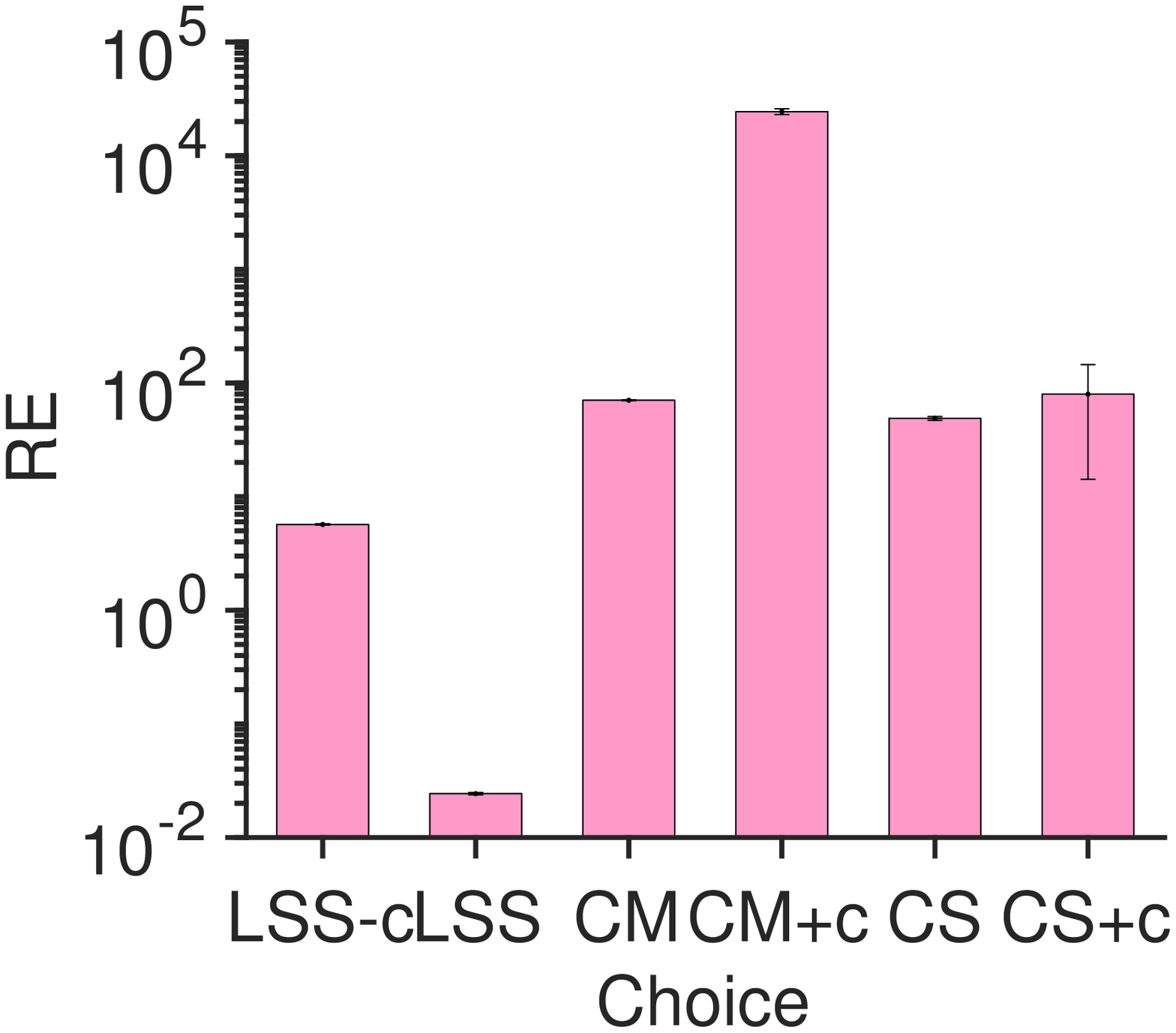}}}
  \subfigure[Entropy]{
  {\includegraphics[width=.14\textwidth]{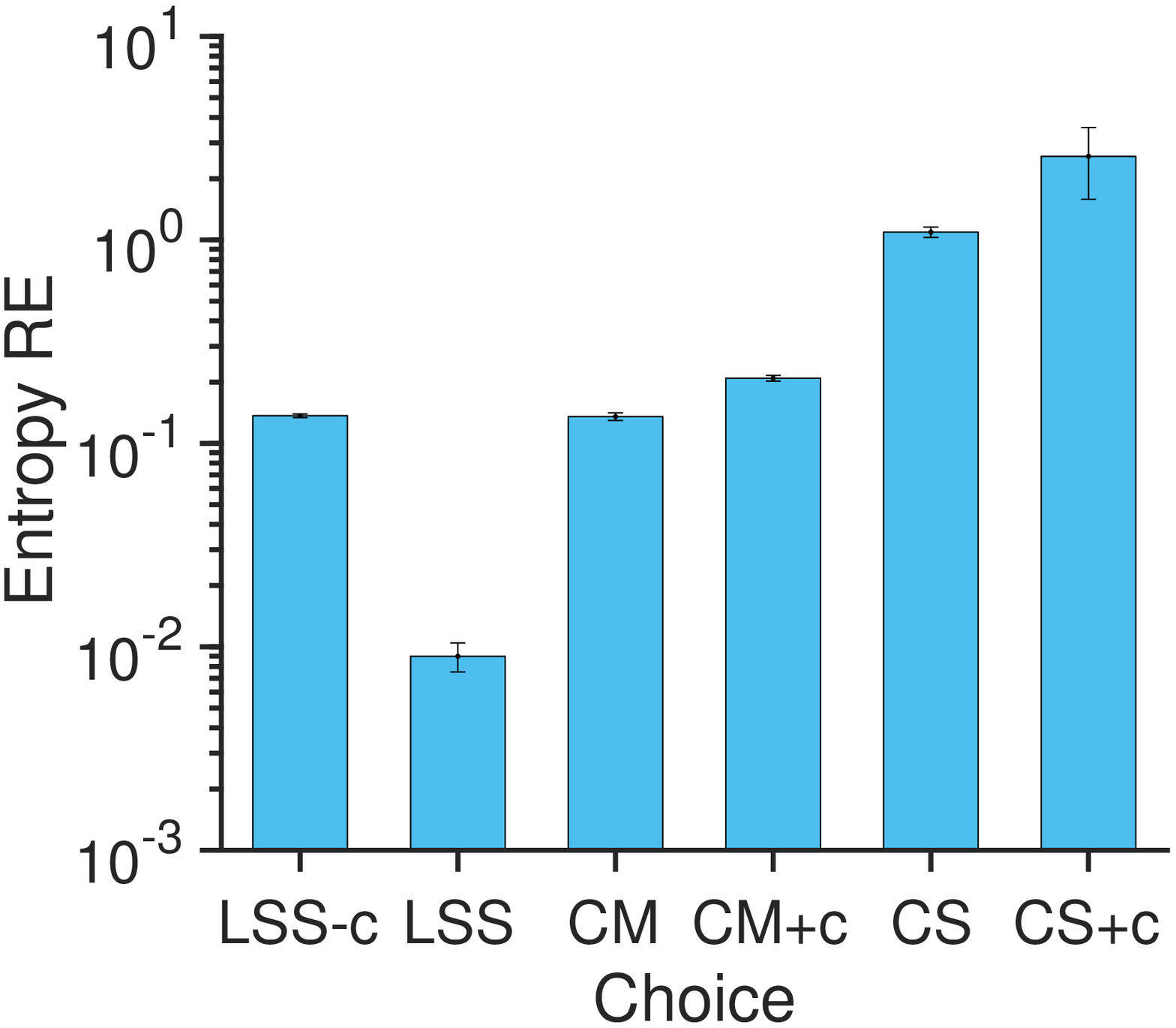}}}
  \subfigure[Heavy hitters]{
  {\includegraphics[width=.14\textwidth]{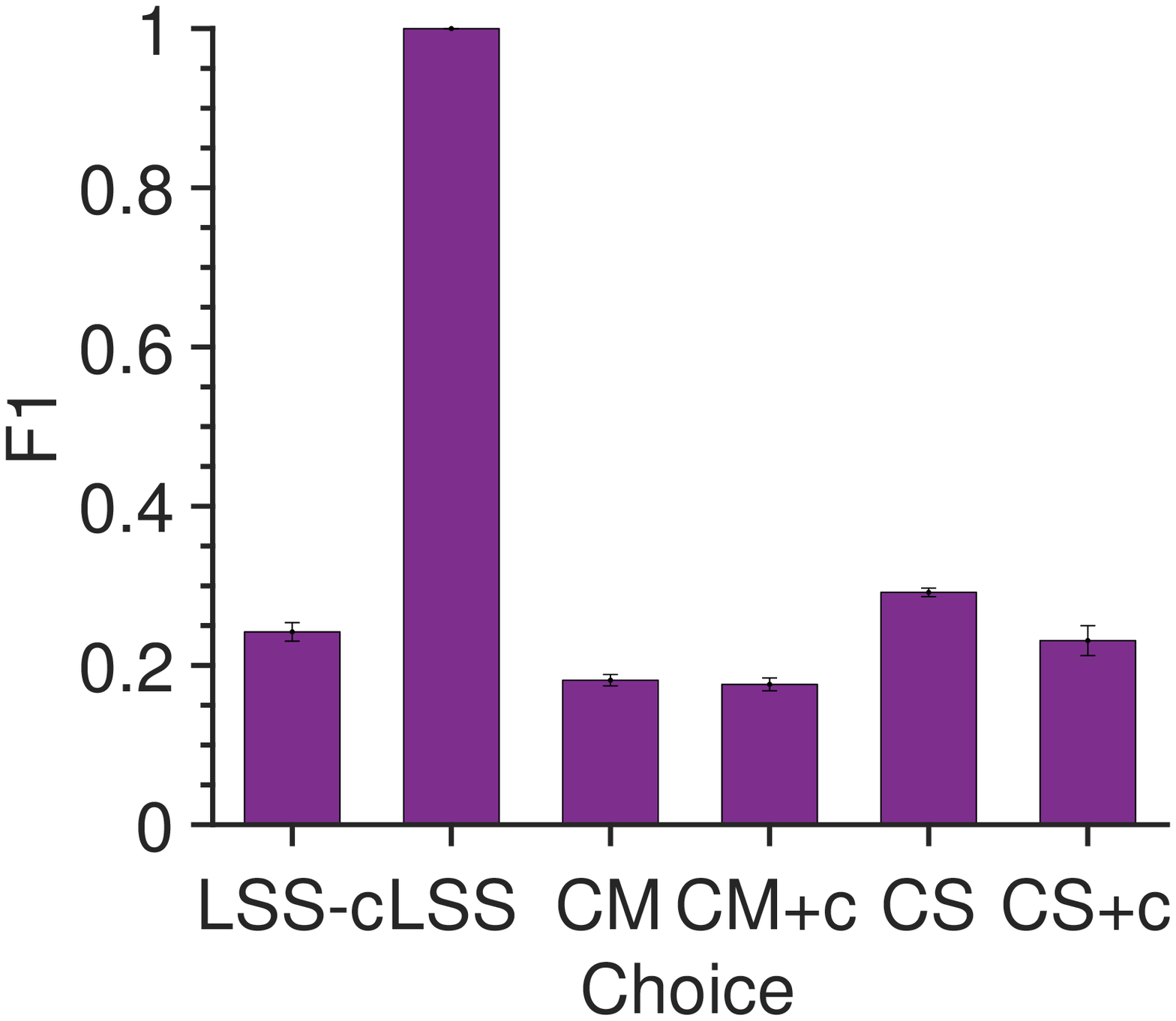}}}
     \caption{LSS performance vs. with or without clustering.}
     \label{fig:DesignChoice}
\end{figure}
}

\begin{figure}[!t]
  \centering
\subfigure[Flow query]{
  {\includegraphics[width=.14\textwidth]{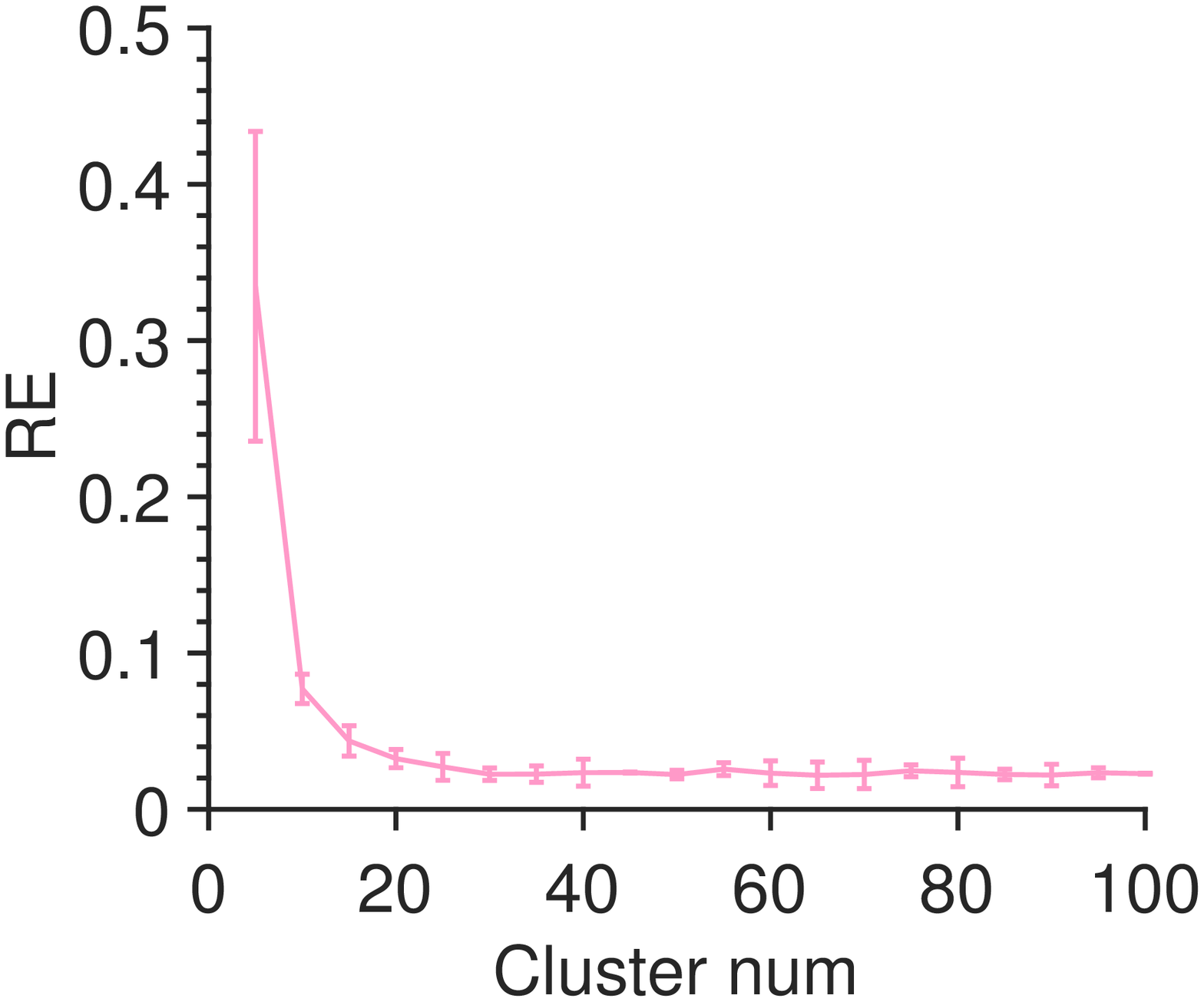}}}
  \subfigure[Entropy]{
  {\includegraphics[width=.14\textwidth]{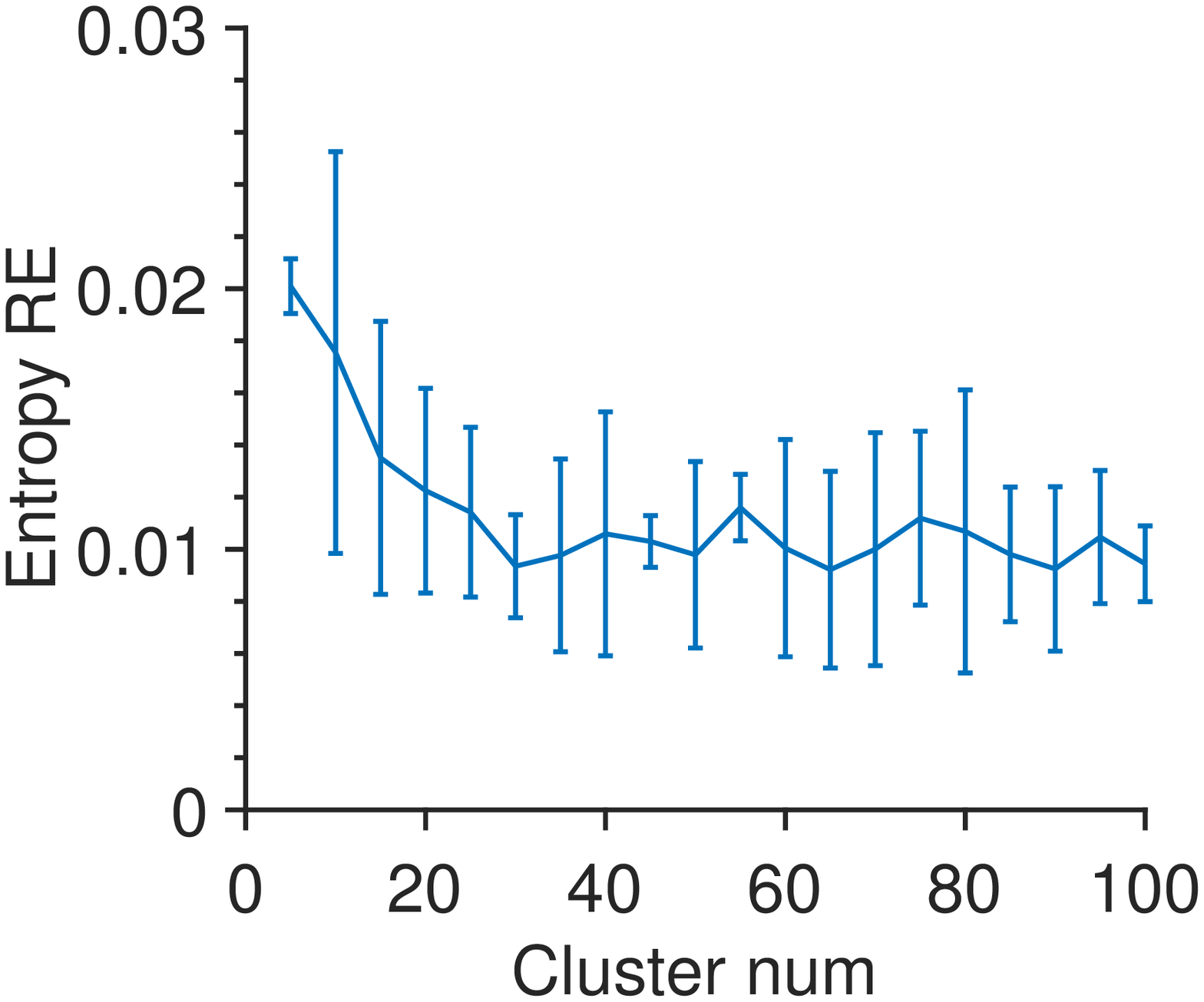}}}
  \subfigure[Heavy hitters]{
  {\includegraphics[width=.14\textwidth]{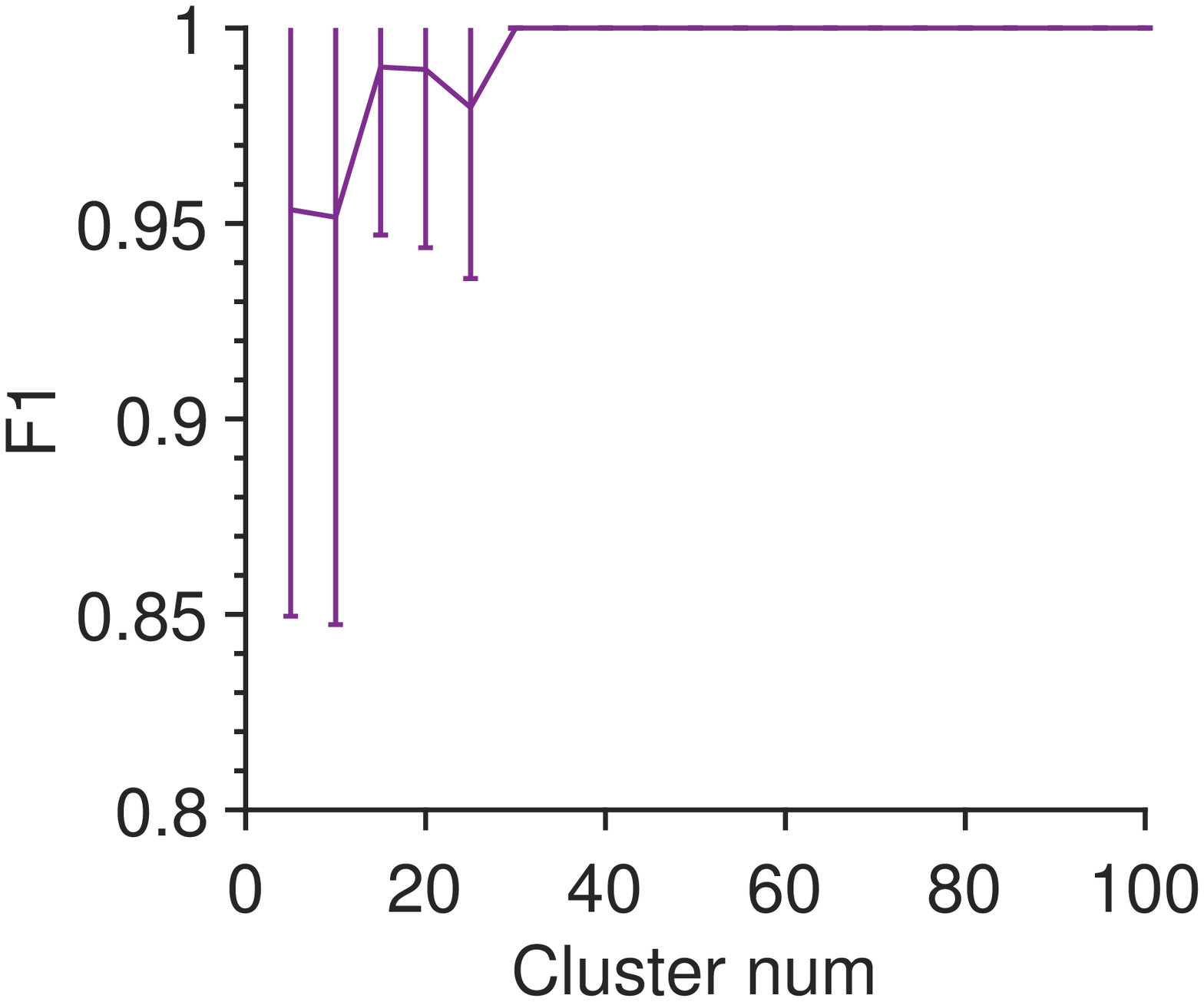}}}
     \caption{LSS performance as a function of the numbers of clusters.}
     \label{fig:FlowClusterNum}
\end{figure}

(iii)  \subsubsubsection{Number of Clusters}: Next, we evaluate LSS' accuracy with respect to the number of clusters. Figure \ref{fig:FlowClusterNum} plots the variation of the estimation accuracy. We see that the estimation accuracy   improves steadily with increasing numbers of clusters from two to ten. The diminishing returns occur when the number of cluster reaches 30.


\co{
(iv)\subsubsubsection{Varying Window Interval}: Next we test LSS' performance as we scale the size of the sliding window with respect to the length of an epoch. Figure \ref{fig:WindowVary} shows that LSS keeps stable performance with increasing sliding windows. This is because by fixing the ratio between the number of buckets and the number of flows to 0.1, the expected number of unique flows per bucket remains the same. 

 \begin{figure}[!t]
  \centering
\subfigure[Flow query]{
  {\includegraphics[width=.14\textwidth]{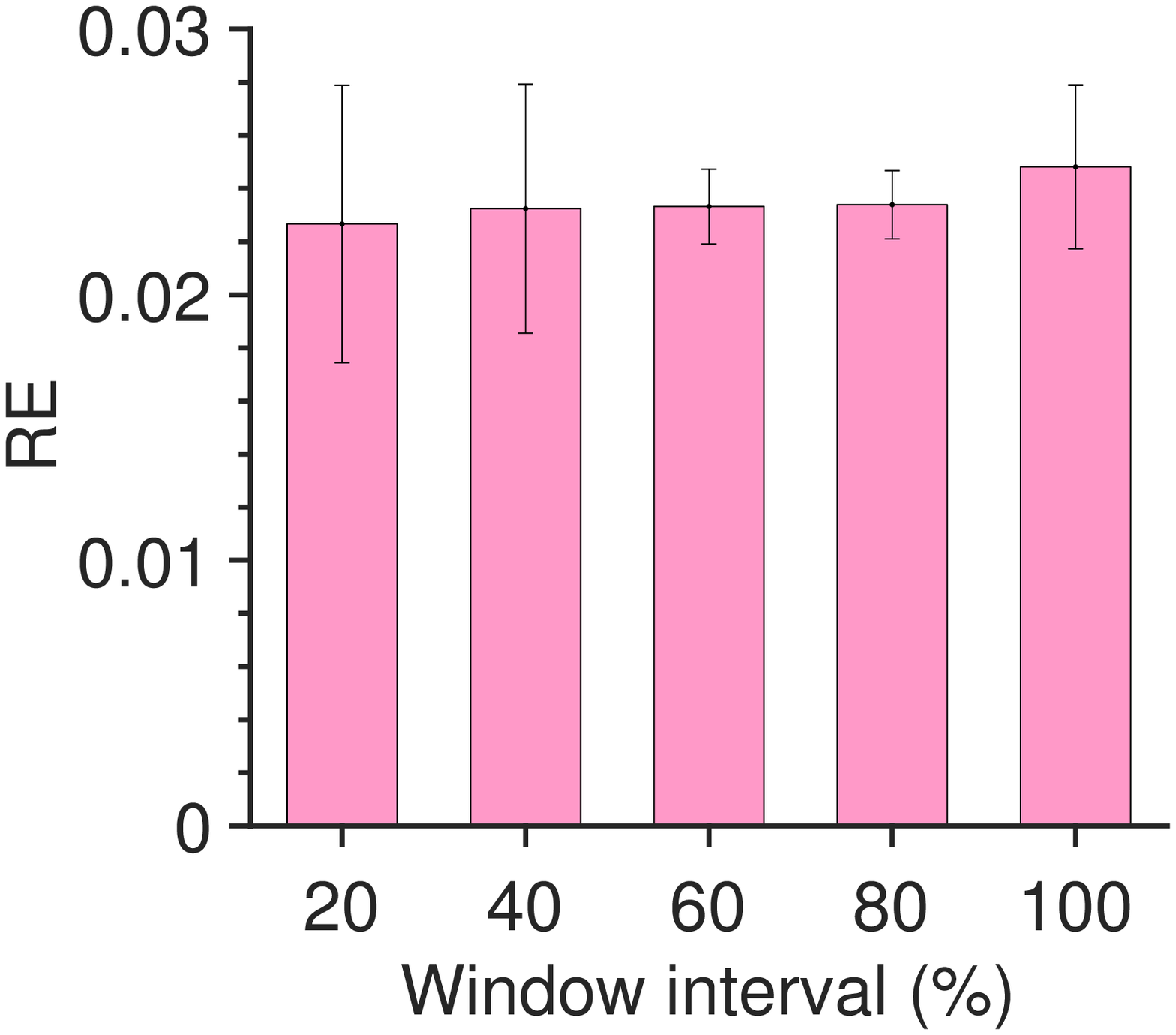}}}
  \subfigure[Entropy]{
  {\includegraphics[width=.14\textwidth]{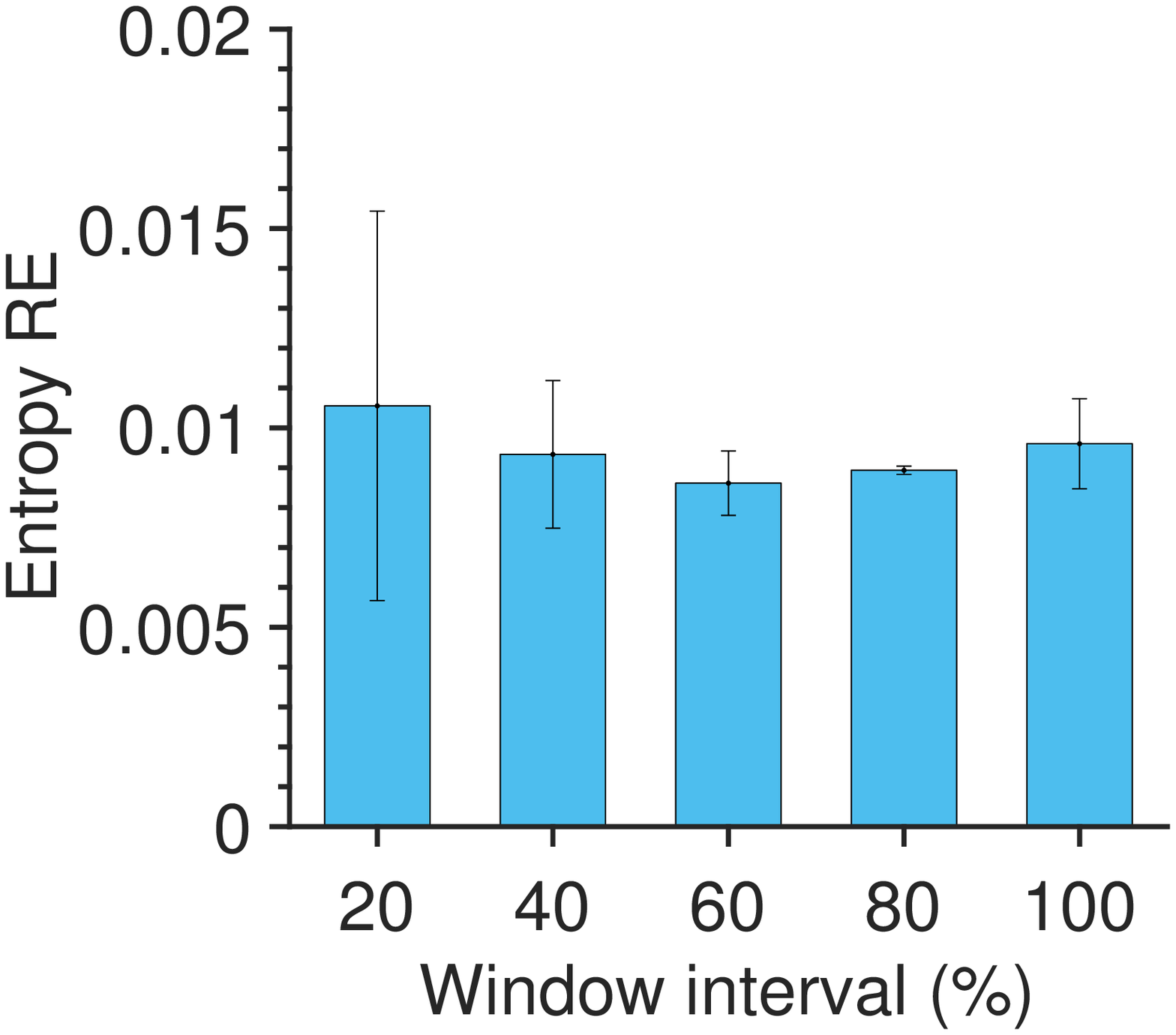}}}
  \subfigure[Heavy hitters]{
  {\includegraphics[width=.14\textwidth]{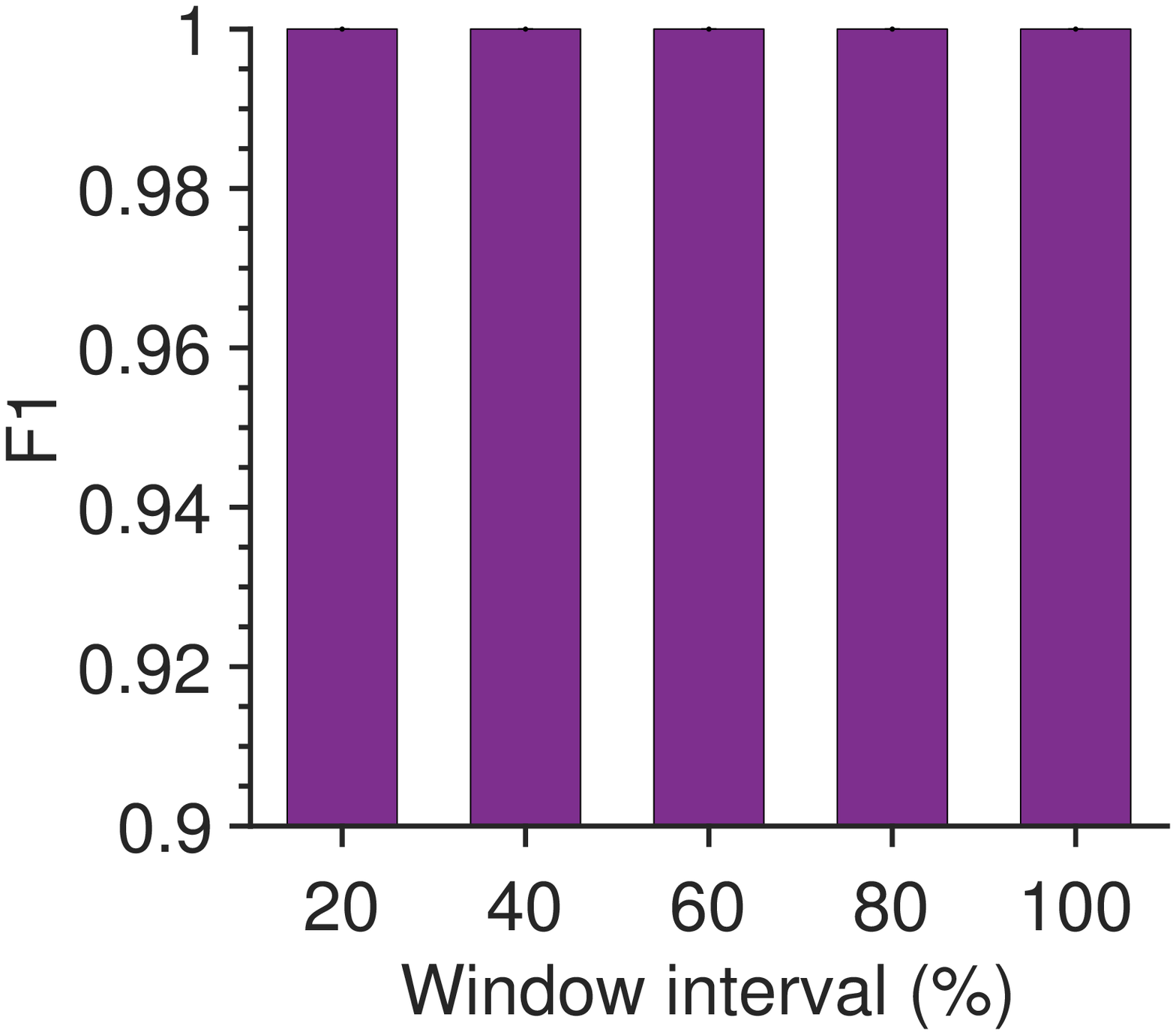}}}
     \caption{LSS performance as a function of the scaled window intervals.}
     \label{fig:WindowVary}
\end{figure}
}

\begin{figure}[!t]
  \centering
{
  {\includegraphics[width=.2\textwidth]{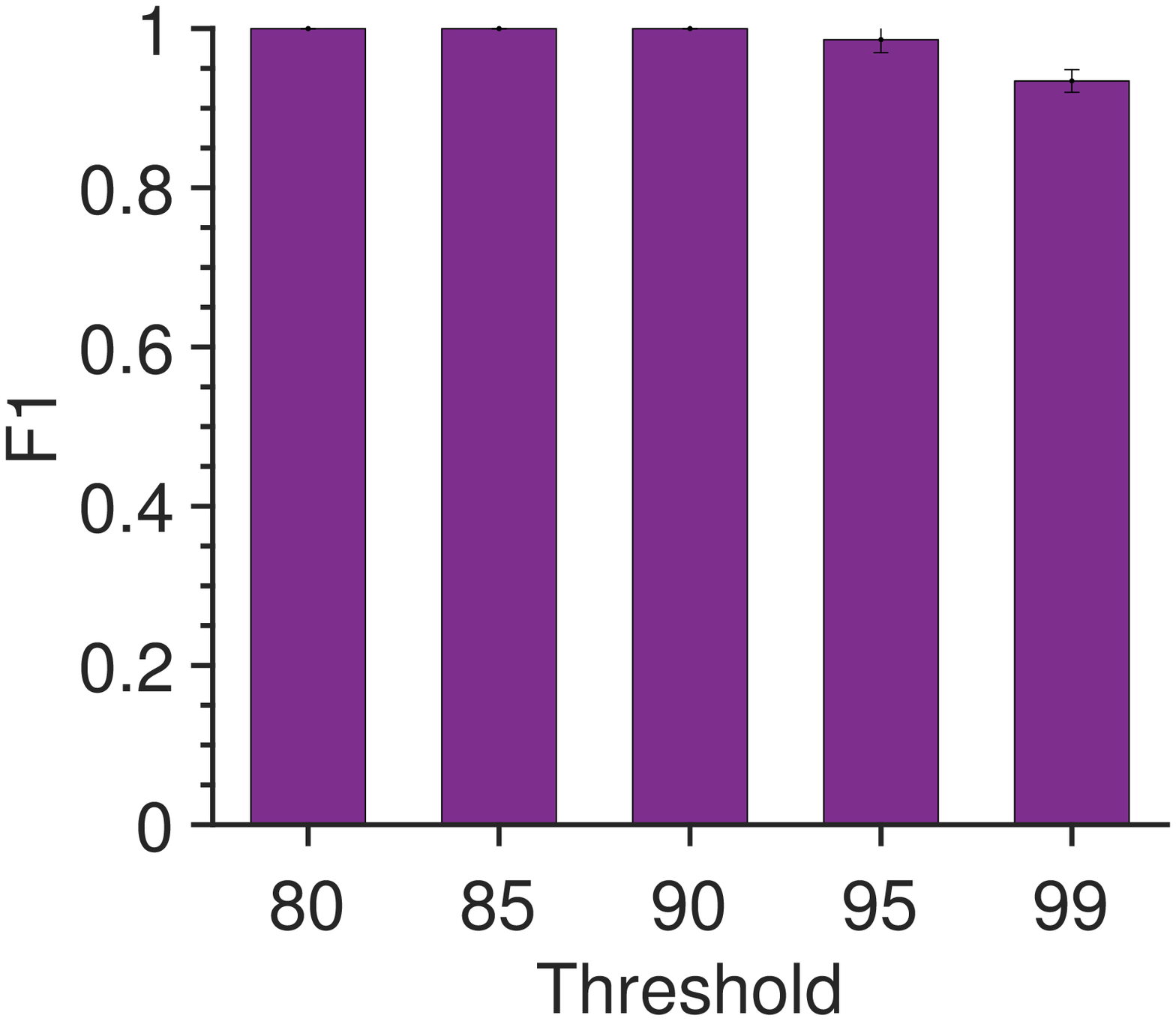}}}
     \caption{F1 scores as a function of  heavy-hitter thresholds.}
     \label{fig:FlowHHThreshold}
\end{figure}

(iv)  \subsubsubsection{Varying Thresholds}: We also tested  LSS'  sensitivity to different heavy-hitter thresholds. Figure \ref{fig:FlowHHThreshold} shows  the heavy-hitter performance degrades gracefully as we change the threshold percentiles from 80 to 99, since heavy hitters are more sensitive to estimation errors as we approach to tighter tails.

 \section{Related Work}

Modern network measurement systems typically rely on programmable frameworks to perform diverse monitoring tasks. First, the end based approach such as Trumpet \cite{DBLP:conf/sigcomm/MoshrefYGV16}, deTector \cite{DBLP:conf/usenix/PengYWGHL17}, Confluo \cite{225998} relies on edge servers to perform end-to-end packet-stream monitoring. To increase  in-network visibility, an in-network based approach such as OpenSketch \cite{DBLP:conf/nsdi/YuJM13}, Planck \cite{DBLP:conf/sigcomm/RasleySDRFABF14},  Marple \cite{DBLP:conf/sigcomm/NarayanaSNGAAJK17},  Everflow \cite{DBLP:conf/sigcomm/ZhuKCGLMMYZZZ15}, FlowRadar \cite{DBLP:conf/nsdi/LiMKY16},  UnivMon \cite{DBLP:conf/sigcomm/LiuMVSB16},  Sonata \cite{DBLP:conf/sigcomm/GuptaHCFRW18} combines the software-defined framework and the  programmability of switches to track fine-grained traffic statistics. A hybrid approach such as PathDump \cite{DBLP:conf/osdi/Tammana0L16}, SwitchPointer \cite{DBLP:conf/nsdi/Tammana0L18} and  \cite{DBLP:conf/sigcomm/RoyZBPS15} combines the resource-intensive end servers and the in-network visibility of  switches.   We present a disaggregated monitoring framework that can be incorporated with end hosts and programmable switch based systems, in order to maximize the network visibility and support non-intrusive monitoring.  
 


\co{ counters indexed by hashing numbers of the key with the same set of hash functions as the insertion proces. select the buckets indexed by these  random numbers from $k$ banks, and finally }


Existing sketches typically choose to tolerate hash collisions with space redundancy. For instance, state-of-the-art sketch structures  \cite{DBLP:conf/icalp/CharikarCF02,DBLP:conf/latin/CormodeM04,DBLP:journals/cacm/CormodeH09,DBLP:conf/sigcomm/0003JLHGZMLU18} choose the least affected bucket from multiple independent bucket arrays. Recently,  ElasticSketch \cite{DBLP:conf/sigcomm/0003JLHGZMLU18}  keeps heavy hitters separately with a  hash table, and puts the rest of items to a count-min sketch. Thus it is less sensitive to heavy hitters compared to prior sketch structures \cite{DBLP:conf/icalp/CharikarCF02,DBLP:conf/latin/CormodeM04,DBLP:journals/cacm/CormodeH09}. Unfortunately, as heavy hitters only represent a small fraction of items, the count-min sketch is still sensitive to hash collisions.  Our work proactively mitigates the downsides of the hash collisions with locality-aware clustering and bucket averaging techniques. 

The sketch structure has been augmented in several dimensions. UnivMon \cite{DBLP:conf/sigcomm/LiuMVSB16} uses an array of count sketch to meet generic monitoring tasks. SketchVisor \cite{DBLP:conf/sigcomm/HuangJLLTCZ17} augments the sketch with a fast-path ingestion path to tolerate bursty traffic. SketchLearn \cite{DBLP:conf/sigcomm/HuangLB18} uses a multi-level array to keep the traffic statistics of specific flow-record bits, and separates large flows from the rest of flows like ElasticSketch \cite{DBLP:conf/sigcomm/0003JLHGZMLU18}   based on inferred flow distributions. Although our work is orthogonal to these studies, the LSS sketch structure can be combined to these frameworks to improve the sketching efficiency. 

Several studies propose to  keep network flow statistics in a hash table \cite{DBLP:conf/sigcomm/MoshrefYGV16,DBLP:conf/sosr/AlipourfardMZ0Y18} at end hosts, whereas the storage requirement is on the order of the number of network flows.  As network flows arrive continuously, the hash table based monitoring application incurs expensive memory costs that reduce the available resources for colocated tenants. Since for multi-tenant data centers, the monitoring application has to control its resource usage to maximize the available resource to meet tenants' needs.   Moreover, the hash table needs to dynamically adjust the data structure when hash collisions occur, i.e., multiple keys are mapped to the same bucket, with linear hashing \cite{DBLP:conf/vldb/Litwin80},  Cuckoo hashing  \cite{DBLP:journals/jal/PaghR04}, or hopscotch hashing \cite{DBLP:conf/wdag/HerlihyST08}. The hash table is agnostic of the self-similarity structure of flow counters, while the sketch can exploit this property to compress the flow counters to a constant-size array.  

\section{Conclusion}

We have proposed a new class of sketch that is resilient to hash collisions, which  groups similar items together to the same bucket array in order to mitigate the error variance, and optimizes the estimation based on an autoencoder model to minimize the estimation error.   We showed that LSS is equivalent to a linear autoencoder that minimizes the  recovery error. To illustrate the feasibility of LSS sketch,  we present a disaggregated monitoring application that decomposes monitoring functions to disaggregated components, which allows for non-intrusive sketch deployment and native network-wide analytics. Extensive evaluation shows that LSS achieves close to optimal performance with a tiny memory footprint, which generalizes to diverse monitoring contexts.

\bibliography{lss}
\bibliographystyle{abbrvnat_noaddr} 

\appendix

\section{Supplementary of Section II}

\subsection{Proof of Lemma 1}
\label{LemmaAppendix}

\textbf{Lemma}:   Let $m$ denote the number of buckets, $N$ the number of unique keys.  For a sketch with $c$ banks of bucket arrays, where each bucket array is of size $\frac{m}{c}$, the expected percent of noisy buckets is $1-{ e }^{ -cN/m }-\frac { cN }{ m } \cdot { e }^{ -c\left( N-1 \right) /m }$.
 
 \begin{proof}
For a sketch with one bucket array that consists of  $m$ buckets, the expected number of keys per bucket amounts to  $\frac{N}{m}$. The expected number of empty buckets is:
$
\sum _{ i }^{  }{ { \left( 1-\frac { 1 }{ m }  \right)  }^{ N } } =m{ \left( 1-\frac { 1 }{ m }  \right)  }^{ N }\approx m{ e }^{ -N/m }
$.
 Similarly, the expected number of buckets  with one key amounts to:
$\sum _{ i }{  \left( \begin{matrix} N{  } \\ 1 \end{matrix} \right) { \left( \frac { 1 }{ m }  \right)  }^{  }{ \left( 1-\frac { 1 }{ m }  \right)  }^{ N{  }-1 } \ } 
 \approx N{ e }^{ -\left( N-1 \right) /m }
$.
 As a result,  the expected percent of buckets that contain at least two keys is $\left( m-m{ e }^{ -N/m }-N{ e }^{ -\left( N-1 \right) /m } \right) \cdot \frac { 1 }{ m } $.    The expected percent of noisy buckets is $1-{ e }^{ -N/m }-\frac { N }{ m } \cdot { e }^{ -\left( N-1 \right) /m }$. 
 
For a sketch with $c$ banks of bucket arrays, where each bucket array is of size $\frac{m}{c}$. We can see that each bucket array still receives $N$ keys. Thus following the same derivation, we have that  the corresponding expected percent of noisy buckets is $1-{ e }^{ -cN/m }-\frac { cN }{ m } \cdot { e }^{ -c\left( N-1 \right) /m }$.
\end{proof}

\subsection{CM and CS Performance Bounds}
\label{BoundAppendix}

\co{Both keep  $k$ banks of arrays of size $m$. The insertion process for Count-Sketch differs a bit, as it  chooses $k$ random sign functions $r_i$ ($i \in \left\{ 1,2,\cdots ,k \right\} $) to weigh the value  by a random sign from $\left\{ +1,-1 \right\} $, and update each selected bucket by the weighted value of the given key, i.e., value $\cdot r_i$(key). For Count-min, it directly increases the counter of the selected bucket by  the value of the incoming key. The query process is similar to the insertion process, for Count-Sketch, it calculates the median of weighted values stored in each bucket, i.e., median value \newline $\left\{ bucket(h_j(key)) \cdot r_j(key) \right\}$; for the Count-min, it approximates the value of a given key by the minimum of mapped buckets. }

A count-min sketch maintains $k$ banks of arrays of size $m$, where $k$ and $m$ are chosen based on the accuracy requirement.  To insert a  key-value pair to the sketch, we  chooses $k$ uniformly-random hash functions $h_j$, $j \in \left\{ 1,2,\cdots ,k \right\} $ to map each key to a randomly chosen bucket from each bank. 

The insertion process for Count-Sketch differs a bit, as it  chooses $k$ random sign functions $r_i$ ($i \in \left\{ 1,2,\cdots ,k \right\} $) to weigh the value  by a random sign from $\left\{ +1,-1 \right\} $, and update each selected bucket by the weighted value of the given key, i.e., value $\cdot r_i$(key). For Count-min, it directly increases the counter of the selected bucket by  the value of the incoming key.

To query a given key,  we use the same set of hash functions to select $k$ buckets from each bank ( for the $j$-th bank, the $ h_j$(key)-th bucket is selected). For Count-Sketch, it calculates the median of weighted values stored in each bucket, i.e., median value \newline $\left\{ bucket(h_j(key)) \cdot r_j(key) \right\}$. While for the Count-min, it approximates the value of a given key by the minimum of mapped buckets. Count-min and Count-sketch needs $k$ hash-function computations and $k$ memory operations when inserting or querying a key-value pair.

For count-min sketch (CM) \cite{DBLP:conf/latin/CormodeM04},  the probability of the minimum of the inserted buckets is greater than the ground-truth value by $\frac { 2 }{ m } { \left\| v\left( x \right)  \right\|  }_{ 1 }$ is:
\[
Pr\left[ min\left( I\left[ i \right] \left[ { h }_{ i }\left( { x }_{ i } \right)  \right]  \right) -v\left( { x }_{ i } \right) \ge \frac { 2 }{ m } { \left\| v\left( x \right)  \right\|  }_{ 1 } \right] \le \frac { 1 }{ { 2 }^{ k } } 
\]
, and the probability of the median of the inserted buckets is 
\[
Pr\left[ { \left( { \hat { x }  }_{ i }-{ x }_{ i } \right)  }^{ 2 }>\frac { t }{ k } \cdot \frac { { \left\| { x }_{  } \right\|  }_{ 2 }^{ 2 } }{ m }  \right] <2{ e }_{  }^{ -\Omega \left( t \right)  }
\]
 for CS \cite{DBLP:conf/icalp/CharikarCF02}

\section{Supplements of Section III}

\subsection{Proof of Theorem 2}
\label{ACAppendix}

\textbf{Theorem} 2: A sketch with one bucket array is equivalent to a linear autoencoder: the insertion process corresponds to an encoding phase $I = \left( { A }^{ T }X \right)$, while the query phase corresponds to  a decoding phase $\hat { X } =A\cdot I$.

\begin{proof}
For each incoming key-value pair ($\kappa(i)$, $X_i$),  the sketch selects only one bucket indexed by a variable $j$ by hashing $\kappa(i)$ with a hash function,  and inserts $X_i$ to this bucket by incrementing the bucket's counter by $X_i$. Equivalently, we set  the $i$-th row vector of $A$ to a 0-1 vector, where only the $j$-th entry $A(i,j) = 1$, and other entries are all set to zeros. Consequently, we can equivalently transform this insertion choice as $I = I + X_i \cdot A(i,:)$.  The insertion process for all key-value pairs can be represented as an encoding phase: $I=\left( { A }^{ T }X \right) $. 

For the query process of a key $\kappa(i)$, the sketch selects the same bucket indexed by $j$ by hashing $\kappa(i)$ with the same hash function as the insertion process, and then returns the bucket's counter $I(j)$ as the approximated value for  $X_i$. Similarly, based on the $i$-th row vector of $A$, denoted as $A(i,:)$,  we can equivalently represent the approximated value as ${\hat { X }} _i = A(i,:) \cdot I$. Therefore, the approximated values for all inserted keys can be calculated as a decoding phase:  $\hat { X } =A\cdot I $.
\end{proof}

\subsection{K-means Model}
\label{KMAppendix}

Specifically,  the K-means clustering method  minimizes the variance of each cluster by finding a set of $k$ points (called centroids) such that the potential function is minimized
\begin{equation}\label{ObjK}
F\left( S \right) = \sum _{ x\in S }{ \min _{ c\in C }{ { \left\| x-c \right\|  }^{ 2 } }  }  
\end{equation}
, and finally outputs  a list of cluster centers that minimizes the variance of within-cluster values.  We choose the Lloyd's algorithm to optimize  Eq. \eqref{ObjK}, which initializes centroids arbitrarily,  partitions points by the nearest centroid, and updates the centroids of each cluster until convergence.  The training process of the K-means clustering  method  has to repetitively update the cluster centers until convergence.

\subsection{Proof of Theorem 3}
\label{Thm3Appendix}

\textbf{Theorem} 3: $Y_j$ is an unbiased estimator for any variable $ { X }_{ i }^{ j }$.    $Pr\left( \left| Y_j -\mu  \right| \ge a \right) \le \frac { { M }^{ 2 } }{ { a }^{ 2 }{ n_j }^{ 2 } } $  for a positive constant $a$.  Moreover, \newline  $Pr\left( \left| Y_j-{ X }_{ i }^{ j } \right| \ge a \right)  \le \frac { { M }^{ 2 } }{ { \left( a-M \right)  }^{ 2 }{ n_j }^{ 2 } } $, for positive constants $a$. 

\begin{proof}
The expectation of  $Y_j$ is exactly the expectation of the variables.
$E\left[ Y_j \right]  = \frac { 1 }{ n_j } E\left[ \sum _{ i }^{  }{ { X }_{ i }^{ j } }  \right] =\frac { 1 }{ n_j } \sum _{ i }^{  }{ E\left[ { X }_{ i }^{ j } \right]  } =\mu $ 

Therefore, $Y_j$ is an unbiased estimator for $\left\{ { X }_{ i }^{ j } \right\} $.  Next, we bound the deviation degree of $Y_j$ from its expectation as follows:
\[
\begin{array}{l}
Var\left[ Y_j \right] =E\left[ { \left( Y_j-\mu  \right)  }^{ 2 } \right] =E\left[ { \left( \frac { { X }^{ j } }{ n_j } -\mu  \right)  }^{ 2 } \right] =\\
E\left[ { \frac { 1 }{ n_j } \left( \sum _{ i }^{  }{ \left( { X }_{ i }^{ j }-\mu  \right)  }  \right)  }^{ 2 } \right] \\
\le \frac { 1 }{ { n_j }^{ 2 } } E\left[ { M }^{ 2 } \right] =\frac { { M }^{ 2 } }{ { n_j }^{ 2 } } 
\end{array}
\]

By Chebyshev's inequality, we bound the range of $Y_j$ as :
$
Pr\left( \left| Y_j-\mu  \right| \ge a \right) \le \frac { Var\left[ Y_j \right]  }{ { a }^{ 2 } } \le \frac { { M }^{ 2 } }{ { a }^{ 2 }{ n_j }^{ 2 } } 
$

Second, the following inequality holds:
\[
\begin{array}{l}
Pr\left( \left| Y_j-{ X }_{ i }^{ j } \right| \ge a \right) =Pr\left( \left| Y_j-\mu +\mu -{ X }_{ i }^{ j } \right| \ge a \right) \\
 \le Pr\left( \left( \left| Y_j-\mu  \right| +\left| { X }_{ i }^{ j }-\mu  \right|  \right) \ge a \right) \\
  =Pr\left( \left( \left| Y_j-\mu  \right|  \right) \ge a -\left| { X }_{ i }^{ j }-\mu  \right| \right) \\ 
  \le Pr\left( \left| Y_j-\mu  \right| \ge a-M \right) \\ 
  \le \frac { { M }^{ 2 } }{ { \left( a-M \right)  }^{ 2 }{ n_j }^{ 2 } } 
\end{array}
\]
The second inequality holds due to the triangle inequality condition $ (\left| Y_j-\mu +\mu -{ X }_{ i }^{ j } \right| \le \left| Y_j-\mu  \right| +\left| { X }_{ i }^{ j }-\mu  \right| )$.
\end{proof}

\co{
\subsection{Cuckoo Filter Details}

A Cuckoo Filter maintains an array of buckets,  where each bucket has several slots, each of which stores one digest calculated as the hash value of the key. A digest  compactly represents the identifier of an incoming item.  The Cuckoo Filter inserts items based on cuckoo hashing, which uses multiple hash functions to map each item to candidate buckets. For an incoming item, if one of candidate buckets has empty slots, then we calculate the digest of this item and put the digest to one empty slot; otherwise, we pick one nonempty slot and displace its digest to its alternative candidate bucket,  then we put the new item to this slot. The displaced digest may further ``kick out'' other digests until no displacements of existing digests, or reaching a maximum number of  displacements. 
}

\end{document}